\documentclass[pdftex,twocolumn,epjc3]{svjour3}
\usepackage{graphicx}% Include figure files
\usepackage{dcolumn}% Align table columns on decimal point
\usepackage{bm}% bold math        % twocolumn
\usepackage{amsmath,multirow}
\usepackage{seqsplit}
\usepackage{amssymb}
\usepackage{subfigure}
\usepackage{mathrsfs}
\usepackage[section]{placeins}
\RequirePackage[T1]{fontenc}
\smartqed  % flush right qed marks, e.g. at end of proof
\journalname{EPJC}
\begin{document}

\title{Shadows of three black holes in static equilibrium configuration}

\author{%\thanksref{editym@163.com}
%        \and
Dan Li$^{1}$ \and
Yuxiang Zuo$^{1}$ \and
Shiyang Hu$^{1,a,\dag}$ \and
Chen Deng$^{2}$ \and
Yu Wang$^{3}$ \and
Wenfu Cao$^{4}$}
%\thankstext[$\star$]{t1}{Thanks to the title}
%\thankstext{e1}{e-mail: danli$@$usc.edu.cn}
\thankstext{e2}{e-mail: husy$\_$arcturus$@$163.com}
%\thankstext{e3}{e-mail: dengchen$@$smail.nju.edu.cn}

\institute{School of Mathematics and Physics, University of South China, Hengyang 421001, People's Republic of China \label{addr1}
      \and School of Astronomy and Space Science, Nanjing University, Nanjing 210023, People's Republic of China \label{addr2}
      \and School of Physical Science and Technology, Guangxi University, Nanning 530004, People's Republic of China \label{addr3}
      \and School of Physics and Technology, University of Jinan, Jinan 250022, People's Republic of China \label{addr4}}
%         \emph{Present Address:} Street, City, Country\label{addr3}
%}

\date{Received: date / Accepted: date}

\maketitle

\begin{abstract}
In this paper, we employ a ray-tracing algorithm to simulate the shadows of three equal-mass black holes in static equilibrium across a wide parameter space. We find that the shadows consist of a larger primary shadow and several distorted, eyebrow-like secondary shadows. The boundaries of these profiles exhibit self-similar fractal structures, which can be attributed to the photon chaotic scattering. In certain parameter spaces, we also observe the ring-like shadows, with the ring diameter associated with the spacing of black holes. Furthermore, when the black holes approach each other sufficiently, their shadows can merge into a standard disk, suggesting a shadow degeneracy between closely arranged triple black holes and a single massive, spherically symmetric black hole. The shadow features of the triple black holes revealed in this study have potential implications for analyzing the shadow formation mechanisms, as well as the gravitational lensing during the merger and inspiral of black holes.
\end{abstract}
\tableofcontents
\section{Introduction}
The images of the supermassive black holes at the center of Virgo (M87$^{*}$) and the Milky Way (Sgr A$^{*}$), released by the Event Horizon Telescope (EHT) Collaboration \cite{Akiyama et al. (2019a),Akiyama et al. (2022a)}, represent a monumental milestone in astronomical exploration. These historic observations not only provide definitive confirmation of black holes' existence but also open a new avenue for studying black hole physics, gravitational theories, and high-energy astrophysics. Consequently, a surge of research has emerged on black hole shadow \cite{Gralla et al. (2019),Chael et al. (2021),Li and He (2021),Peng et al. (2021),Hu et al. (2022),Zeng et al. (2022),Hou et al. (2022),Heydari-Fard et al. (2023a),Hu et al. (2023),Meng et al. (2023),Yang et al. (2023),Gao et al. (2023),Hu et al. (2024),Sui et al. (2024),He et al. (2025a),He et al. (2025b),Zeng et al. (2025b),He et al. (2025c),Hu et al. (2025),Hou et al. (2025)}, gravitational lensing \cite{Lu and Xie (2021),Gao and Xie (2021),Kuang et al. (2022),Qi et al. (2023),He et al. (2024)}, holographic images of the Einstein ring \cite{Aslam et al. (2024),Zeng et al. (2024),Zeng et al. (2025a)}, and accretion disk radiation \cite{Heydari-Fard and Sepangi (2021),Liu et al. (2021),Boshkayev et al. (2022),Heydari-Fard et al. (2023b),Boshkayev et al. (2024)}. However, current observational limitations constrain the EHT's targets to only M87$^{*}$ and Sgr A$^{*}$, and the astrophysical information extractable from these images remains restricted.

The next-generation Event Horizon Telescope (ngEHT) project aims to significantly extend interferometric baselines by incorporating space-borne antennas, thereby achieving unprecedented imaging resolution \cite{Johnson et al. (2023),Peter et al. (2023),Shavelle and Palumbo (2024)}. This technological leap will not only enable more precise measurements of shadow geometry in M87$^{*}$ and Sgr A$^{*}$, but also potentially expand observational capabilities to other black holes, compact objects, and even multiple black hole systems. Notably, observational evidence has already hinted at the existence of supermassive binary black holes and black hole clusters \cite{Sudou et al. (2003),Sundelius et al. (1988),Vitral and Mamon (2021)}. Investigating the imaging signatures of such systems could provide the scientific community with direct insights into the gravitational interactions, structural configurations, and evolution of multiple black hole systems.

In fact, numerous studies have investigated the shadow characteristics of binary black holes. Nitta et al. employed the Kastor-Traschen solution to model the contracting spacetime of a binary black hole system with a cosmological constant, numerically simulating the shadow evolution of colliding black holes driven by spacetime contraction \cite{Nitta et al. (2011)}. The results demonstrated that as the binary black holes converge, their individual shadows undergo deformation and develop distinctive eyebrow-like secondary shadow. Building on this work, Yumoto et al. conducted further investigations by deriving analytical expressions for photon motion in binary black hole spacetimes, thereby elucidating the formation mechanism of the eyebrow-like shadow. Through extensive numerical simulations across a broader parameter space, they additionally identified ring-shaped shadow structures \cite{Yumoto et al. (2012)}. The authors in \cite{Bohn et al. (2015)} numerically simulated photon trajectories in binary black hole spacetime. By employing a fisheye projection technique, they visually reconstructed the black hole shadows and revealed self-similar fractal structures at the shadow boundaries. To further investigate how black hole spins influence binary black holes' shadow, Cunha et al. employed a ray-tracing algorithm to simulate shadows based on the double-Kerr solution. Their work demonstrated that spin alignment between the two black holes significantly modifies the resulting image features \cite{Cunha et al. (2018)}. Meanwhile, the study in \cite{Chernov (2024)} focused specifically on photon ring formation in binary black hole system and its dependence on system parameters. Additional studies on binary black hole shadows can be found in \cite{Shipley and Dolan (2016),Moreira et al. (2025)}. Overall, the shadow characteristics of such system have been nearly exhaustively explored. This comprehensive understanding naturally leads us to pose the next fundamental question: What distinctive shadow features would manifest in a triple black hole system? Addressing this unexplored problem constitutes the primary motivation of the present study.

The triple black hole system resembles the classic and enigmatic three-body problem---without any constraints, the geodesic equations for particles in its spacetime become non-integrable due to violations of Liouville's integrability theorem. This poses significant challenges for studying such system. Moreover, constructing an exact analytical metric describing triple black holes is virtually impossible. Fortunately, when additional constraints are imposed on the triple black hole system, the target spacetime can be described by the Majumdar-Papapetrou solution \cite{Majumdar (1947),Papaetrou (1947)}. In this scenario, the gravitational attraction between black holes is theoretically balanced by electrostatic repulsion, thereby allowing the three black holes to maintain a static equilibrium configuration. Hartle and Hawking demonstrated that when the black holes in the Majumdar-Papapetrou solution are extremal Reissner-Nordstr\"{o}m (RN) black holes (i.e., $Q=M$, where $Q$ and $M$ represent the charge and mass of the black hole, respectively), the solution satisfies the source free Einstein-Maxwell field equations \cite{Hartle and Hawking (1972)}. Moreover, as the Majumdar-Papapetrou metric is an exact analytical solution, it provides a tractable framework for investigating the shadows of triple black holes.

The remainder of this paper is organized as follows. Section 2 provides a concise review of the Majumdar-Papapetrou metric and derives the equations of motion for photons in this spacetime. Section 3 introduces the essential ray-tracing algorithm for simulating black hole shadows. In Section 4, we present the shadows of triple black holes across different parameter spaces, with detailed analysis of their distinctive features. The final section summarizes our findings. Throughout this work, we adopt geometric units with $G=c=M_{s}=1$, where $M_{s}$ denotes the total mass of three black holes.
\section{Spacetime and geodesics}
Theoretically, maintaining stable configurations of three static black holes in vacuum is extraordinarily challenging. However, such equilibrium becomes possible when all three black holes carry like-sign charges. In this scenario, the gravitational attraction between black holes is precisely counterbalanced by the electrostatic repulsion. The Majumdar-Papapetrou solution originates from this principle, describing spacetimes containing $n$ static equilibrium black holes. In isotropic coordinates, the Majumdar-Papapetrou metric reads as \cite{Chernov (2024),Majumdar (1947),Papaetrou (1947),Wang et al. (2023)}
\begin{eqnarray}\label{1}
\textrm{d}s^{2} = - \frac{\textrm{d}t^{2}}{\mathcal{U}^{2}} + \mathcal{U}^{2}\left(\textrm{d}x^{2}+\textrm{d}y^{2}+\textrm{d}z^{2}\right).
\end{eqnarray}
Here, $\mathcal{U}$ denotes the spacetime gravitational potential, given by
\begin{eqnarray}\label{2}
\mathcal{U} = 1 + \sum_{i=1}^{n}\frac{m_{i}}{r_{i}},
\end{eqnarray}
where $m_{i}$ represents the mass of the $i$th black hole, and $r_{i}$ takes the form
\begin{eqnarray}\label{3}
r_{i} = \sqrt{\left(x-x_{i}\right)^{2}+\left(y-y_{i}\right)^{2}+\left(z-z_{i}\right)^{2}}.
\end{eqnarray}
Here, $(x_{i},y_{i},z_{i})$ denote the spatial coordinates of the black holes. Remarkably, these black holes maintain static equilibrium regardless of their spatial configuration. For our subsequent analysis, we set $n=3$ and fix equal masses for all three black holes, i.e., $m_{1}=m_{2}=m_{3}=M_{s}/3$. Consequently, we have the metric potential for three static black hole spacetime in geometric units,
\begin{eqnarray}\label{4}
\mathcal{V} = 1 + \sum_{i=1}^{3}\frac{1}{3r_{i}}.
\end{eqnarray}
Hence, we have
\begin{eqnarray}\label{5}
\textrm{d}s^{2} = -\frac{\textrm{d}t^{2}}{\mathcal{V}^{2}}+\mathcal{V}^{2}\left(\textrm{d}x^{2}+\textrm{d}y^{2}+\textrm{d}z^{2}\right).
\end{eqnarray}

In spacetime \eqref{5}, the geodesic equations for photons are governed by canonical equations in the Hamiltonian frame. To this end, we first define the Lagrangian as
\begin{eqnarray}\label{6}
\mathcal{L} = \frac{1}{2}\left[-\frac{\dot{t}^{2}}{\mathcal{V}^{2}}+\mathcal{V}^{2}\left(\dot{x}^{2}+\dot{y}^{2}+\dot{z}^{2}\right)\right],
\end{eqnarray}
where the overdot denotes the derivatives of generalized coordinates with respect to an affine parameter $\lambda$, i.e., $\dot{t} = \textrm{d}t/\textrm{d}\lambda$. Then, through the Euler-Lagrange equations, we introduce the conjugate momenta
\begin{equation}\label{7}
p_{t} = -\frac{1}{\mathcal{V}^{2}}\dot{t},
\end{equation}
and
\begin{equation}\label{8}
p_{\alpha} = \mathcal{V}^{2}\dot{\alpha},
\end{equation}
where $\alpha$ runs from $x$ to $z$. Applying the Legendre transformation $\mathcal{H}=\vec{p}\cdot\vec{v}-\mathcal{L}$ yields the Hamiltonian for photon motion in spacetime \eqref{5},
\begin{eqnarray}\label{9}
\mathcal{H} = \frac{1}{2}\left[-\mathcal{V}^{2}p_{t}^{2}+\frac{1}{\mathcal{V}^{2}}\left(p_{x}^{2}+p_{y}^{2}+p_{z}^{2}\right)\right].
\end{eqnarray}
For photons, the Hamiltonian constraint is $\mathcal{H} = 0$, while for massive particles it becomes $\mathcal{H} = -1/2$. We note that the Hamiltonian \eqref{9} does not explicitly depend on the coordinate $t$. In fact, due to the time-translational invariance, $p_{t}$ is a motion constant during the propagation of light rays, with its relation to the photon's specific energy $E$ given by $p_{t} = -E$. From the Hamiltonian constraint, the photon's $p_{t}$ can be determined given its three-momentum components,
\begin{eqnarray}\label{10}
p_{t}^{2} = \frac{p_{x}^{2}+p_{y}^{2}+p_{z}^{2}}{\mathcal{V}^{4}},
\end{eqnarray}
which is important in the determination of initial conditions during ray-tracing method described in the following section.

According to canonical equations, the propagation of light rays is governed by
\begin{eqnarray}\label{11}
\dot{t} = \frac{\partial\mathcal{H}}{\partial p_{t}},\ \dot{x} = \frac{\partial\mathcal{H}}{\partial p_{x}},\ \dot{y} = \frac{\partial\mathcal{H}}{\partial p_{y}},\ \dot{z} = \frac{\partial\mathcal{H}}{\partial p_{z}},
\end{eqnarray}
\begin{eqnarray}\label{12}
\dot{p_{t}} = -\frac{\partial\mathcal{H}}{\partial t},\ \dot{p_{x}} = -\frac{\partial\mathcal{H}}{\partial x},\ \dot{p_{y}} = -\frac{\partial\mathcal{H}}{\partial y},\ \dot{p_{z}} = -\frac{\partial\mathcal{H}}{\partial z}.
\end{eqnarray}
Given the initial coordinates and momentum of a light ray $(t,x,y,z,p_{t},p_{x},p_{y},p_{z})$, we employ a fifth- and sixth-order Runge-Kutta-Fehlberg integrator with adaptive step sizes to integrate Eqs. \eqref{11} and \eqref{12}, the photon path can be obtained.
\section{Ray-tracing method}
The study of black hole shadows necessitates the application of ray-tracing algorithms. While numerous sophisticated ray-tracing methods exist in the scientific literature \cite{Vincent et al. (2011),Bambi (2012),Cunha et al. (2016),Hu et al. (2021),Velsquez-Cadavid et al. (2022),Huang et al. (2024)}, this investigation adopts the framework of \verb"ODYSSEY" \cite{Pu et al. (2016)}. We employ a collinear three static black hole system to elucidate the ray-tracing procedure. As illustrated in Fig. 1, the spacetime containing three static black holes is parameterized using a local coordinate system $xyz$, with the three black holes designated as $m_{1}$, $m_{2}$, and $m_{3}$, respectively. Here, $m_{2}$ is positioned at the coordinate origin $o$, while the other two black holes are symmetrically distributed along the $y$-axis relative to $m_{2}$. The blue sphere in the figure represents the observer, whose local coordinate system is described by $x^{\prime}y^{\prime}z^{\prime}$. Here, $\overline{oo^{\prime}}$ coincides with $\overline{o^{\prime}z^{\prime}}$. We further define the observation angle $\Theta$ as the angle between $\overline{oo^{\prime}}$ and the $z$-axis, and the observation azimuth $\Phi$ as the angle between the projection of the observer onto the $\overline{xoy}$ plane and the $x$-axis. The observed image is described by the $\overline{x^{\prime}o^{\prime}y^{\prime}}$ plane. Notably, for the black hole configuration depicted in Fig. 1---where three black holes are aligned along the $y$-axis---the resulting image depends on both the azimuthal angle $\Phi$ and the inclination $\Theta$.
\begin{figure}%[tbph]
\center{
\includegraphics[width=6cm]{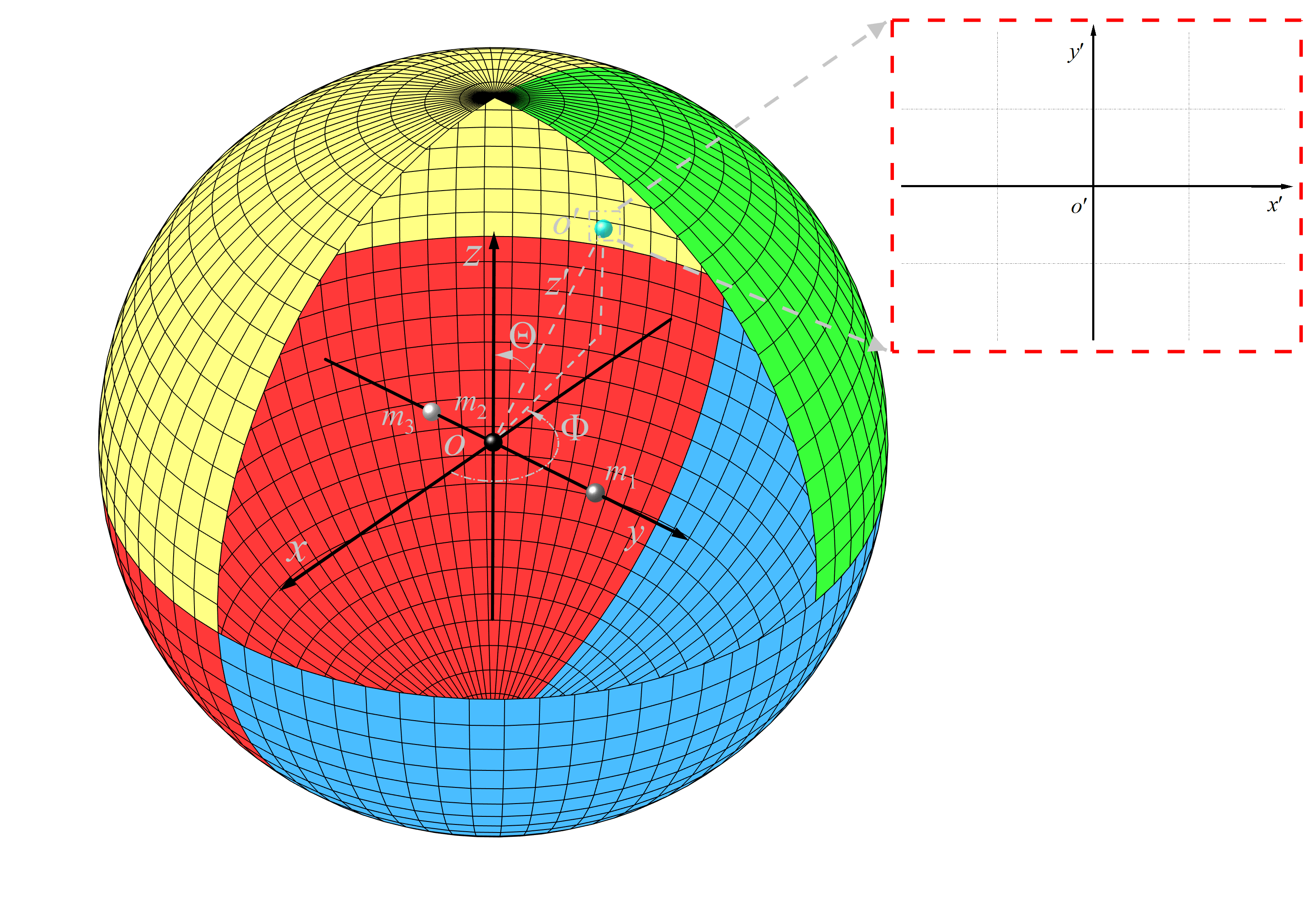}
\caption{Ray-tracing map and light source configuration in the triple static black holes' spacetime. The spatial configuration of the three black holes, $m_{1}$, $m_{2}$, and $m_{3}$, is described by the coordinate system $xyz$. The observer is represented by a blue sphere, with his local frame and field of view defined by the coordinate system $x^{\prime}y^{\prime}z^{\prime}$ and $\overline{x^{\prime}o^{\prime}y^{\prime}}$ plane (see zoomed-in view), respectively. $\Phi$ and $\Theta$ are azimuthal angle and inclination angle of the observer, respectively. We employ a spherical light source with a radius of $1500$ $M_{s}$ encompassing both the three black holes and the observer. Light rays emitted from the observation plane are captured by either $m_{1}$, $m_{2}$, or $m_{3}$, with their corresponding pixel coordinates filled in dark gray, black, or light gray, respectively. For rays that escape capture and subsequently strike the spherical light source, their pixel coordinates are color-coded according to intersection location: yellow ($z>0$, $y<0$), green ($z>0$, $y>0$), red ($z<0$, $y<0$), and blue ($z<0$, $y>0$).}}\label{fig1}
\end{figure}

To visualize the black hole shadows more intuitively, we employ a celestial light source with radius $r_{\textrm{source}} = 1500$ $M_{s}$ enveloping both the observer and the three black holes, as represented by the colored sphere in Fig. 1. Notably, the sphere consists of four distinct colors, while the three black holes are depicted in dark gray, black, and light gray, respectively. Consequently, when light rays emitted from the observation plane $\overline{x^{\prime}o^{\prime}y^{\prime}}$ are captured by a black hole, their corresponding pixels are filled with either dark gray, black, or light gray. For light rays that graze the black holes and subsequently hit the celestial sphere, their observation coordinates are marked with green, yellow, red, or blue, according to the hitting location.

Subsequently, we trace light rays emitted from each pixel on the observation plane within the black holes' local coordinate system. This requires determining the initial conditions of the light rays in the $xyz$ frame. Through coordinate transformation, we have \cite{Pu et al. (2016),Younsi et al. (2016)}
\begin{eqnarray}\label{13}
x = \mathscr{T}\cos\Phi - x^{\prime}\sin\Phi,
\end{eqnarray}
\begin{equation}\label{14}
y = \mathscr{T}\sin\Phi + x^{\prime}\cos\Phi,
\end{equation}
and
\begin{equation}\label{15}
z = \left(r_{\textrm{obs}} - z^{\prime}\right)\cos\Theta + y^{\prime}\sin\Theta,
\end{equation}
in which $\mathscr{T}$ takes the form
\begin{equation}\label{16}
\mathscr{T} = \left(r_{\textrm{obs}}-z^{\prime}\right)\sin\Theta-y^{\prime}\cos\Theta.
\end{equation}
Here, $r_{\textrm{obs}}$ denotes the observation distance, and we fix it at $1000$ $M_{s}$.

Since spacetime \eqref{5} is asymptotically flat, we can reasonably assume that light rays hit the observation plane orthogonally. In other word, the initial velocity of light rays satisfies $(\dot{x^{\prime}},\dot{y^{\prime}},\dot{z^{\prime}})=(0,0,1)$. Substituting this condition into the differential forms of \eqref{13}-\eqref{15} yields:
\begin{eqnarray}\label{17}
\dot{x} = -\sin\Theta\cos\Phi,
\end{eqnarray}
\begin{equation}\label{18}
\dot{y} = -\sin\Theta\sin\Phi,
\end{equation}
\begin{equation}\label{19}
\dot{z} = -\cos\Theta.
\end{equation}
From these, we derive the photon's three-momentum using definition \eqref{8}, and subsequently determine $p_{t}$ via \eqref{10}. Thus, we obtain the initial conditions for light rays corresponding to each pixel on the observation plane. According to the canonical equations \eqref{11} and \eqref{12}, the fate of each light ray can be precisely determined. Those rays that plunge into the black holes contribute to the black hole shadow.
\section{Results}
We focus on the shadows of triple black holes in two configurations: (i) A collinear configuration where all three black holes lie on a straight line. Here, black hole $m_{2}$ is positioned at the coordinate origin, while the other two black holes are symmetrically distributed along the $y$-axis at distances $l$ from $m_{2}$. Their coordinates are $(0,l,0)$, $(0,0,0)$, and $(0,-l,0)$, respectively. (ii) An equilateral triangle configuration with side length $l$, where $m_{2}$ is located at $(\sqrt{3}l/2,0,0)$ and other two black holes are placed on the $y$-axis at $(0,l/2,0)$ and $(0,-l/2,0)$. For both configurations, the free parameters of the system are $\Theta$, $\Phi$, and $l$.
\begin{figure*}%[tbph]
\center{
\includegraphics[width=5cm]{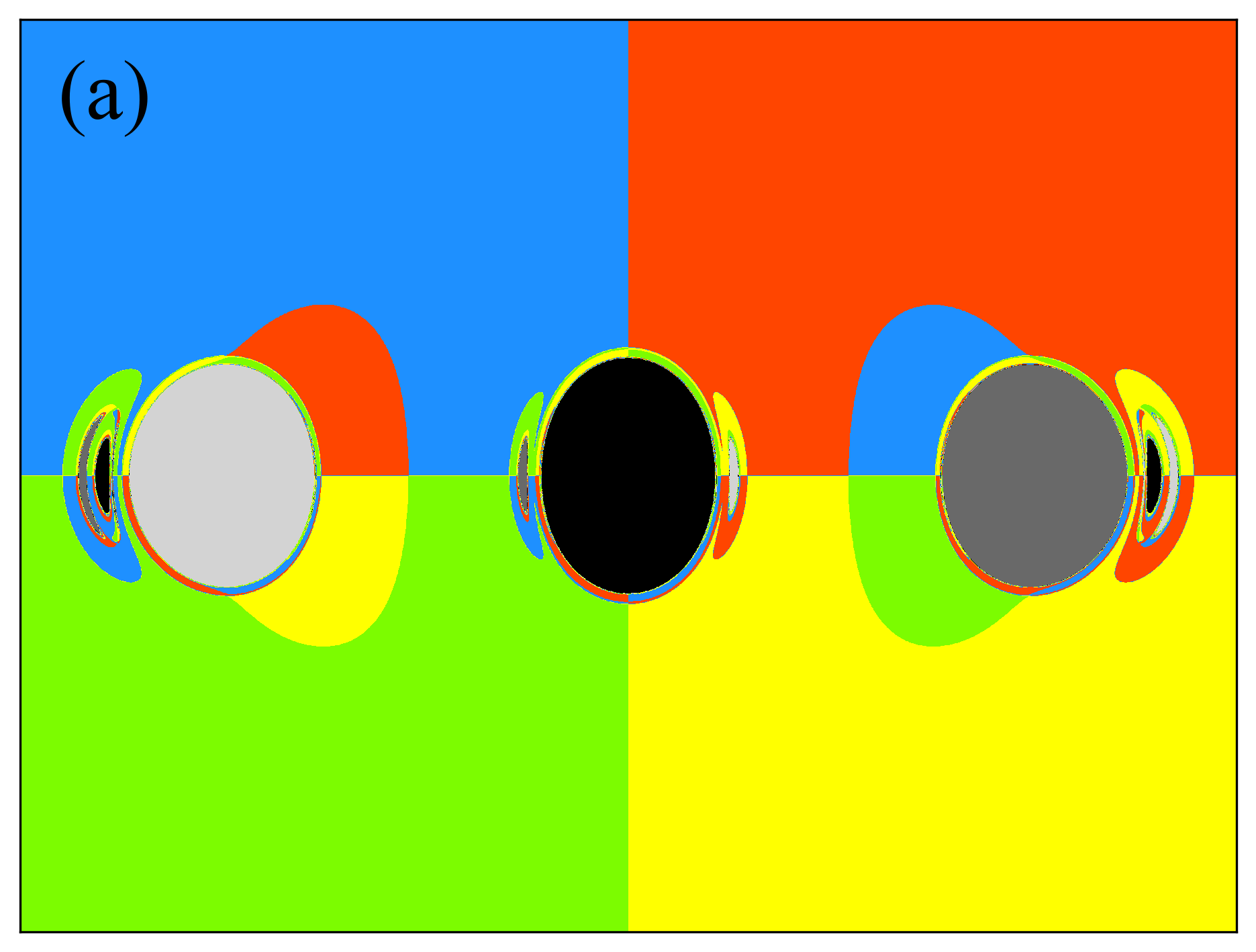}
\includegraphics[width=5cm]{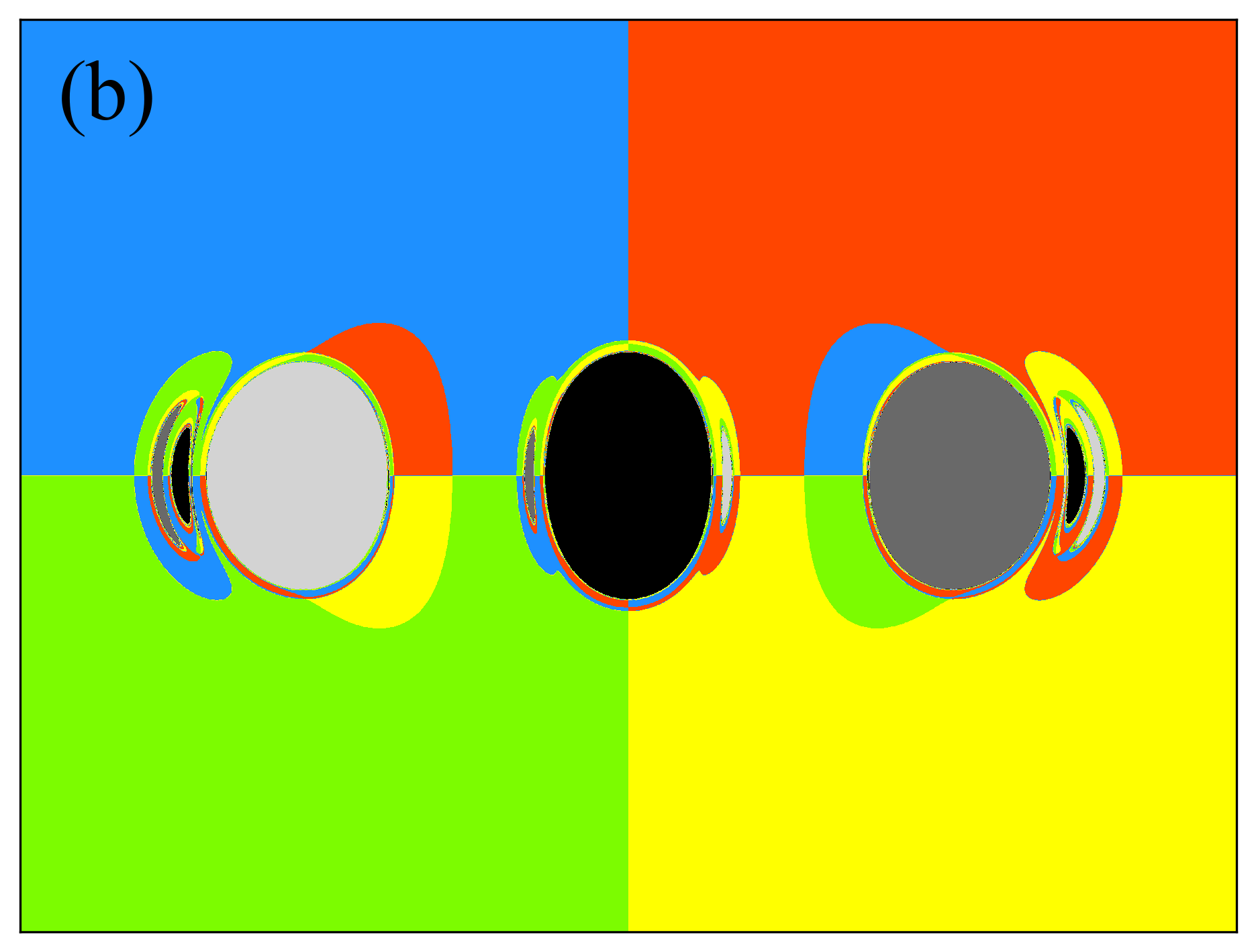}
\includegraphics[width=5cm]{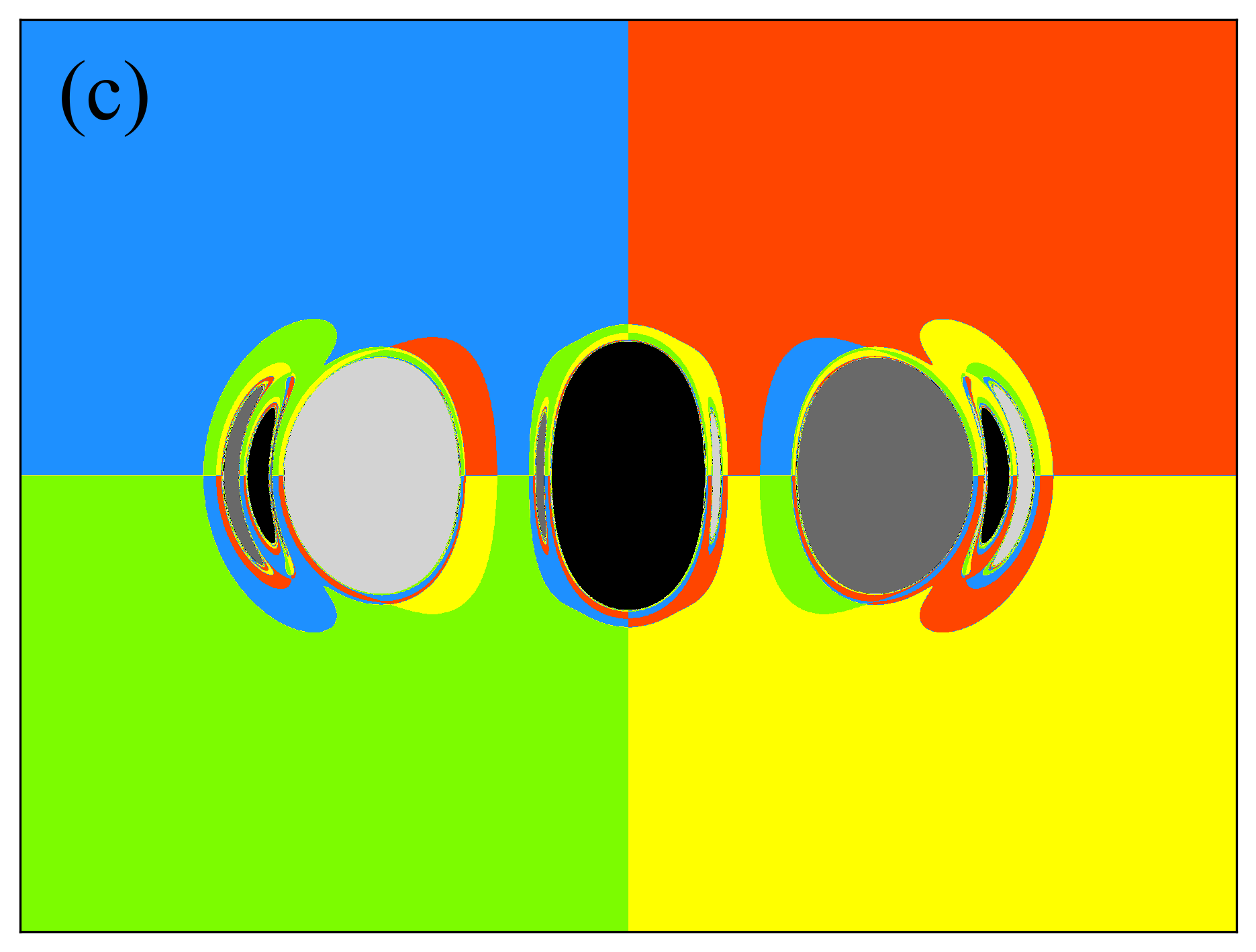}
\includegraphics[width=5cm]{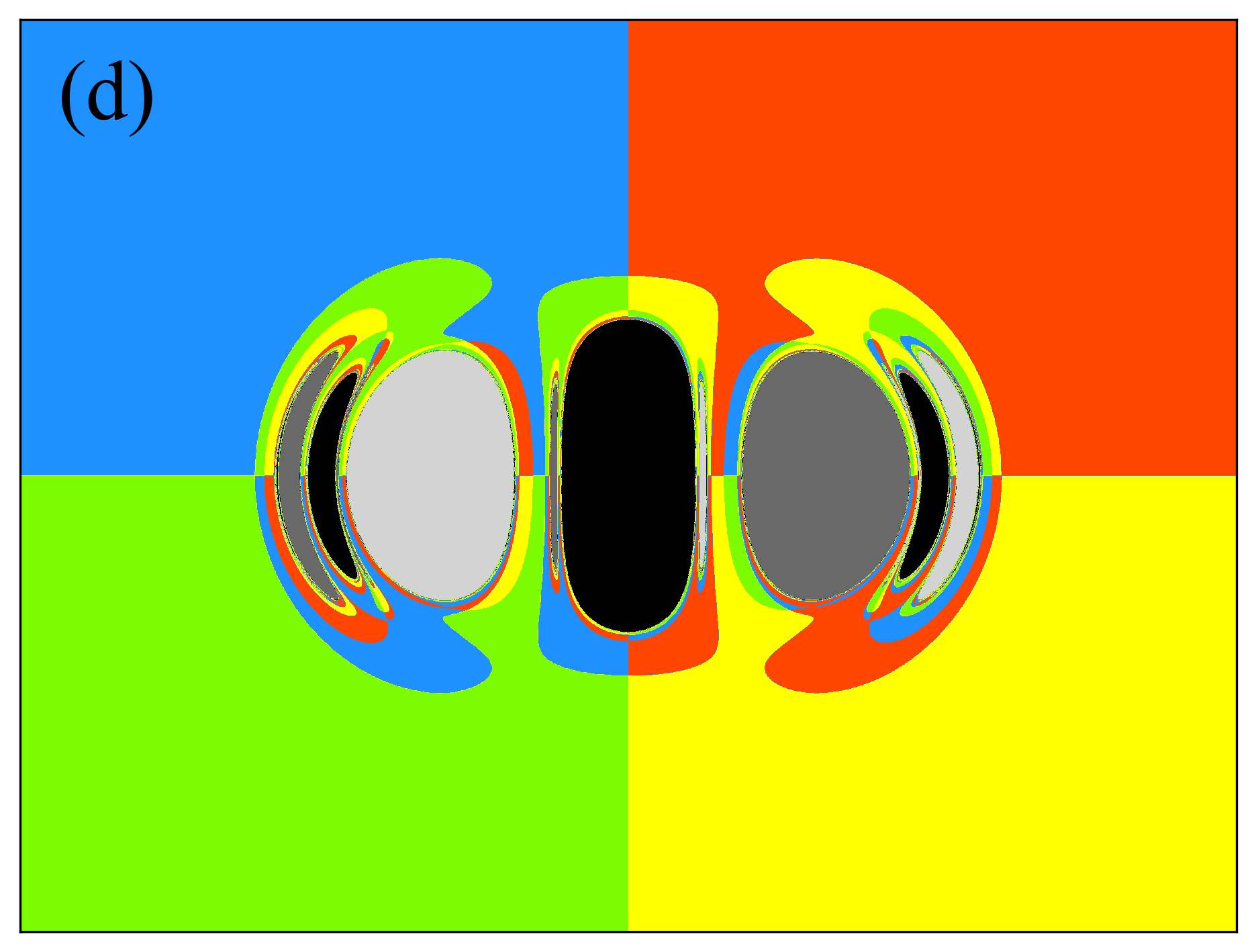}
\includegraphics[width=5cm]{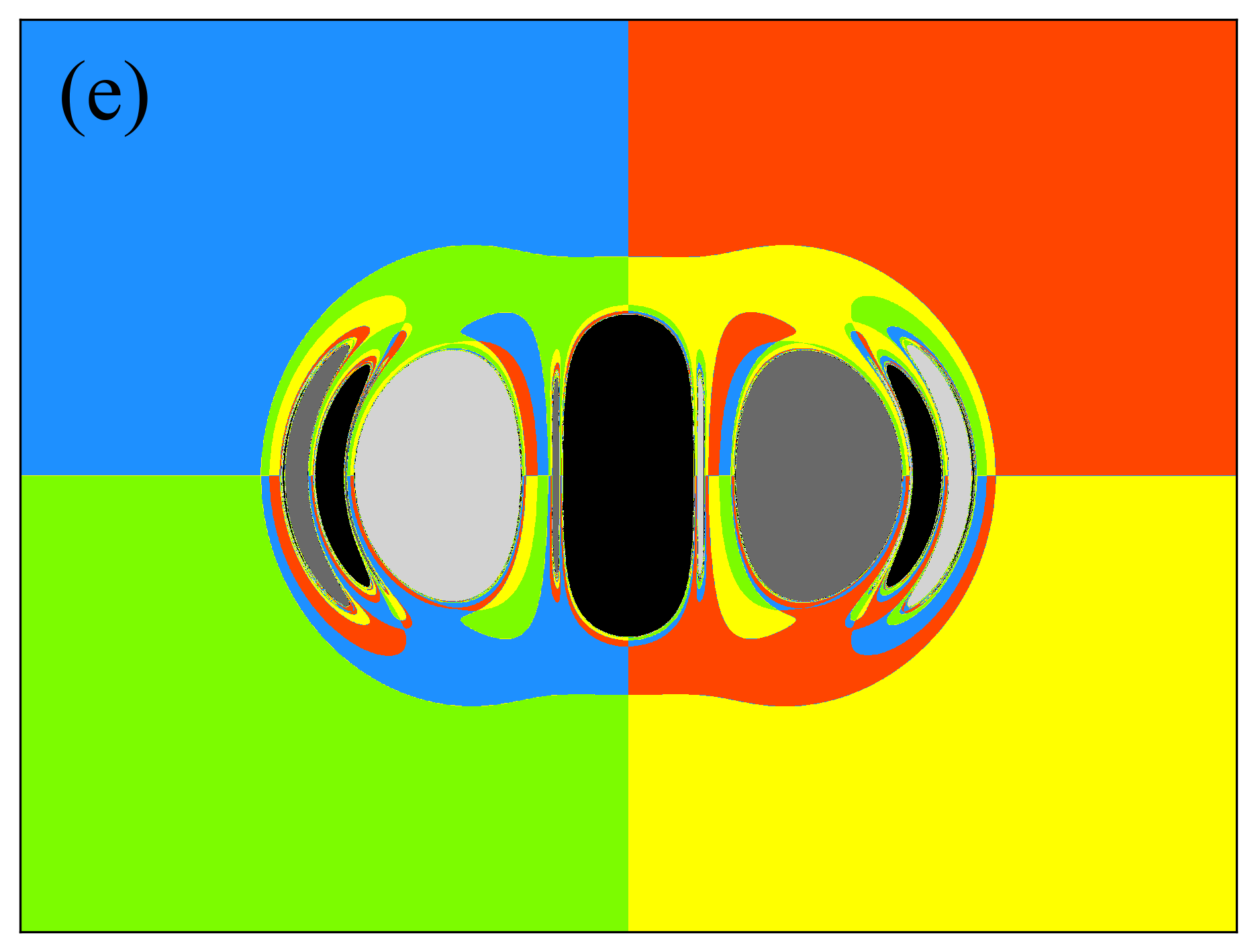}
\includegraphics[width=5cm]{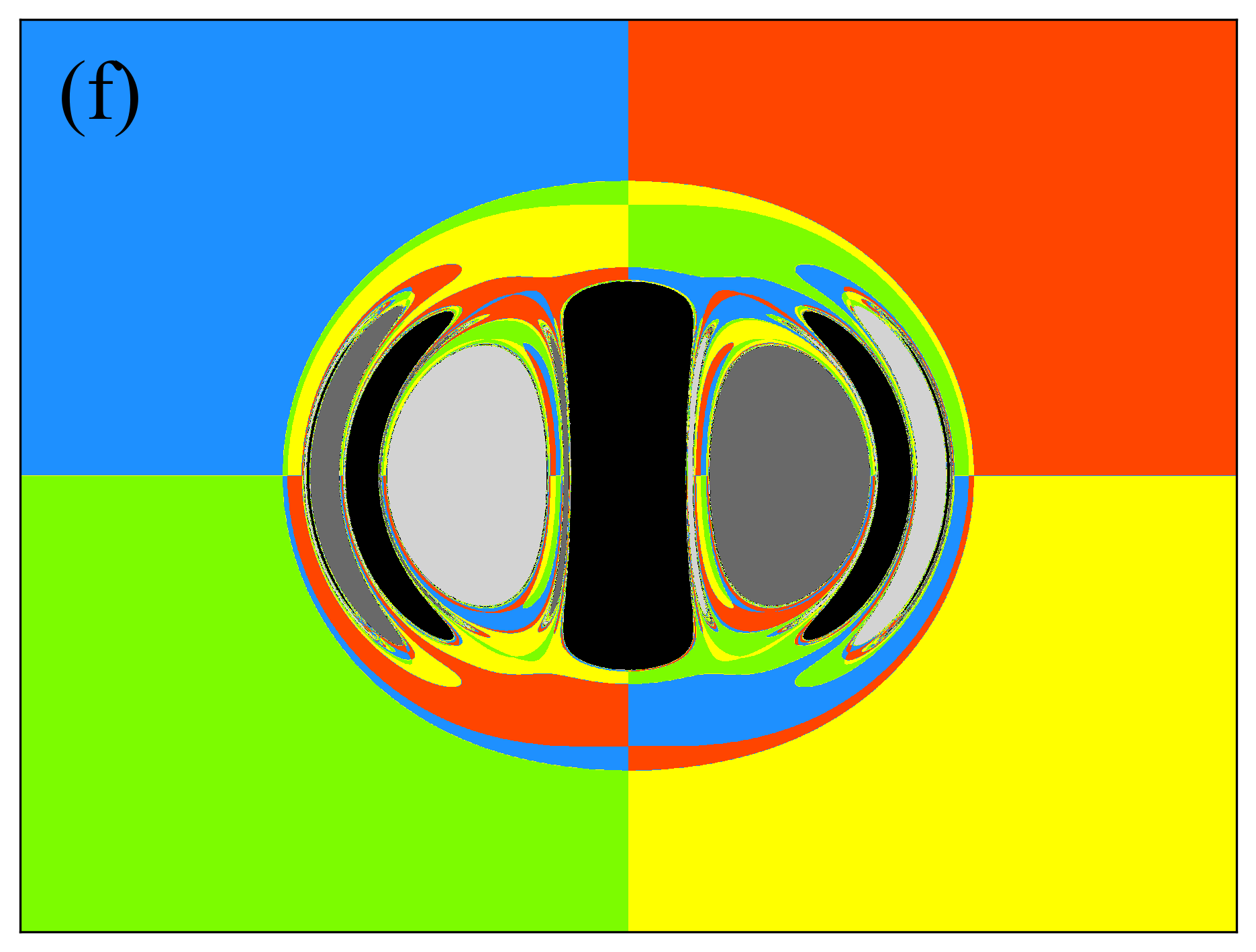}
\includegraphics[width=5cm]{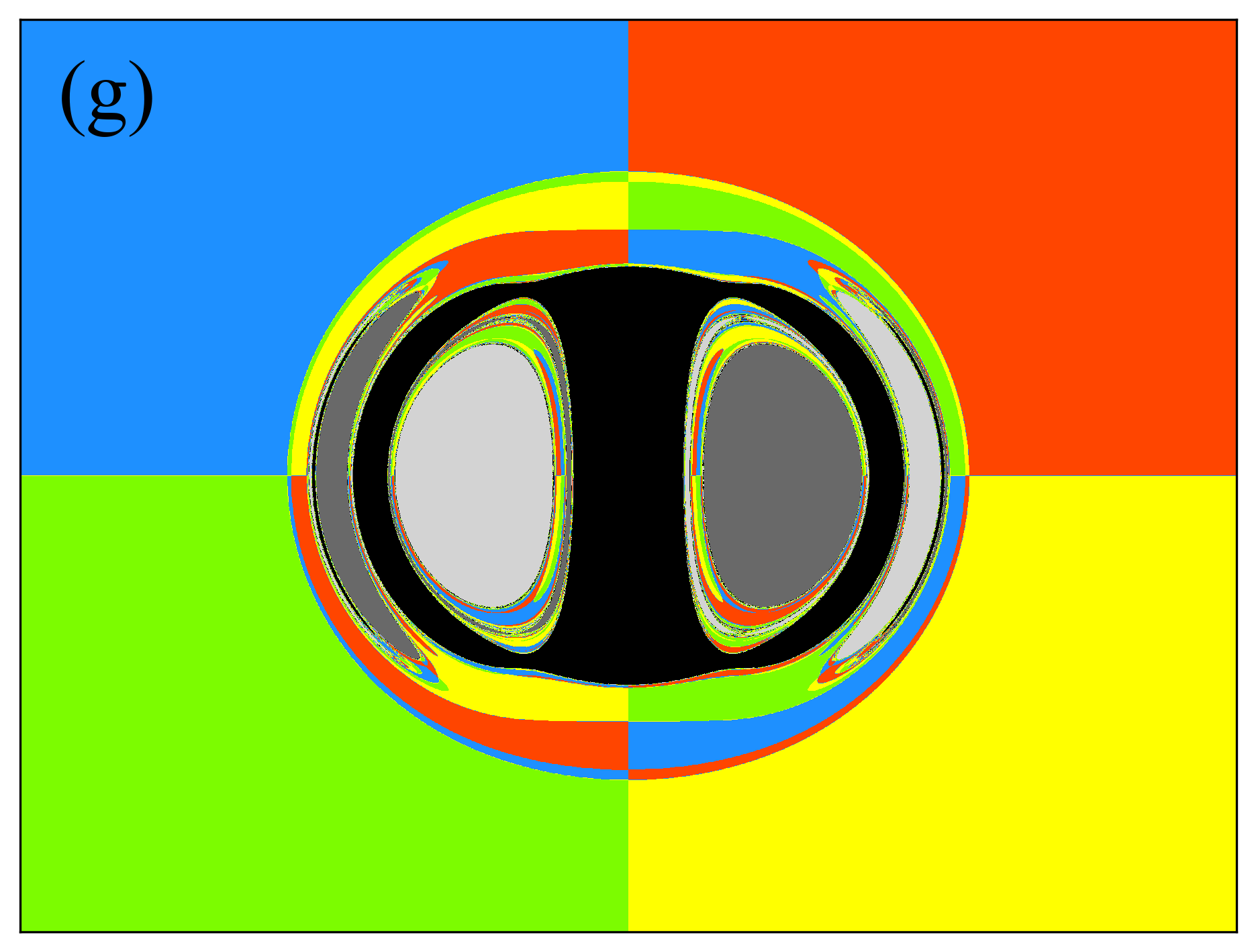}
\includegraphics[width=5cm]{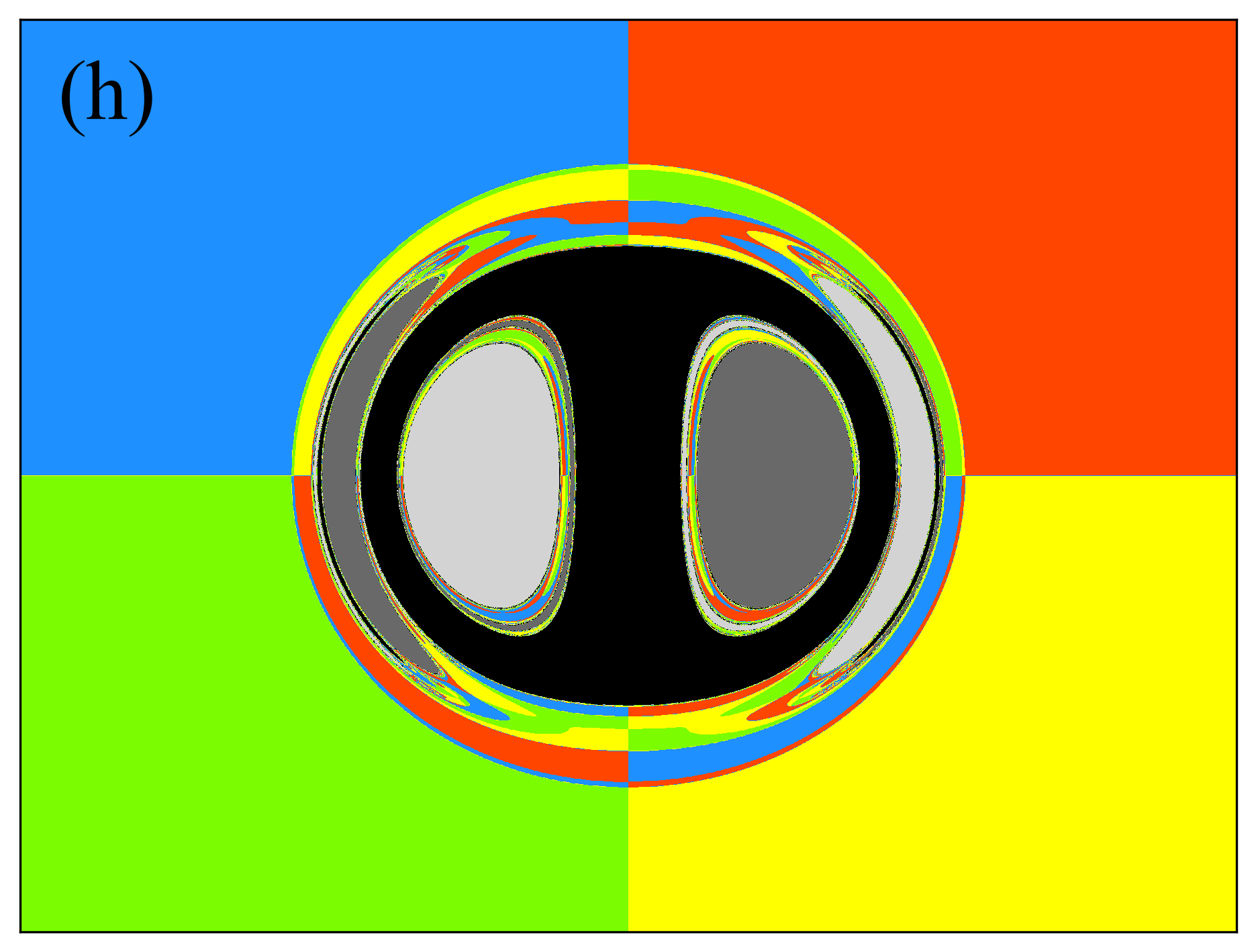}
\includegraphics[width=5cm]{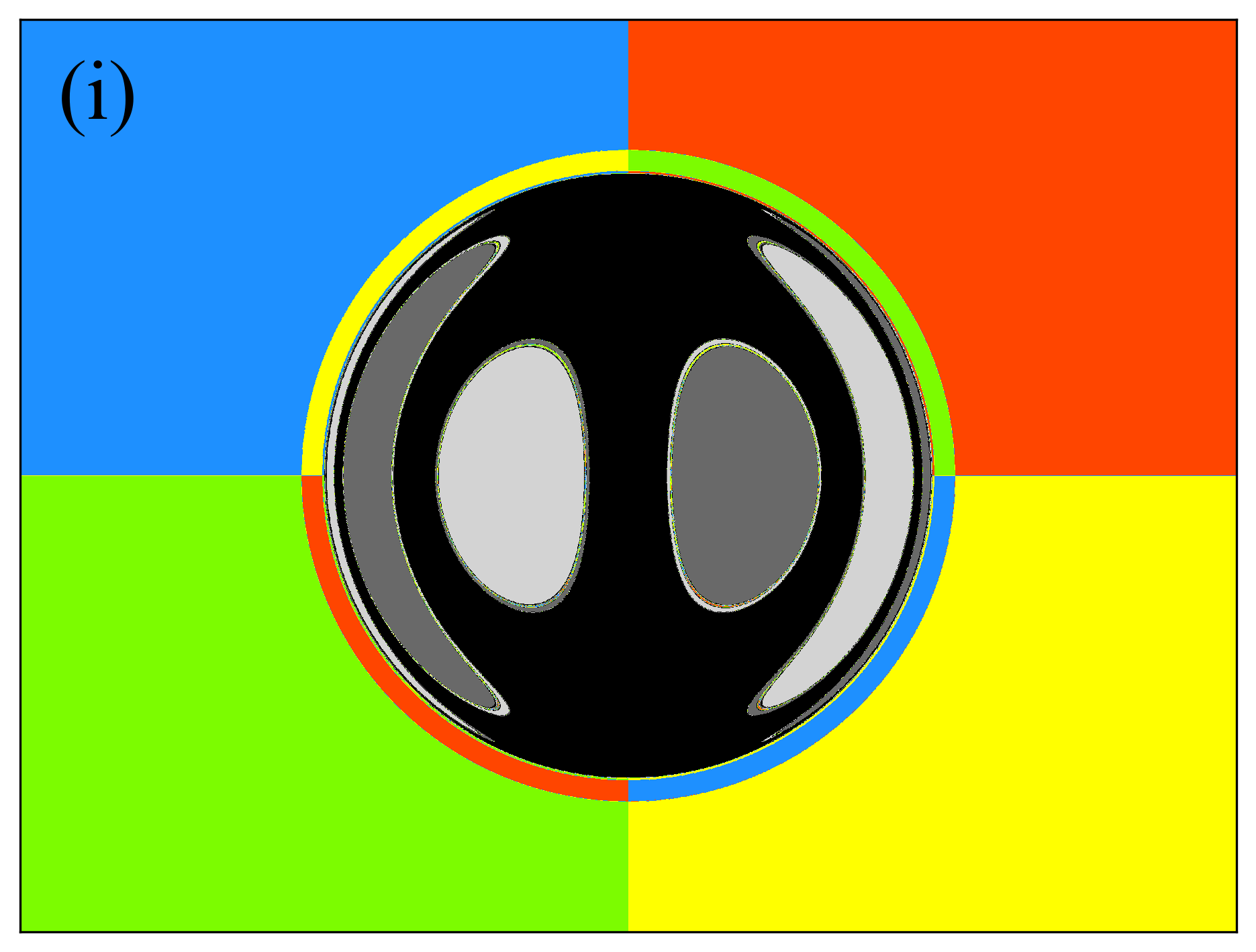}
\caption{Shadows of triple static black holes with varying parameter $l$. From panel (a) to panel (i), $l$ are $4$, $3$, $2$, $1.2$, $1.1$, $0.7$, $0.6$, $0.5$, and $0.1$, respectively. Here, we have the observation angle $\Theta = 90^{\circ}$ and azimuthal angle $\Phi = 0^{\circ}$. In most panels, beyond the three prominent primary shadows, we clearly identify eyebrow-like secondary shadows generated by gravitational lensing. As the black holes approach each other, these secondary shadows exhibit increasing distortion and elongation. When the three black holes become sufficiently close ($l\leqslant 0.1$), their primary shadows merge into a single larger shadow. This behavior suggests a fundamental degeneracy between three tightly packed black holes and a single massive black hole in terms of their photon-capture capabilities.}}\label{fig2}
\end{figure*}
\subsection{Collinear configuration}
We set the observation plane with $x^{\prime} \in [-8,8]$ $M_{s}$ and $y^{\prime} \in [-6,6]$ $M_{s}$ at a resolution of $2400 \times 1800$ pixels. Fig. 2 demonstrates the variation of triple black holes' shadow with $l$, where the dark gray, black, and light gray shadows are cast by $m_{1}$, $m_{2}$, and $m_{3}$, respectively. As clearly observed when the black holes are widely separated ($l=4$), each shadow consists of multiple disjoint components, with the largest structure classified as the ``primary shadow'' and the eyebrow-like features termed ``secondary shadows''. For $m_{2}$, its primary shadow exhibits elongation toward both ends of the $y^{\prime}$-axis, forming an elliptical shape, while its secondary shadows appear symmetrically outside the primary shadows of the other two black holes---this feature originate from photons grazing $m_{1}$ or $m_{3}$ before be captured by $m_{2}$. The $m_{1}$ and $m_{3}$ shadows form mirror images with respect to the $y^{\prime}$-axis. Their primary shadows exhibit a tendency to converge toward the $y^{\prime}$-axis, while the secondary shadows near the primary shadows of the other two black holes. The deformation of the primary shadows of $m_{1}$ and $m_{3}$ arises because the strong gravitational field on the left (right) side of $m_{1}$ ($m_{3}$) disrupts its inherent spacetime symmetry.

As $l$ decreases, the shadows of the three black holes progressively approach each other while undergoing significant morphological changes. Notably for $m_{2}$, its primary shadow elongates from an elliptical shape into a columnar form, ultimately expanding into a perfect circle at $l=0.1$ that envelops the other two black holes' shadows. In contrast, the primary shadows of $m_{1}$ and $m_{3}$ undergo relatively minor modifications---they morph into arch by $l=0.7$ and subsequently remain nearly unchanged. However, their secondary shadows exhibit progressive elongation and distortion as $l$ decreases. Interestingly, we find that when the three black holes are sufficiently close, their combined shadow becomes virtually indistinguishable from that of a single spherically symmetric black hole, as exemplified in panel (i). This demonstrates that the photon capture capability of three tightly packed black holes can be effectively mimicked by a single black hole. This degeneracy suggests a novel perspective on black hole internal structure: the event horizon of a single black hole might originate from multiple tightly packed extremal RN black holes in its interior. Moreover, in each panel, we detect image features near the shadow boundary that originate from distinct black holes. Although resolution limitations render these features only as discrete points or line segments, their alternating pattern remains identifiable. This phenomenon arises from chaotic scattering of photon due to the non-integrable nature of spacetime \eqref{5}, generating self-similar fractal structures along the shadow boundary.

By varying the observer's azimuthal angle $\Phi$ to $45^{\circ}$ and $90^{\circ}$, we further investigate the influence of $l$ on the shadows of triple black holes, with the corresponding results displayed in Figs. 3 and 4. At an observation azimuth of $45^{\circ}$, the position, size, and shape of $m_{1}$'s primary shadow remain largely unaffected by the parameter $l$, while its two secondary shadows located leftward of the $m_{2}$ and $m_{3}$ exhibit progressive elongation as $l$ decreases. Notably, at $l=1$ (panel (d)), $m_{1}$ develops a third secondary shadow on the image's right side in addition to the two prominent left-side secondaries. This newly emerged shadow feature ultimately merges with the left-side secondaries to form a shadow ring as $l$ further decreases, as demonstrated in panel (g). Similar evolutionary patterns occur for $m_{2}$ and $m_{3}$---their primary and secondary shadows gradually coalesce into rings with diminishing $l$. Particularly for $m_{3}$, Fig. 3 clearly documents this ring formation process.
\begin{figure*}%[tbph]
\center{
\includegraphics[width=4cm]{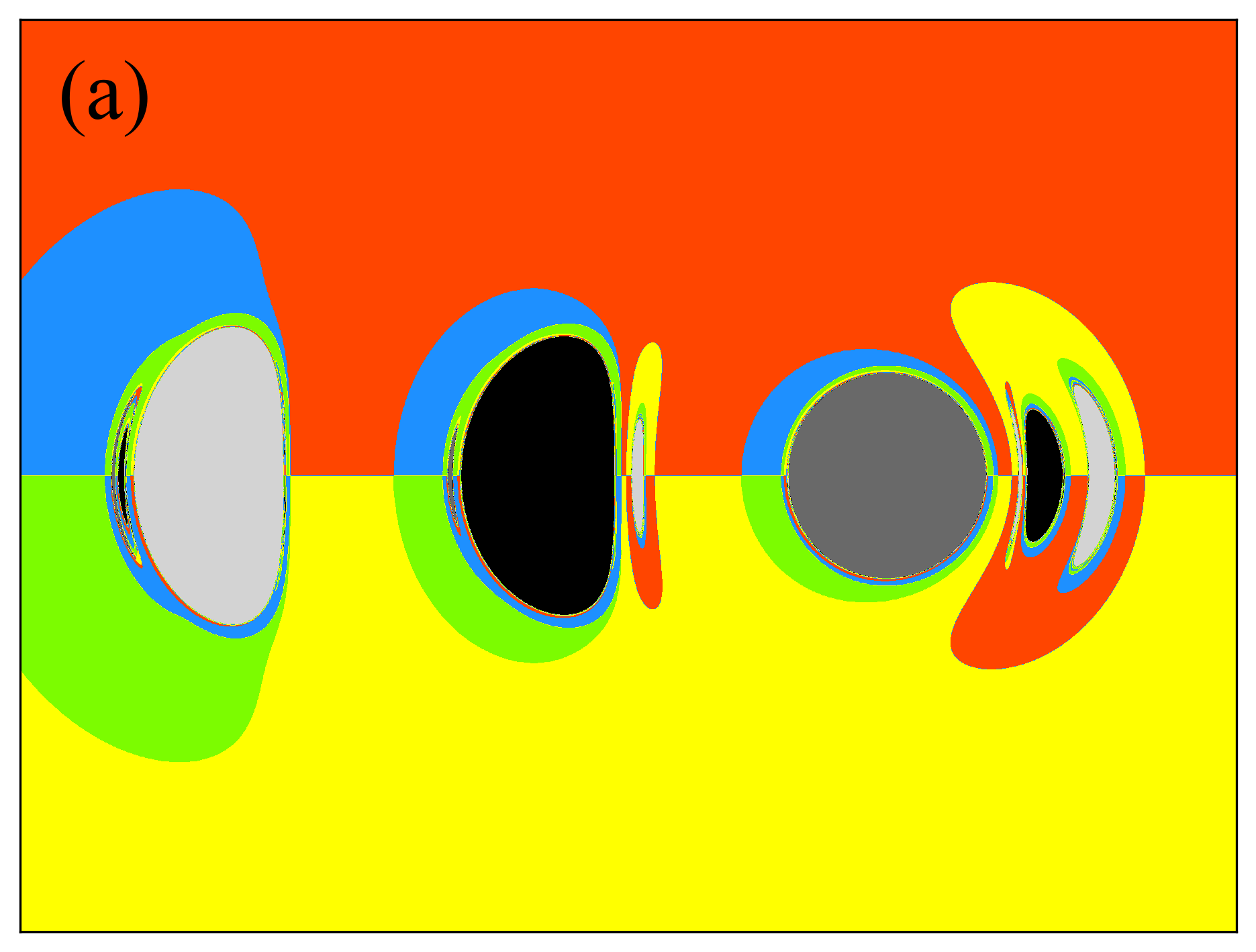}
\includegraphics[width=4cm]{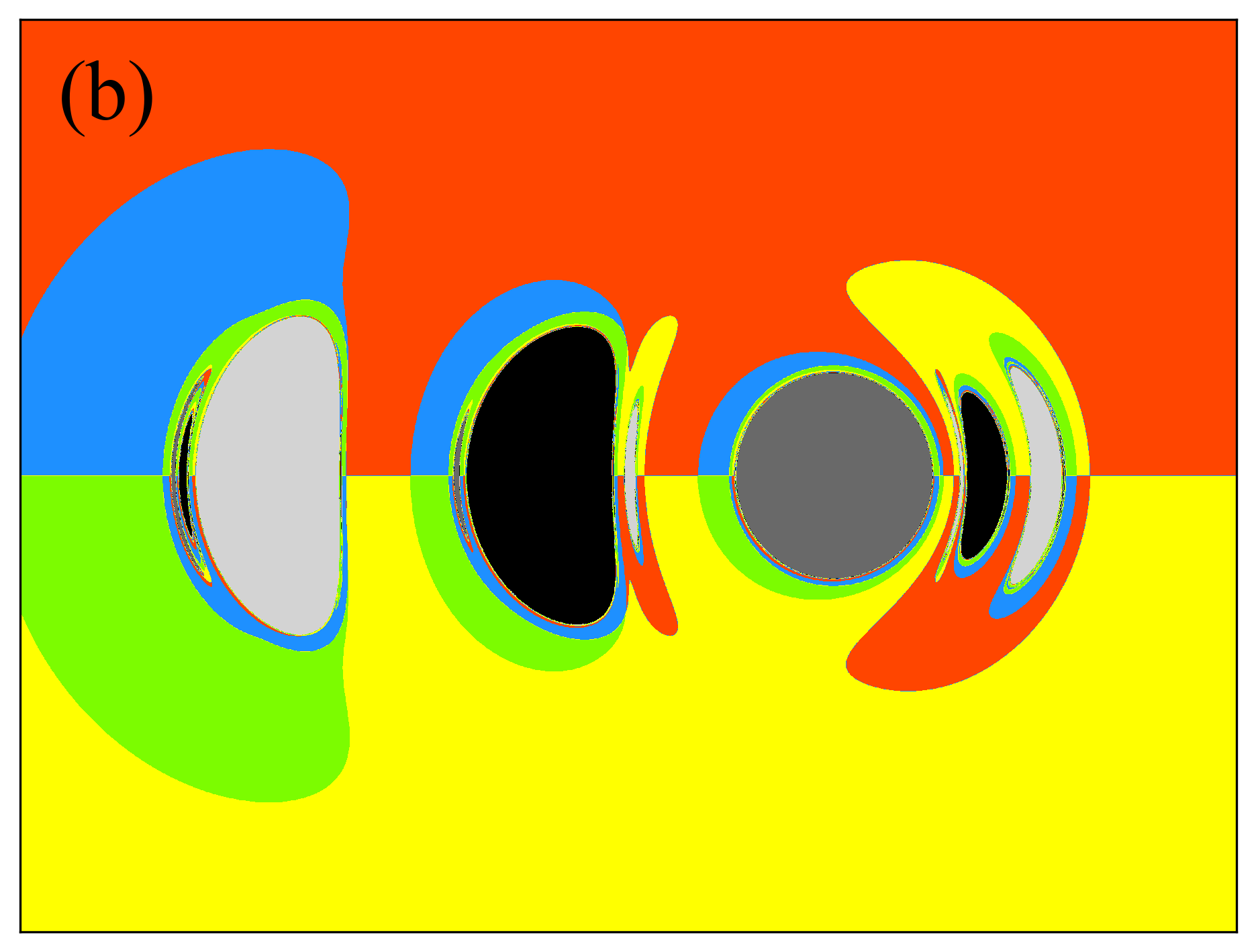}
\includegraphics[width=4cm]{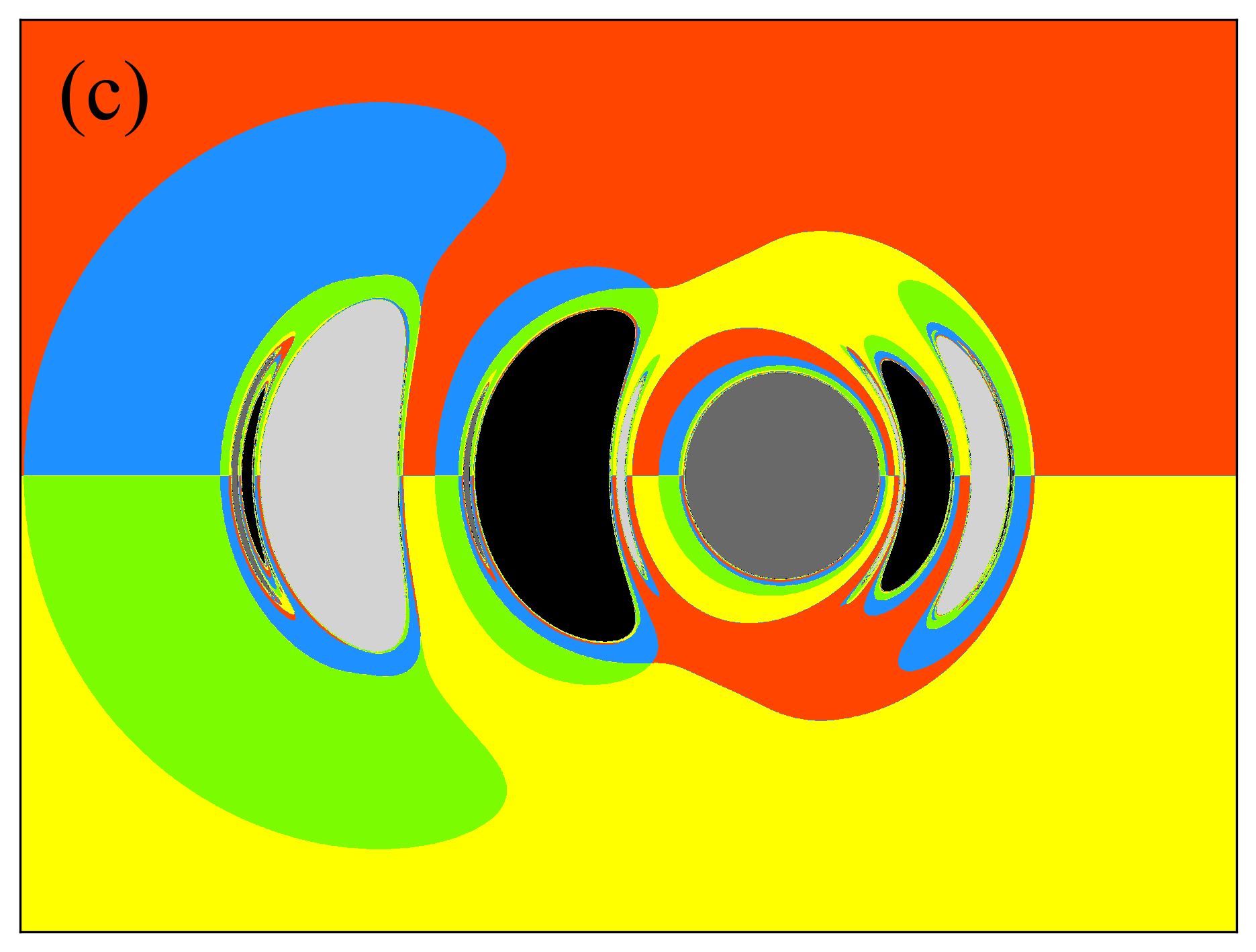}
\includegraphics[width=4cm]{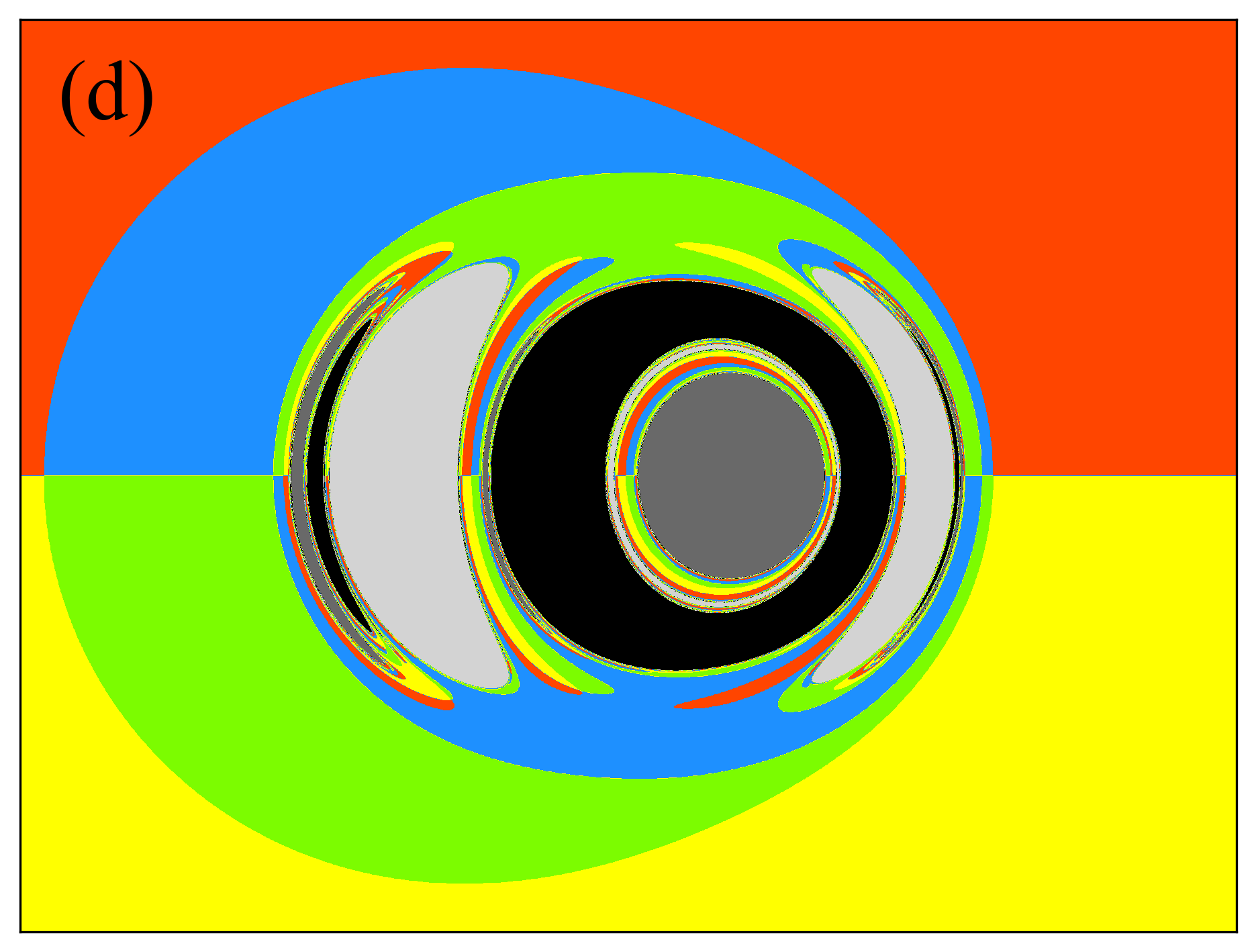}
\includegraphics[width=4cm]{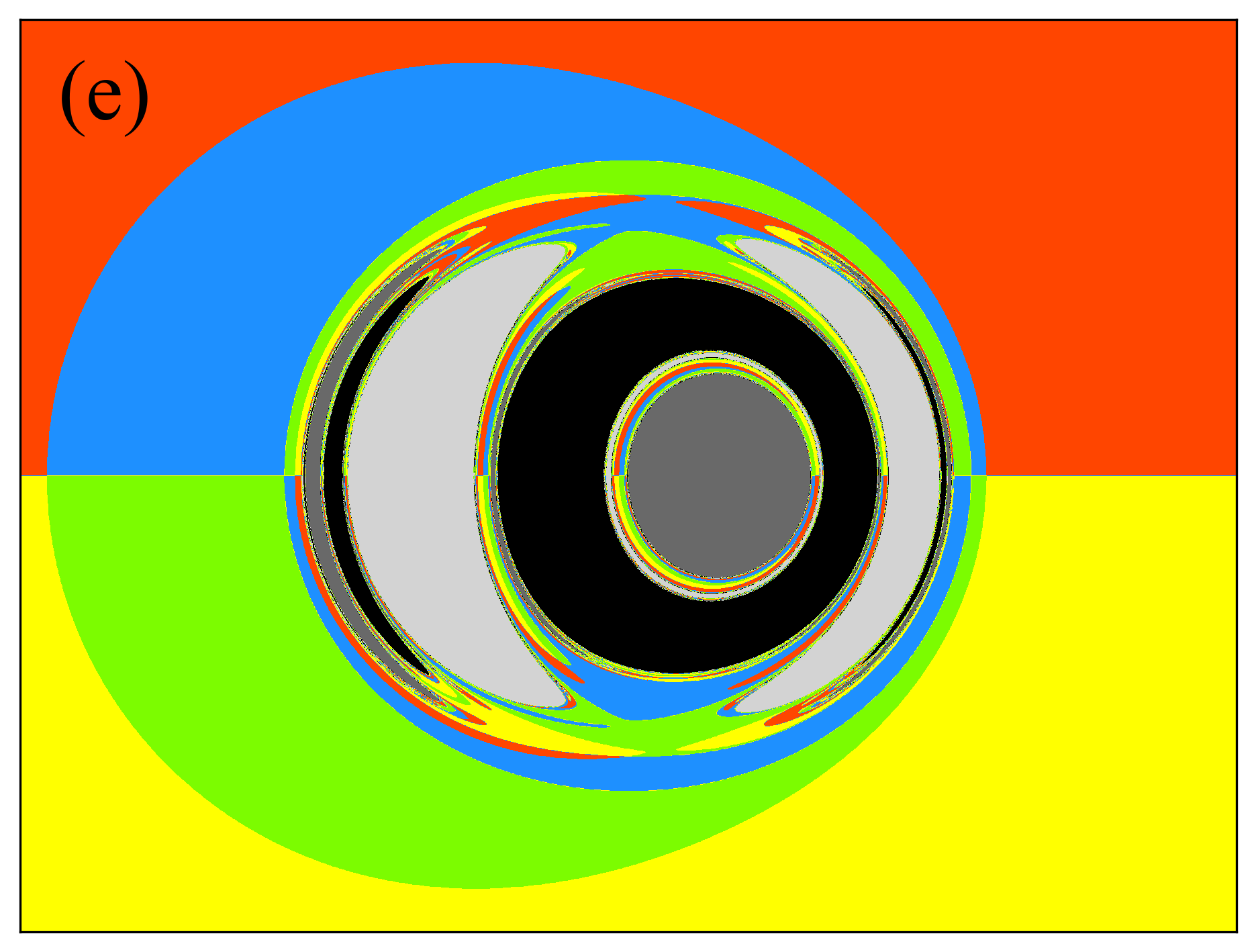}
\includegraphics[width=4cm]{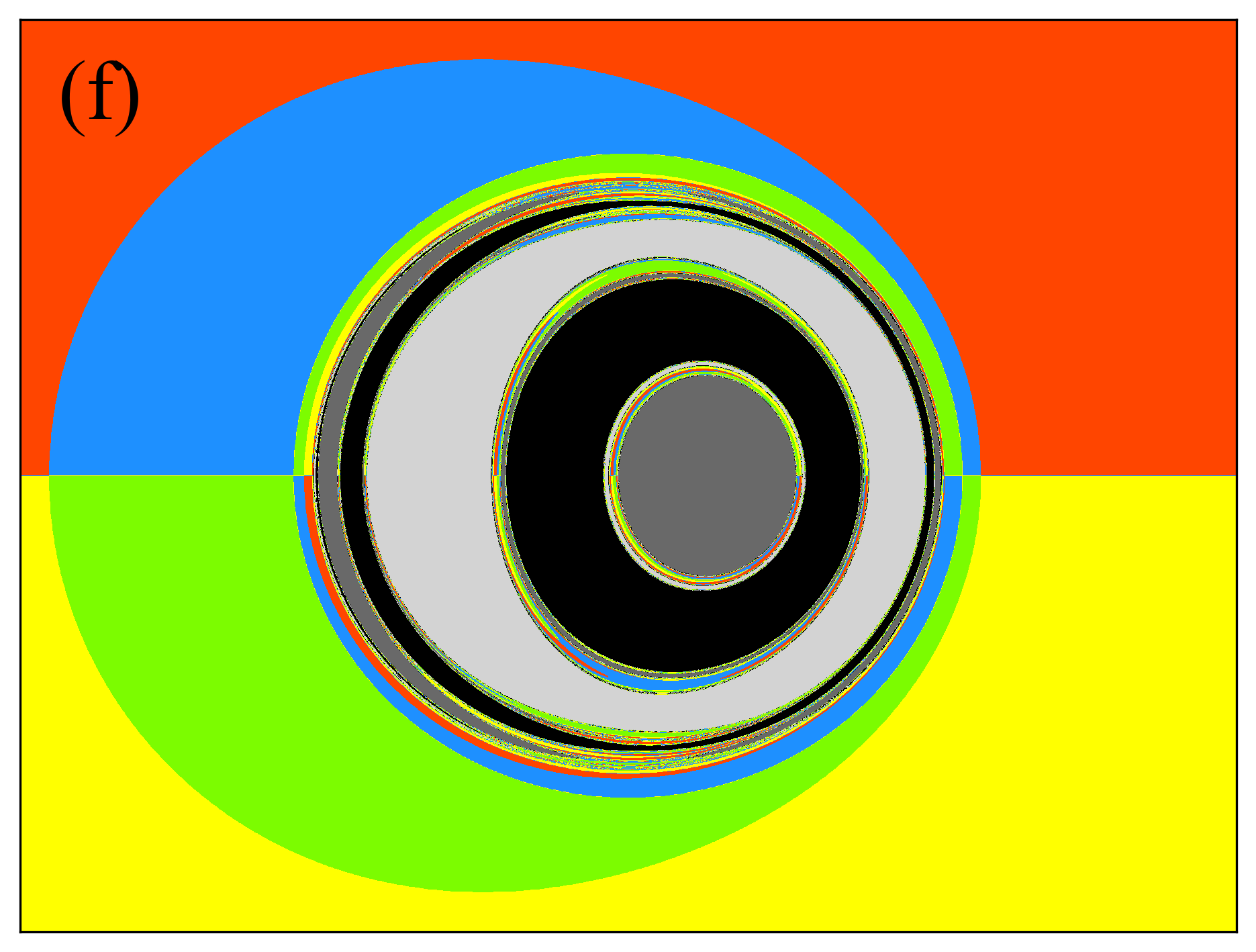}
\includegraphics[width=4cm]{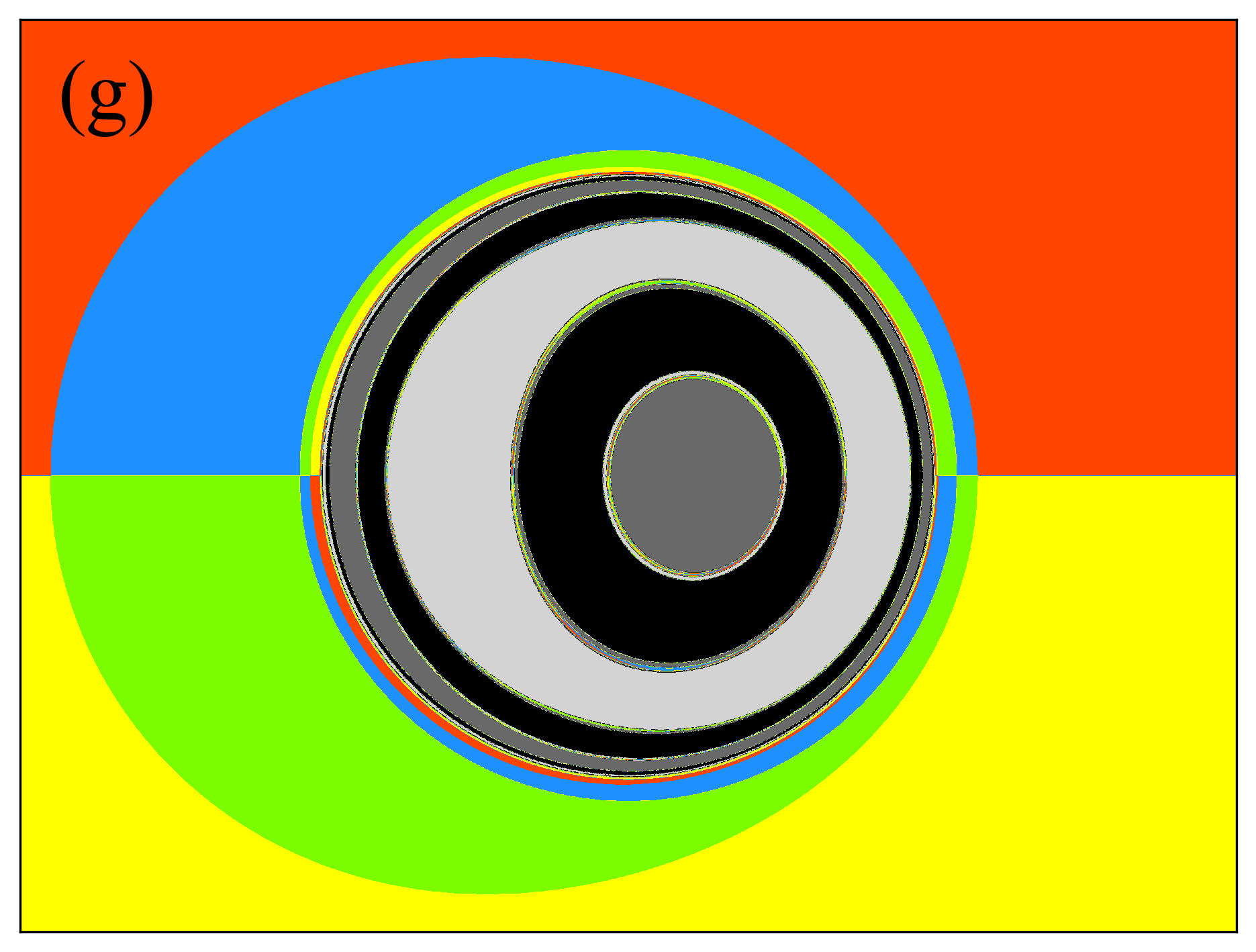}
\includegraphics[width=4cm]{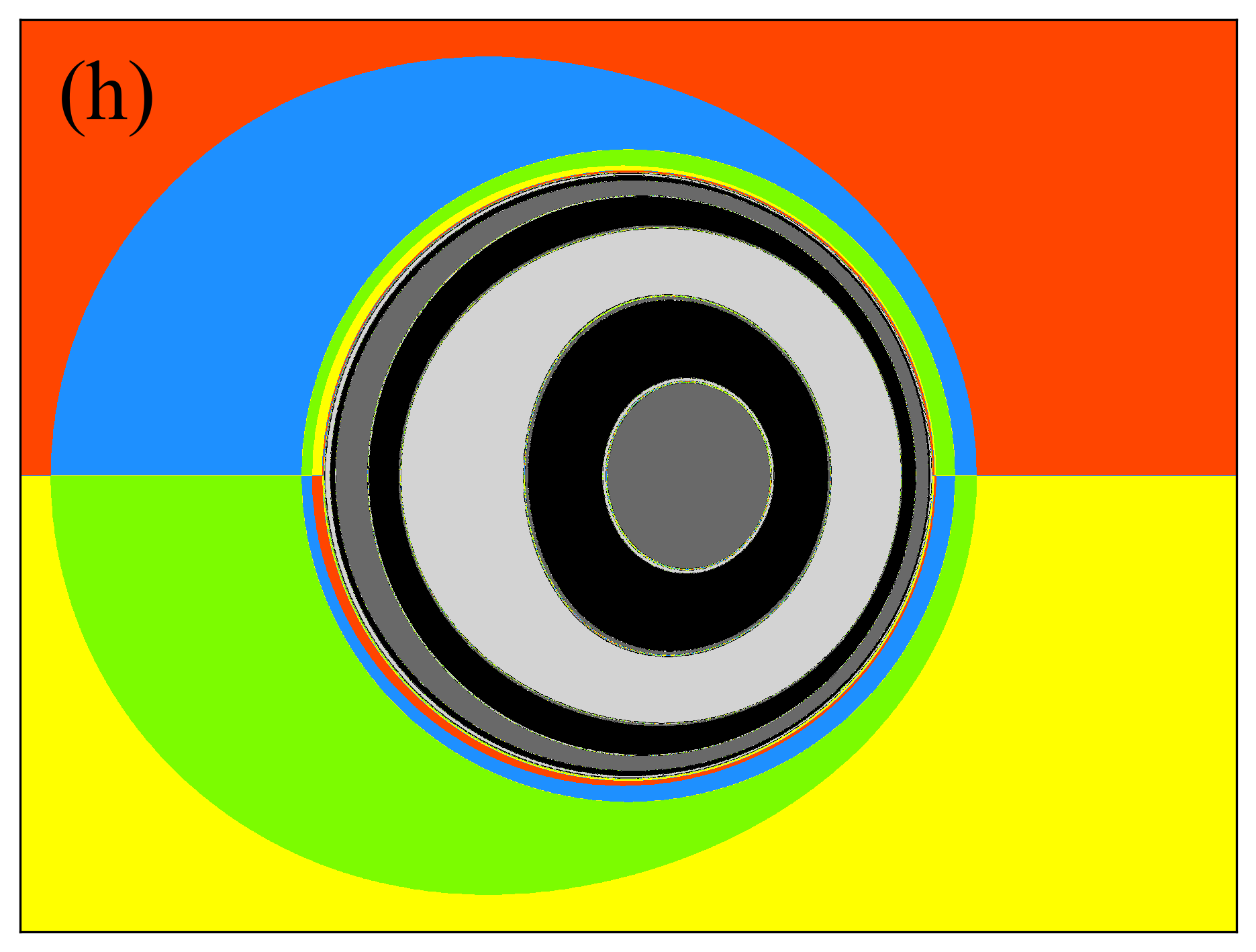}
\caption{Shadows of triple static black holes with different $l$. From panel (a) to panel (h), the distance $l$ are $4$, $3$, $2$, $1$, $0.75$, $0.5$, $0.25$, and $0.1$, respectively. Here, we have the observation angle $\Theta = 90^{\circ}$ and azimuthal angle $\Phi = 45^{\circ}$. The primary shadow of $m_{1}$ remains nearly invariant throughout the evolution, maintaining its quasi-circular morphology. In contrast, the primary shadows of $m_{2}$ and $m_{3}$ undergo a progressive transformation from arched to crescent-shaped profiles as $l$ decreases, ultimately merging with their respective secondary shadows to form asymmetric rings. Notably, similar to Fig. 2, when $l$ becomes sufficiently small ($l\leqslant 0.1$), all three shadows coalesce into a unified structure.}}\label{fig3}
\end{figure*}
\begin{figure*}%[tbph]
\center{
\includegraphics[width=5cm]{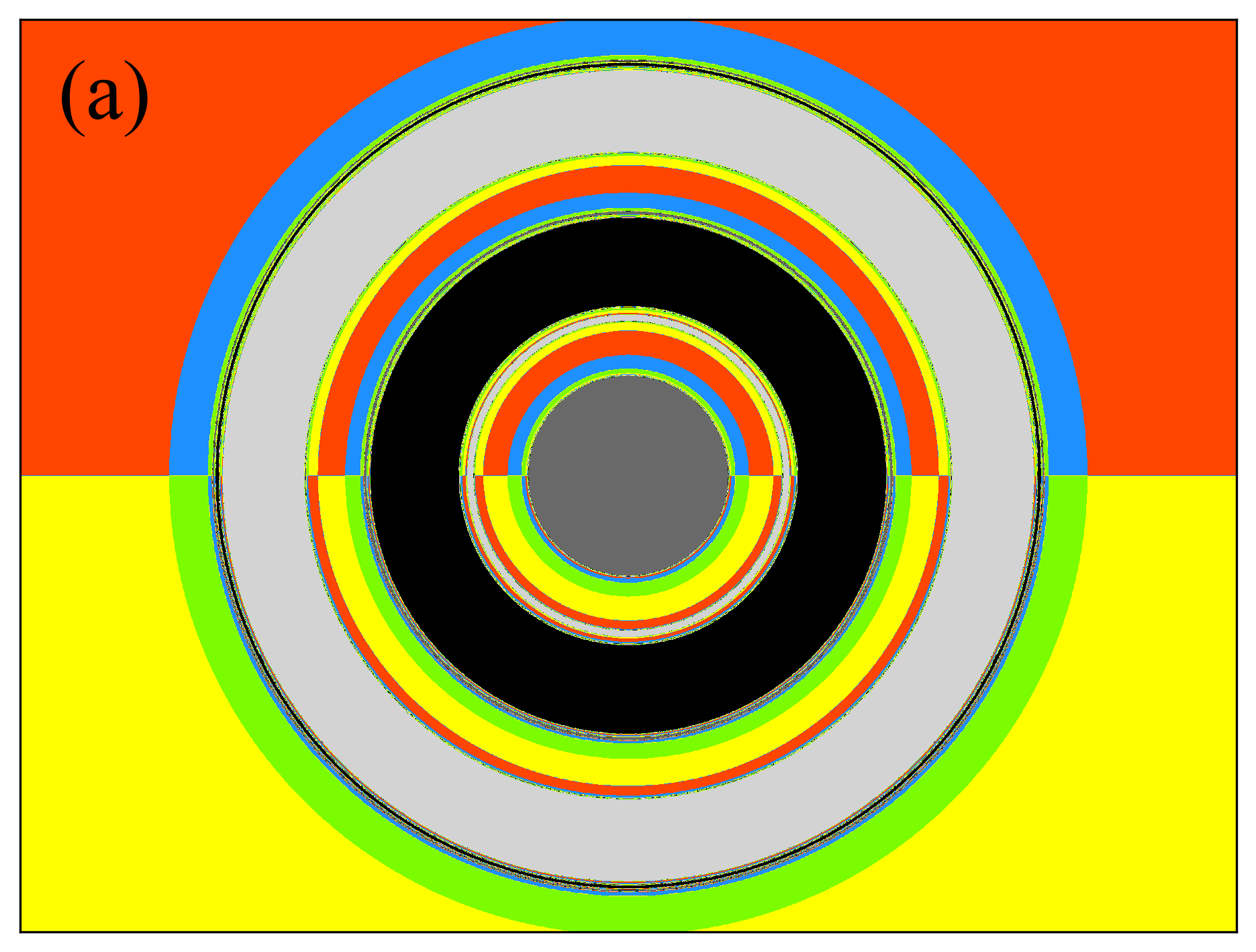}
\includegraphics[width=5cm]{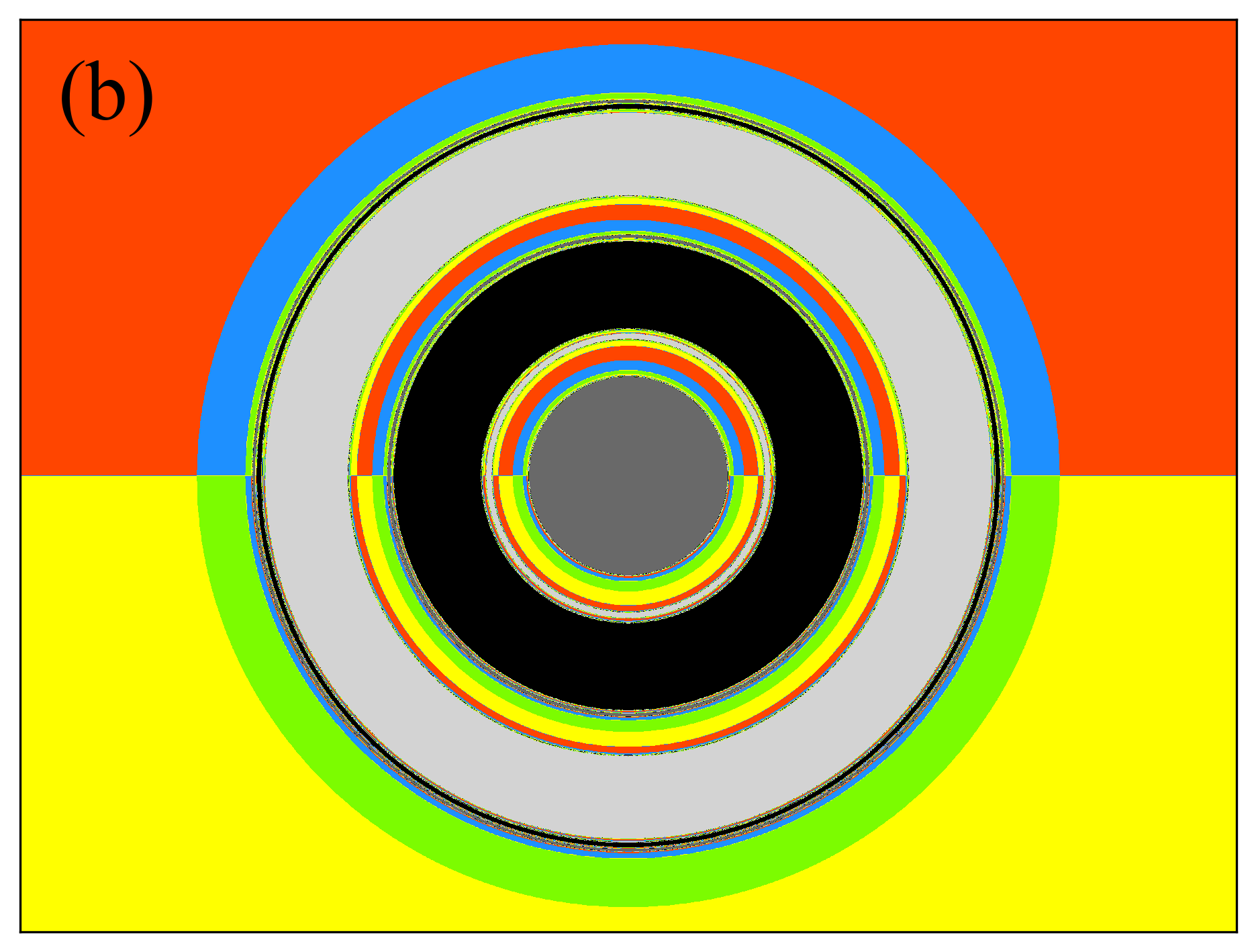}
\includegraphics[width=5cm]{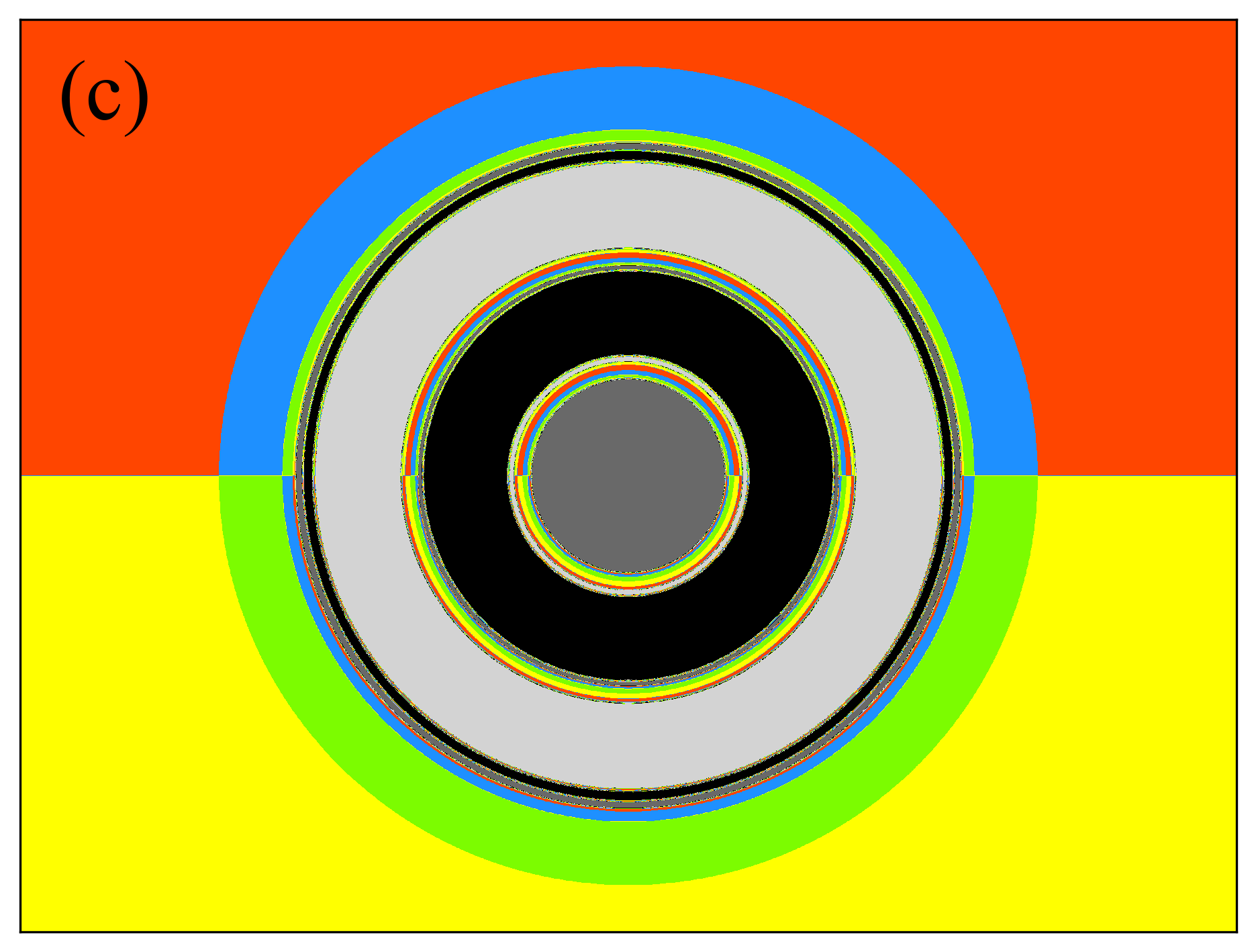}
\includegraphics[width=5cm]{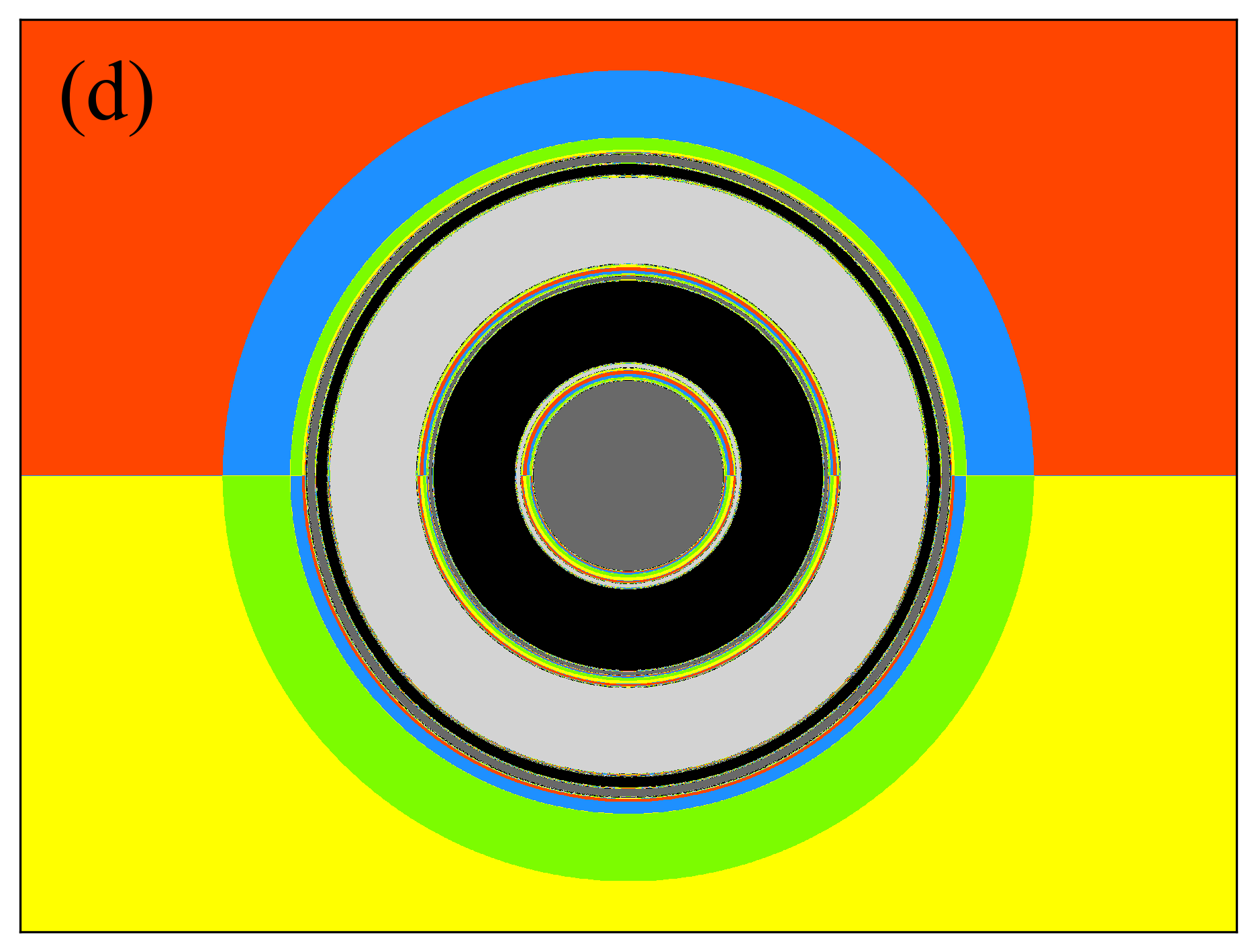}
\includegraphics[width=5cm]{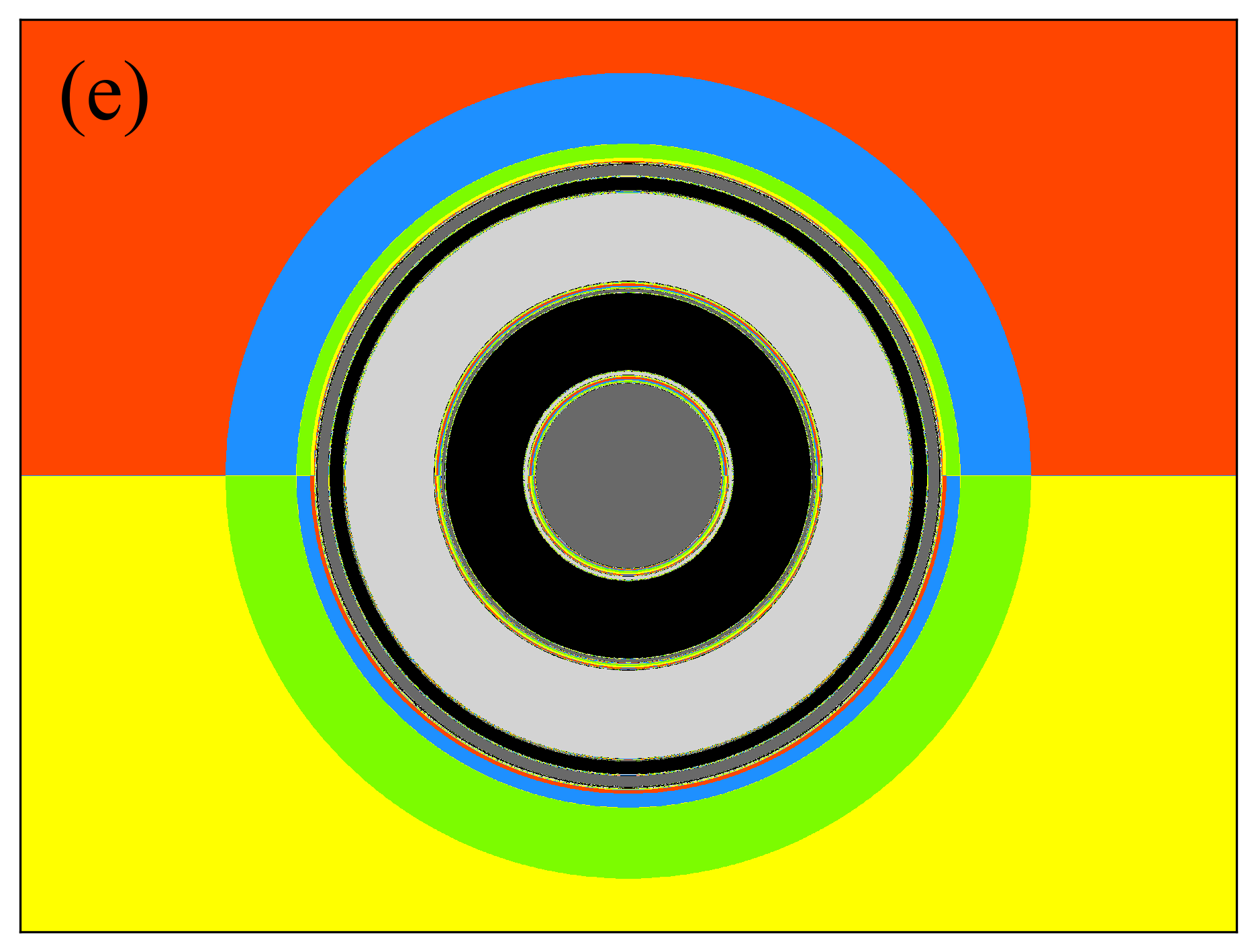}
\includegraphics[width=5cm]{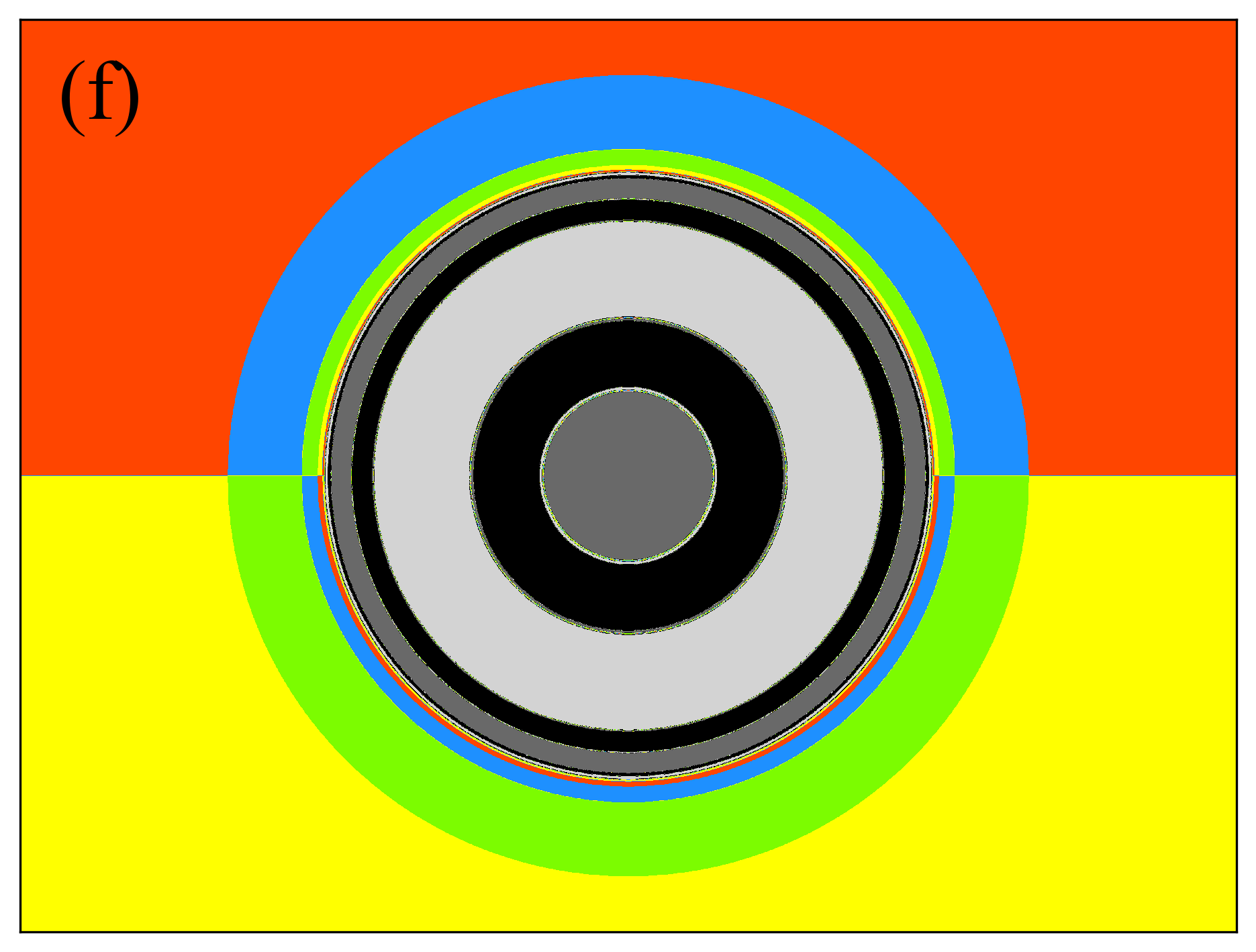}
\caption{Evolution of shadows of triple static black holes with parameter $l$. From panel (a) to panel (f), the distance $l$ are $3$, $2$, $1$, $0.75$, $0.5$, and $0.1$, respectively. Here, we have the observation angle $\Theta = 90^{\circ}$ and azimuthal angle $\Phi = 90^{\circ}$. we observe sequentially from the image center outward: the circular shadow of $m_{1}$, followed by the ring-like shadow of $m_{2}$, and finally the ring-like shadow of $m_{3}$. As $l$ decreases, the ring shadows of both $m_{2}$ and $m_{3}$ exhibit significant contraction. At the critical separation $l=0.1$, all three shadows coalesce into a perfect disk profile.}}\label{fig4}
\end{figure*}

When the azimuthal angle reaches $\Phi = 90^{\circ}$, the three black holes become colinearly aligned with the line of sight. In this configuration, $m_{1}$ produces a perfectly disk shadow, while $m_{2}$ and $m_{3}$ generate thick ring-like shadows through an Einstein-ring formation mechanism, with their ring radii showing a positive correlation with $l$. Since $m_{3}$ lies behind $m_{2}$, its ring shadow undergoes magnification by both $m_{1}$ and $m_{2}$, making it the largest among the three black holes. Additionally, we clearly identify thin ring-like shadows from all three black holes, which correspond to higher-order secondary shadows created by gravitational lensing. Notably, when $l$ becomes sufficiently small, we observe the convergence of all three shadows into a single faint region.
\begin{figure*}%[tbph]
\center{
\includegraphics[width=5cm]{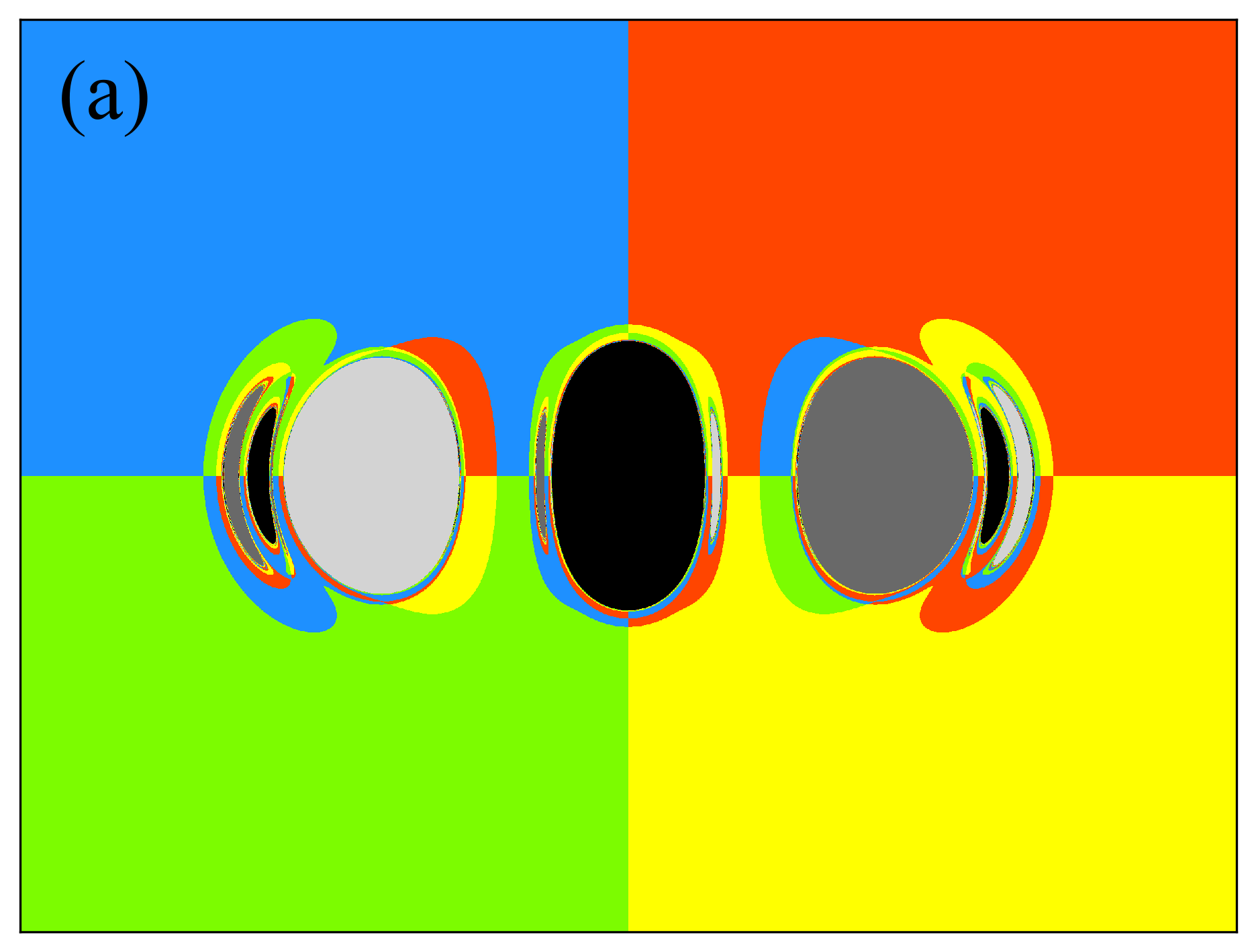}
\includegraphics[width=5cm]{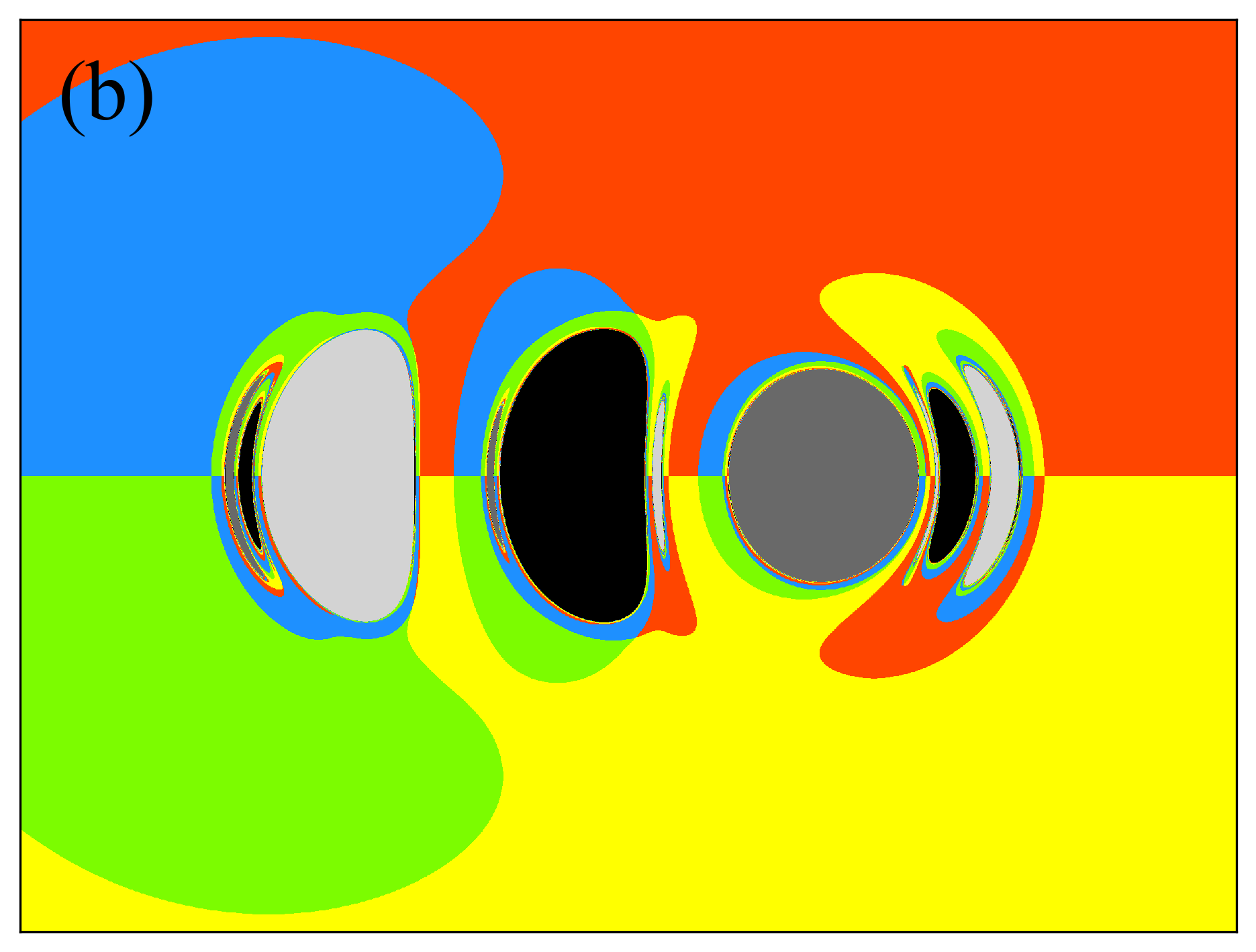}
\includegraphics[width=5cm]{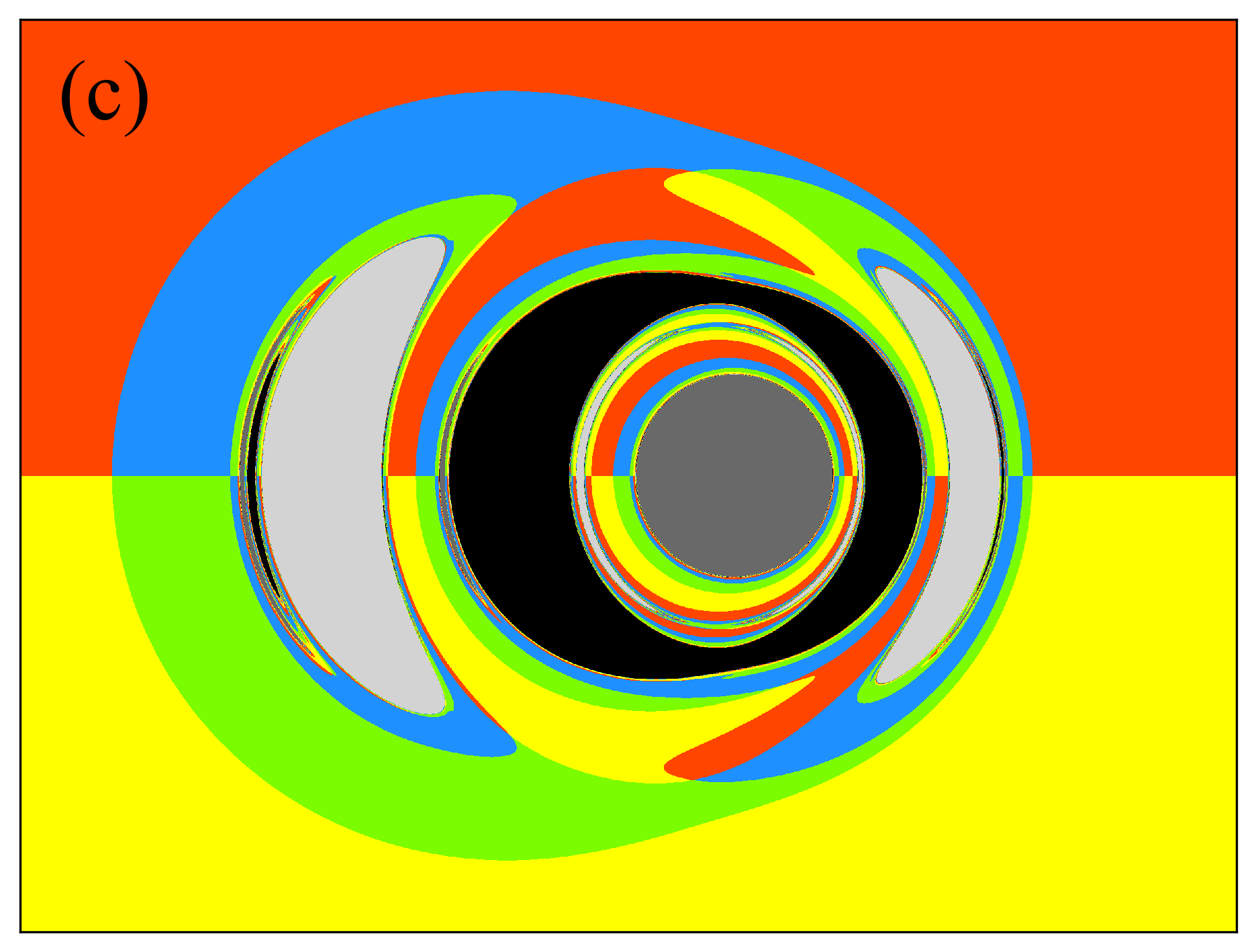}
\includegraphics[width=5cm]{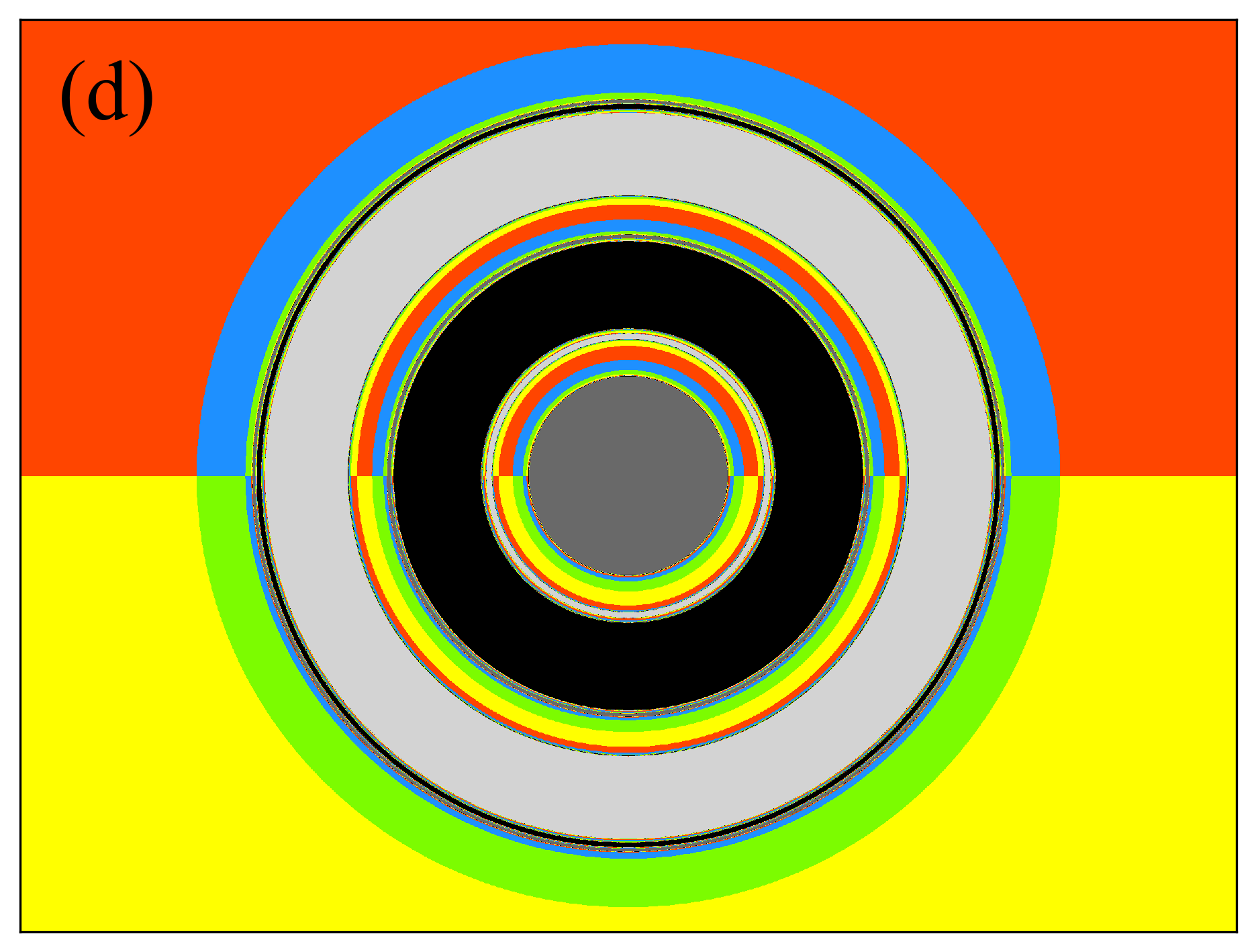}
\includegraphics[width=5cm]{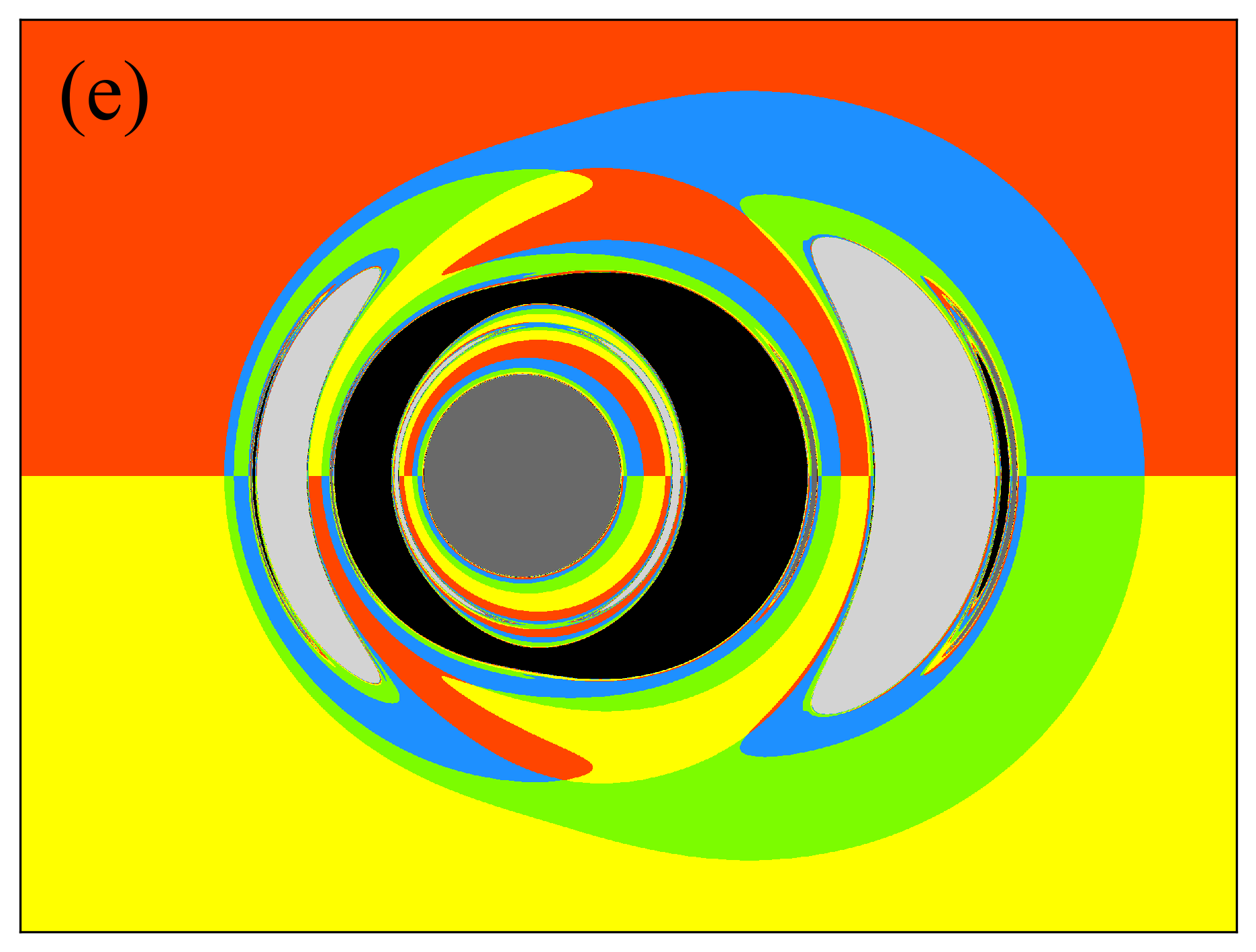}
\includegraphics[width=5cm]{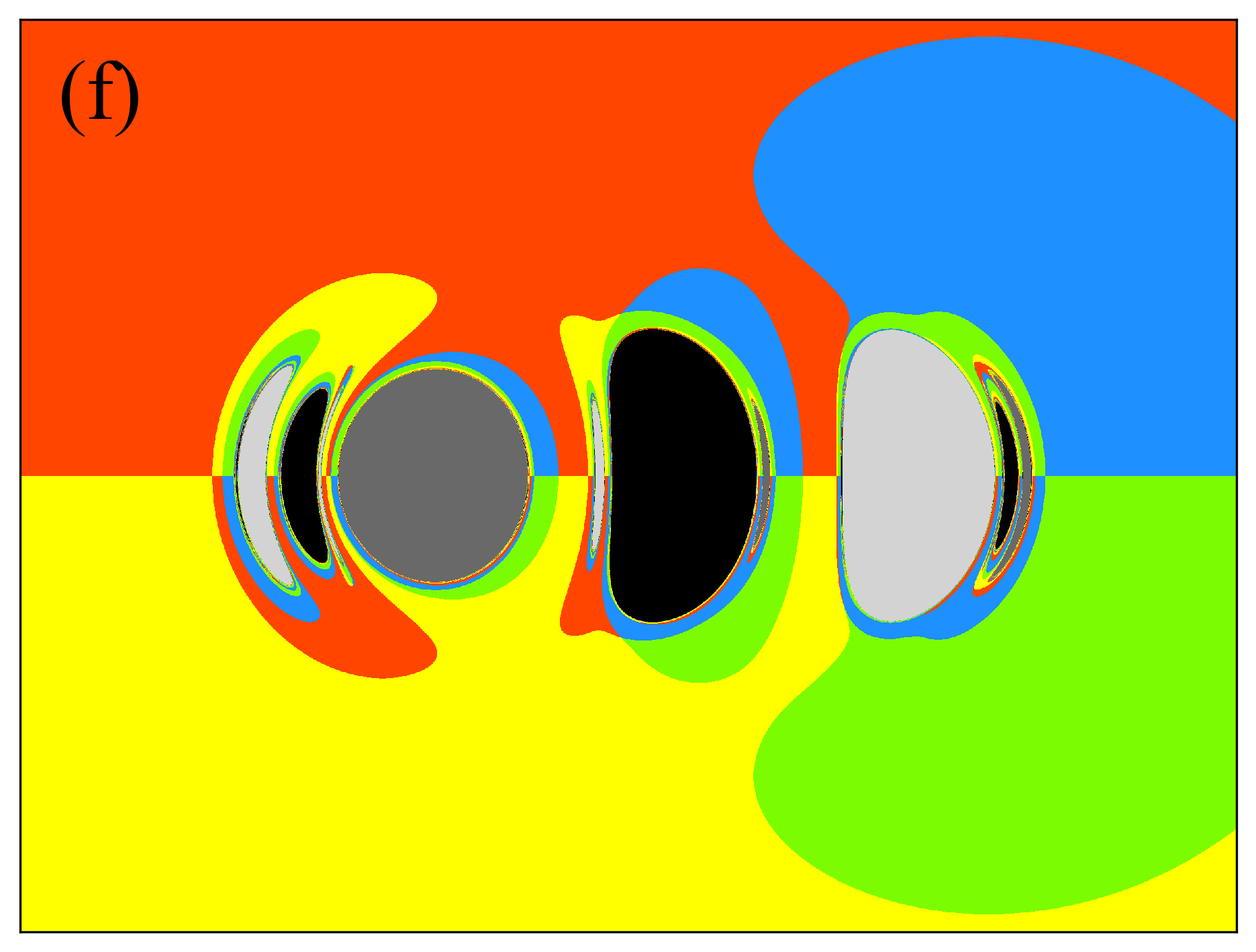}
\includegraphics[width=5cm]{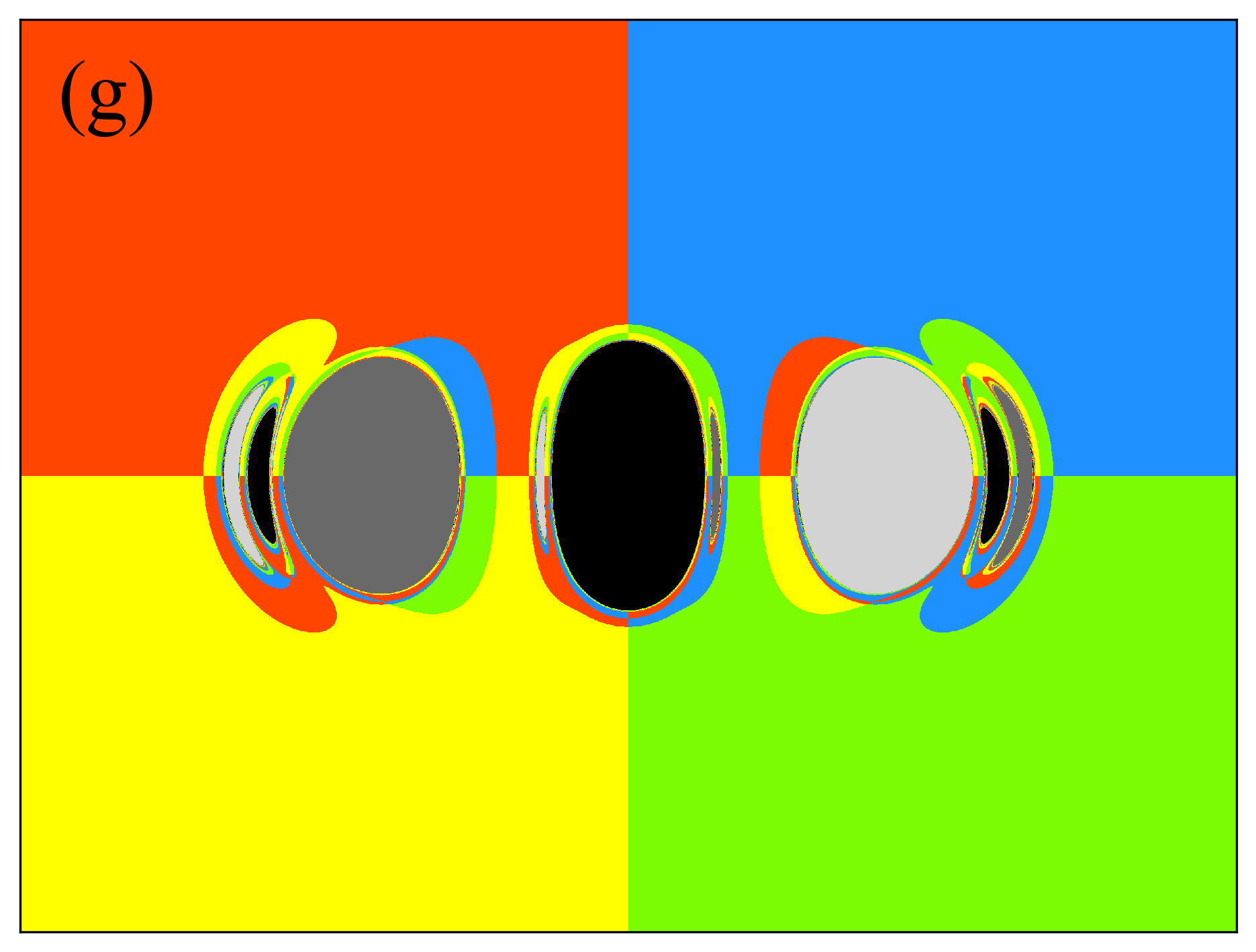}
\includegraphics[width=5cm]{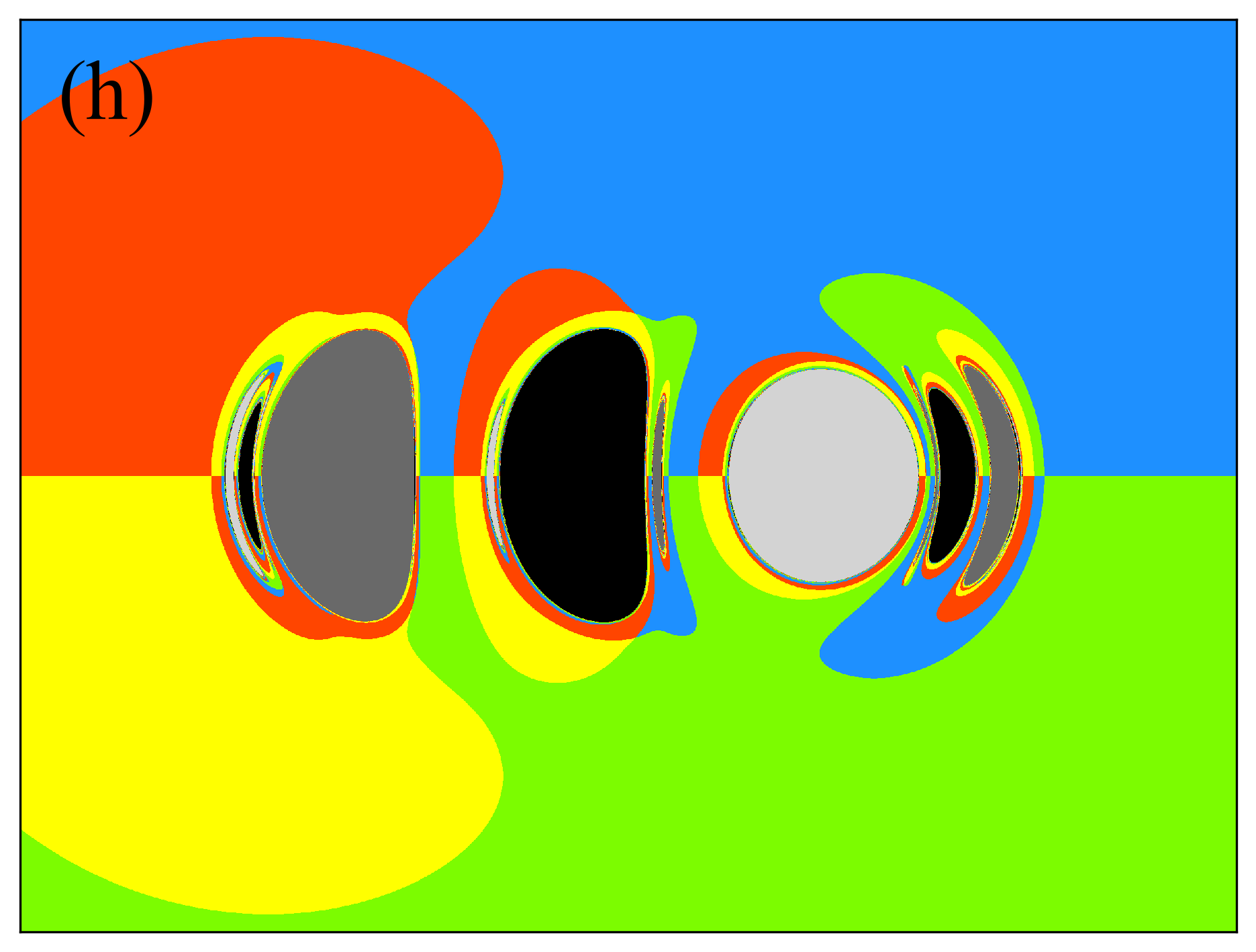}
\includegraphics[width=5cm]{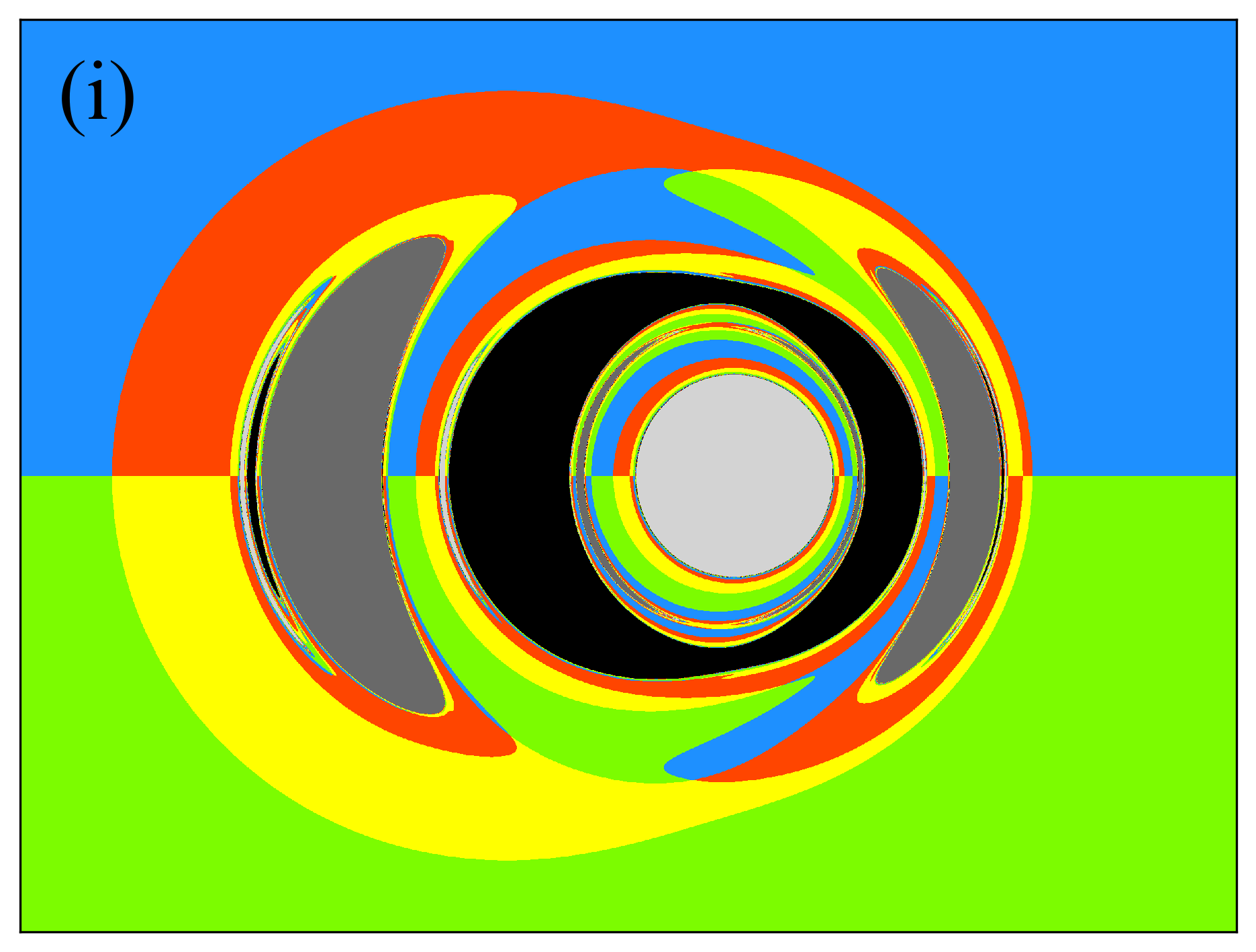}
\includegraphics[width=5cm]{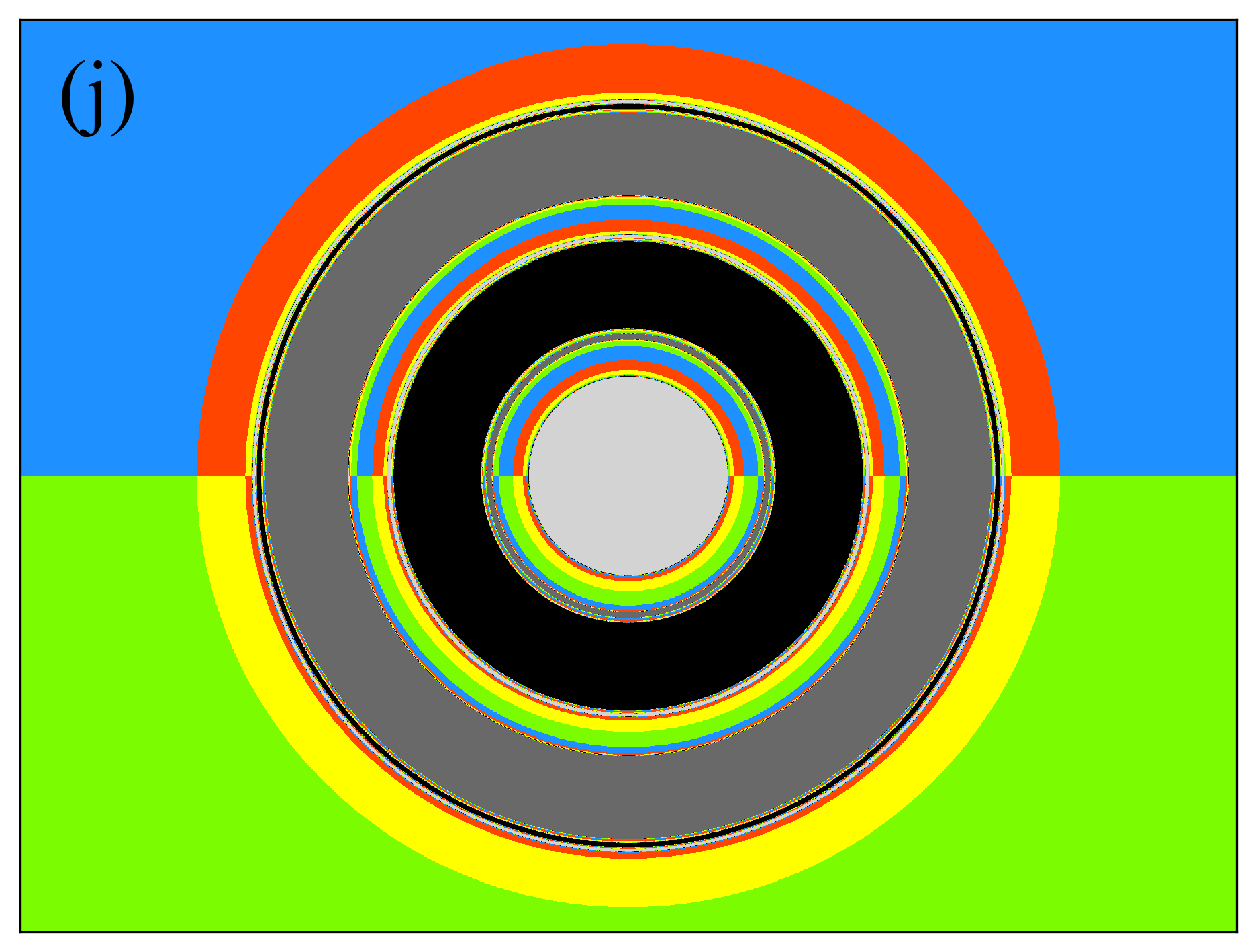}
\includegraphics[width=5cm]{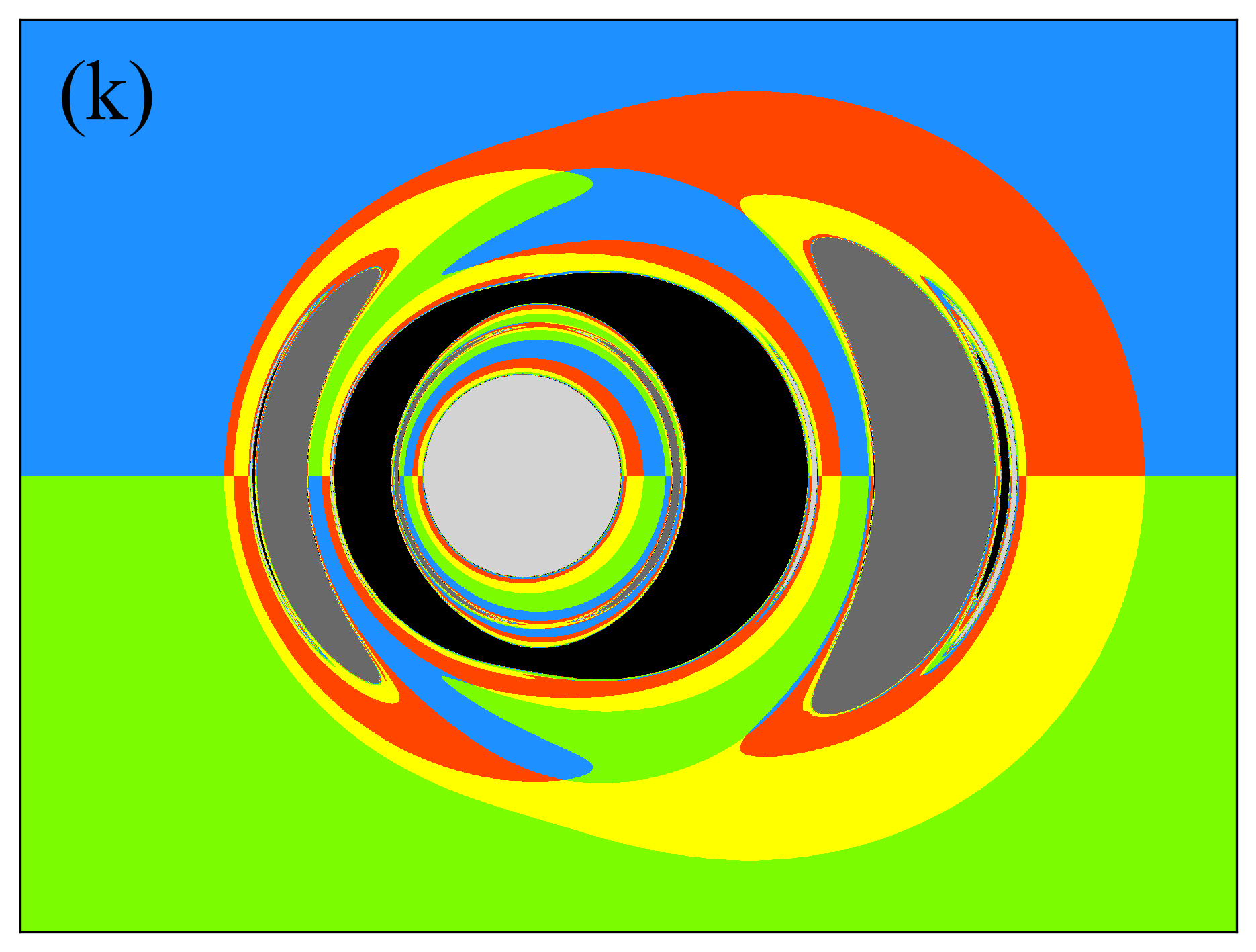}
\includegraphics[width=5cm]{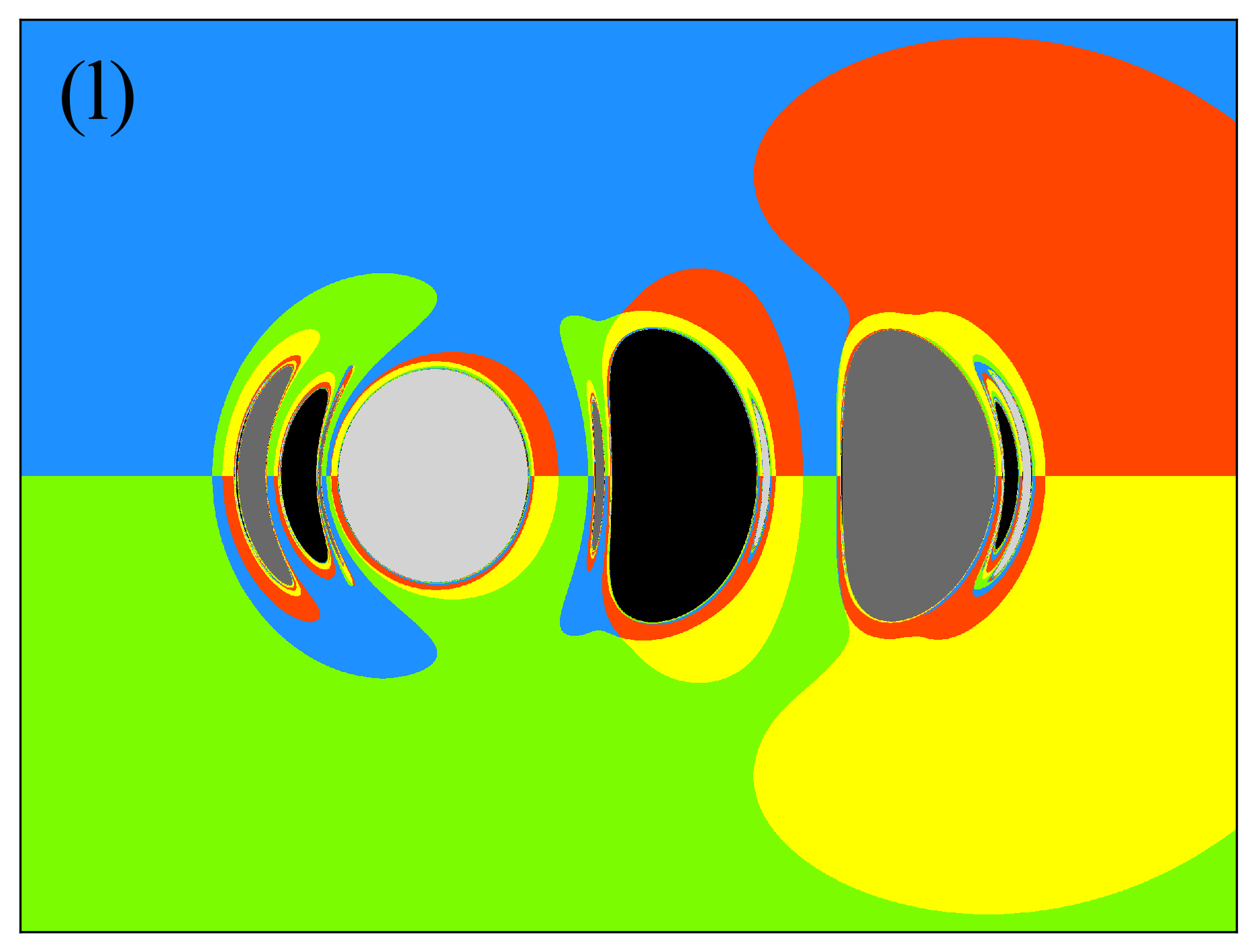}
\caption{Shadow evolution of triple static black holes under varying azimuthal angles $\Phi$ at fixed separation $l=2$. Panels (a)-(l) display configurations from $\Phi=0^{\circ}$ to $\Phi=330^{\circ}$ in $30^{\circ}$ increments, with a constant observation inclination of $\Theta=90^{\circ}$.}}\label{fig5}
\end{figure*}

Through Figs. 2-4, we observe distinctive features in the shadows of triple black holes: ring-shaped shadow and eyebrow-like shadow. Although similar features have been identified in black hole binaries \cite{Nitta et al. (2011),Yumoto et al. (2012),Bohn et al. (2015),Cunha et al. (2018),Shipley and Dolan (2016),Moreira et al. (2025)}, scalar hairy black holes \cite{Cunha et al. (2015),Cunha et al. (2016),Gyulchev et al. (2024)}, and the single black hole with quadrupole or octupole distortions \cite{Abdolrahimi et al. (2015),Grover et al. (2018)}, their quantitative occurrence is lower than revealed in our study. Specifically, the number of these fine structures depends on both the count and spatial distribution of compact objects. This implies that multiple concentric ring-shaped shadows or clustered eyebrow-like shadows may serve as unique observational probes for triple black hole system. Furthermore, Figs. 2-4 not only display shadows at varying spatial separations but also effectively trace the evolutionary sequence of shadows during the triple black holes merger process.

It is worth emphasizing that when spacetime perturbations induced by black hole motion are neglected, model \eqref{5} can qualitatively characterize snapshots of the dynamic triple black holes at arbitrary instants. This approach conveniently enables preliminary investigation of shadow evolution during the rotation or inspiral of such system. Accordingly, we (i) simulated the rotation of the triple black holes by continuously varying the observer's azimuthal angle while maintaining a fixed separation ($l = 2$) in the collinear configuration, with the resulting shadow evolution presented in Fig. 5; and (ii) simultaneously varied both the black hole separation and azimuthal angle to model shadow evolution during the inspiral phase, as shown in Fig. 6. In the former case, we find that when $\Phi$ changes from $0^{\circ}$ to $30^{\circ}$, the primary shadow of the right black hole ($m_{1}$) becomes more circular, while the primary shadows of $m_{2}$ and $m_{3}$ exhibit crescent shapes eroded by $m_{1}$. This occurs because the right side of the observation screen represents the approaching side while the left corresponds to the receding side, placing $m_{2}$ and $m_{3}$ behind $m_{1}$ in panel (b). Simultaneously, we observe elongated multiple secondary shadows from all three black holes, with the most pronounced changes occurring in $m_{3}$'s counterparts. As the azimuthal angle further increases, the primary and secondary shadows of $m_{2}$ become vertically connected and merge into a ring-like structure enveloping $m_{1}$'s primary shadow due to lensing effects, while $m_{3}$'s shadow continues to exhibit progressive elongation. When the three black holes rotate into alignment with the line of sight, ring-shaped shadows emerge---a phenomenon we have detailed in Fig. 4. As the orbital motion continues, we obtain an image that is symmetric with panel (c) about the $y^{\prime}$-axis, resulting from $m_{1}$ having moved to the observer's left while $m_{3}$ has shifted to the right. At $\Phi=180^{\circ}$, the three black holes complete a half-orbit, exchanging the positions of $m_{1}$ and $m_{3}$ relative to panel (a). For $\Phi>180^{\circ}$, the shadow configurations become mirror images about the $y^{\prime}$-axis of their counterparts in panels (a)-(f), due to the system's intrinsic symmetry.

In Fig. 6, we find that during the early inspiral phase, such as panels (a)-(m), the image morphology depends more strongly on $\Phi$ than on separation $l$, as evidenced by shadow deformations nearly identical to those in Fig. 5. During this process, $m_{1}$'s primary shadow remains nearly circular as it migrates from the right to the left side of the image, while its faint secondary shadows are only detectable upon magnification. Meanwhile, $m_{2}$'s primary shadow transitions through a crescent phase, temporarily merges with its secondary shadows to form a coherent ring structure, and eventually fragments back into an elongated shape. Concurrently, $m_{3}$'s primary shadow exhibits similar ring-like snapshots, completing its left-to-right positional shift during the ring fragmentation phase. During the late inspiral phase, the influence of the separation $l$ on the image becomes pronounced. As $l$ decreases, $m_{2}$'s primary and secondary shadows first converge and envelop $m_{1}$'s primary shadow. Subsequently, $m_{3}$'s shadow replicates $m_{1}$'s structural behavior along $m_{1}$'s periphery. Ultimately, as shown in panel (r), the shadows of all three black holes become nearly indistinguishable, separated only by thin rings that gradually vanish with further decreasing $l$. In summary, Figs. 5 and 6 present the evolving shadow features of the triple black holes during their orbital, inspiral, and merger phases. These results characterize the deformation and motion of both primary and secondary shadows for each black hole, providing important insights for future observations of multi-black-hole systems.
\begin{figure*}%[tbph]
\center{
\includegraphics[width=4cm]{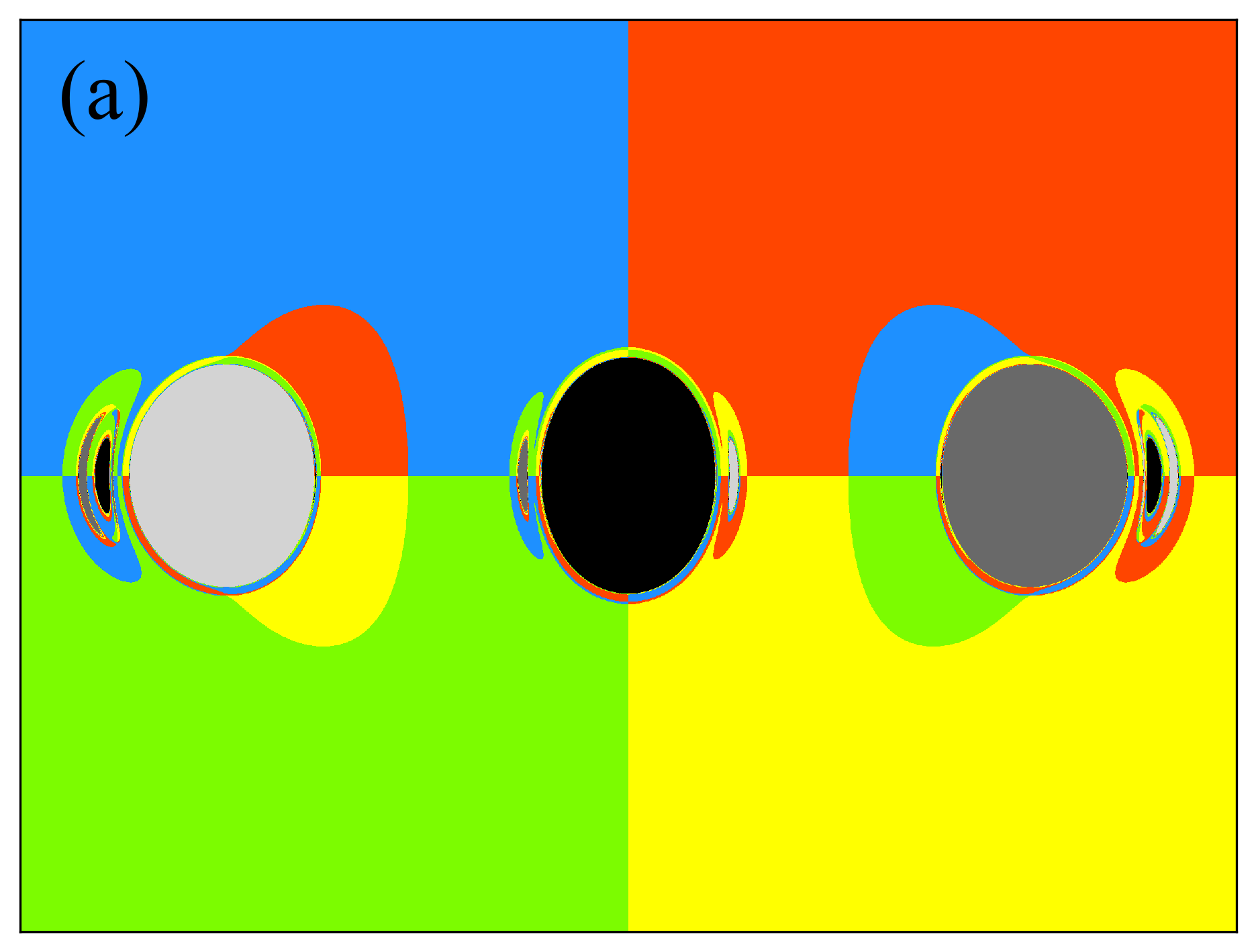}
\includegraphics[width=4cm]{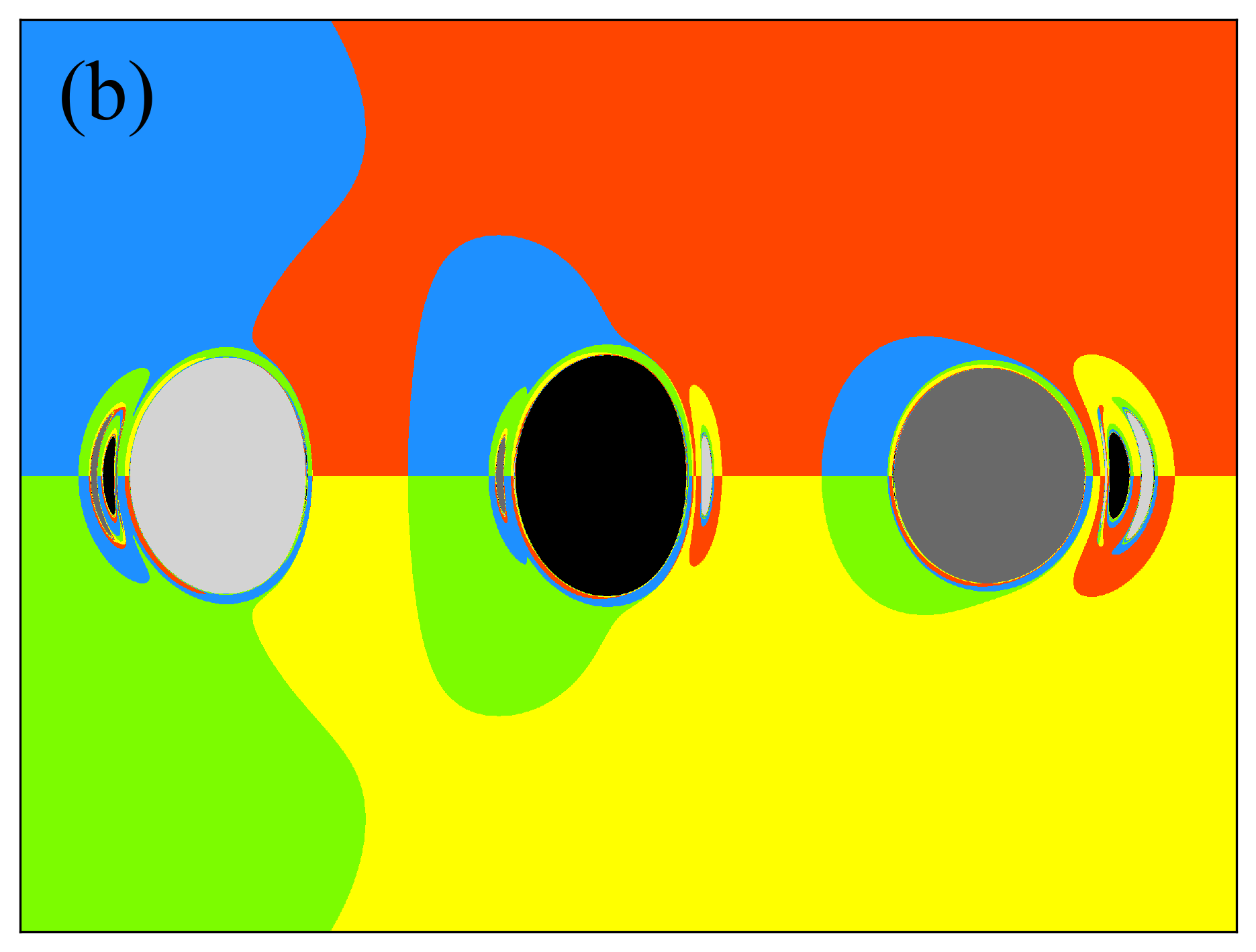}
\includegraphics[width=4cm]{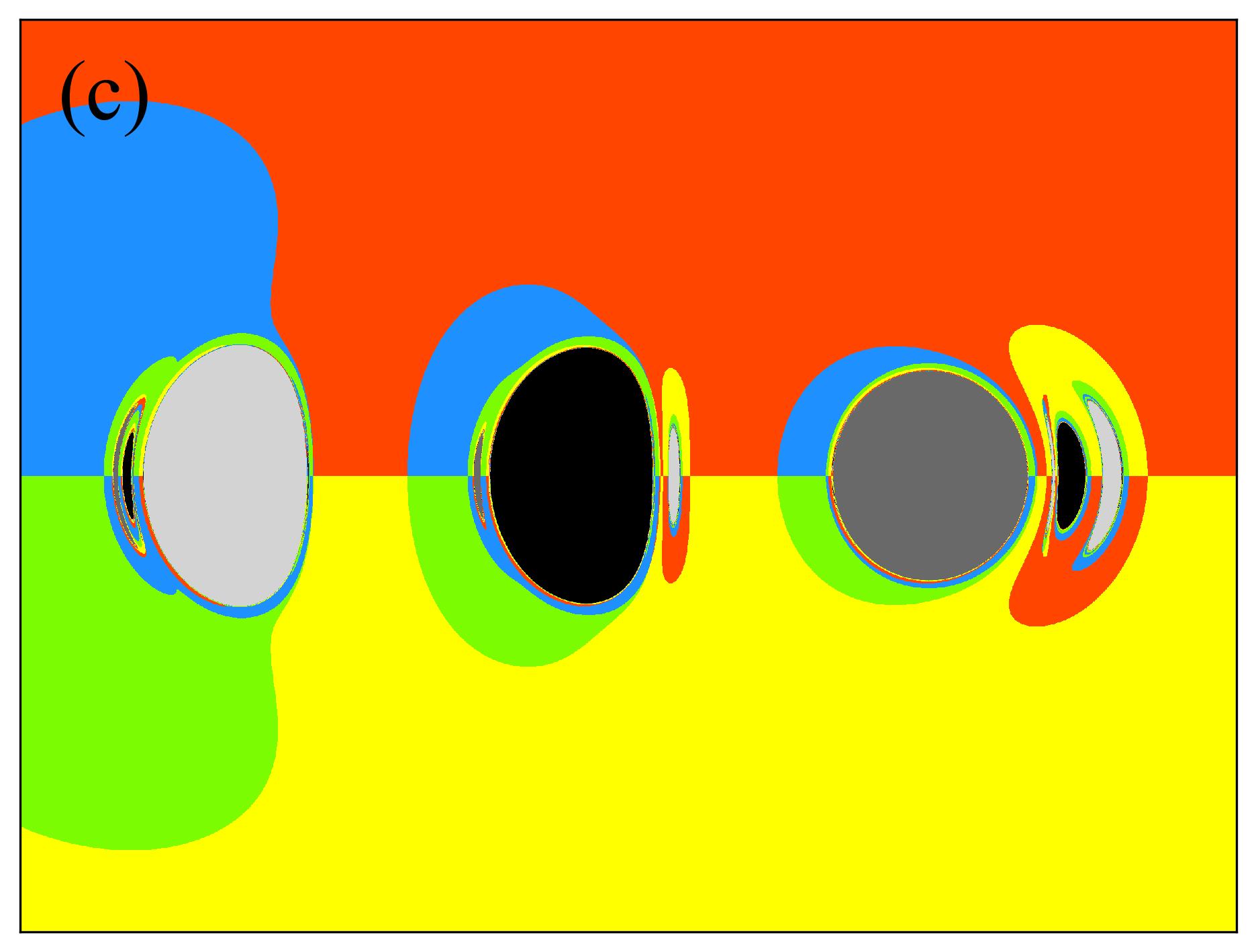}
\includegraphics[width=4cm]{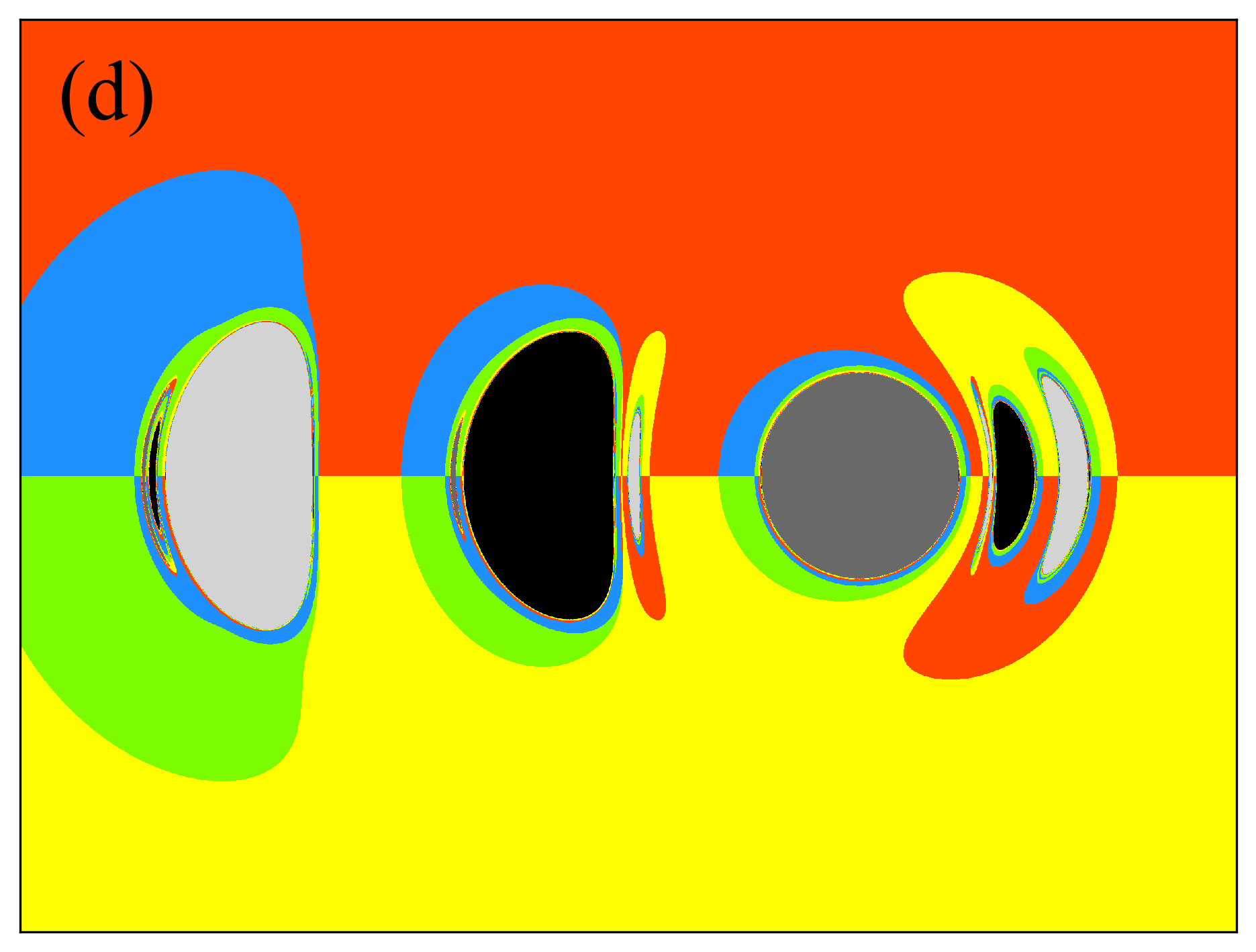}
\includegraphics[width=4cm]{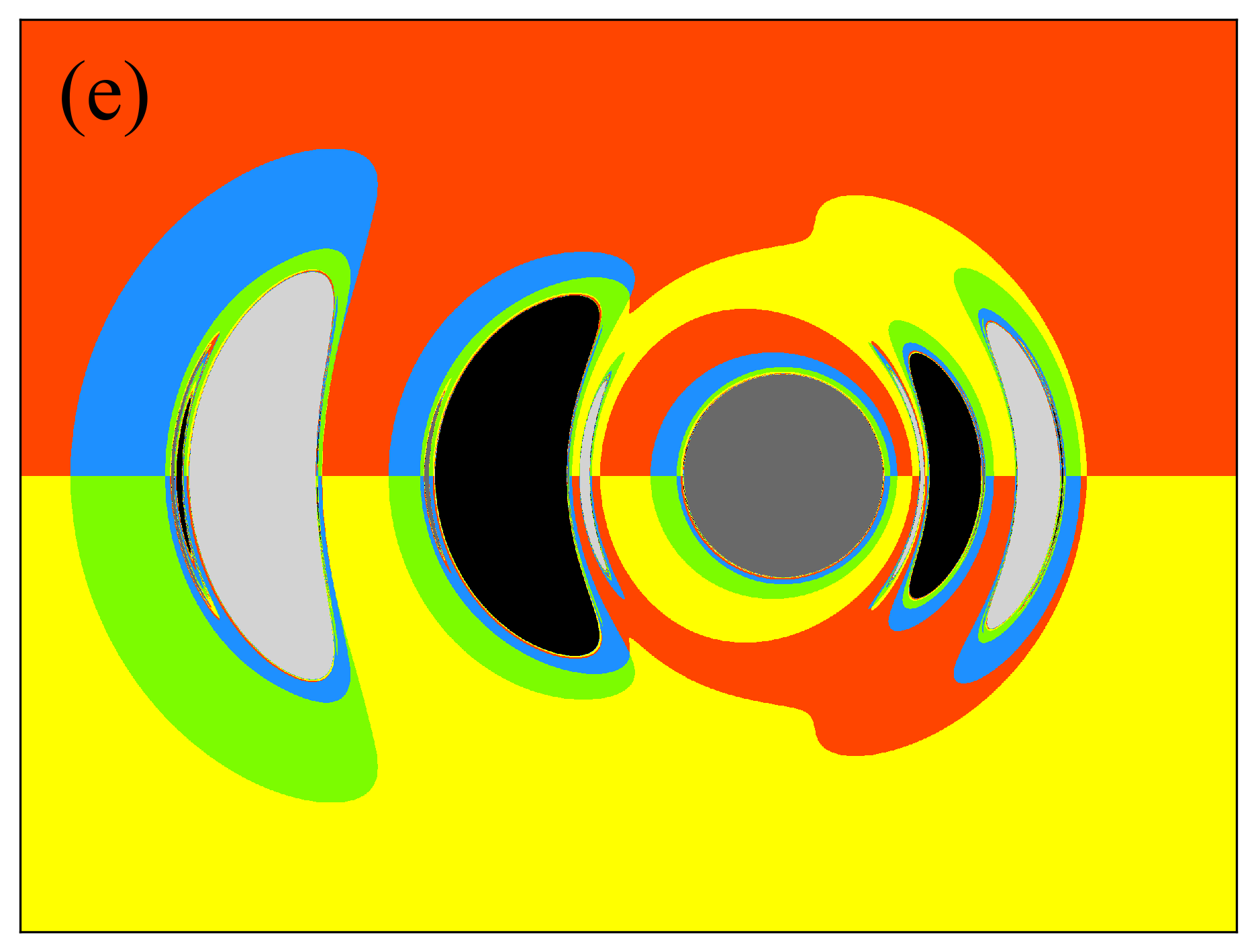}
\includegraphics[width=4cm]{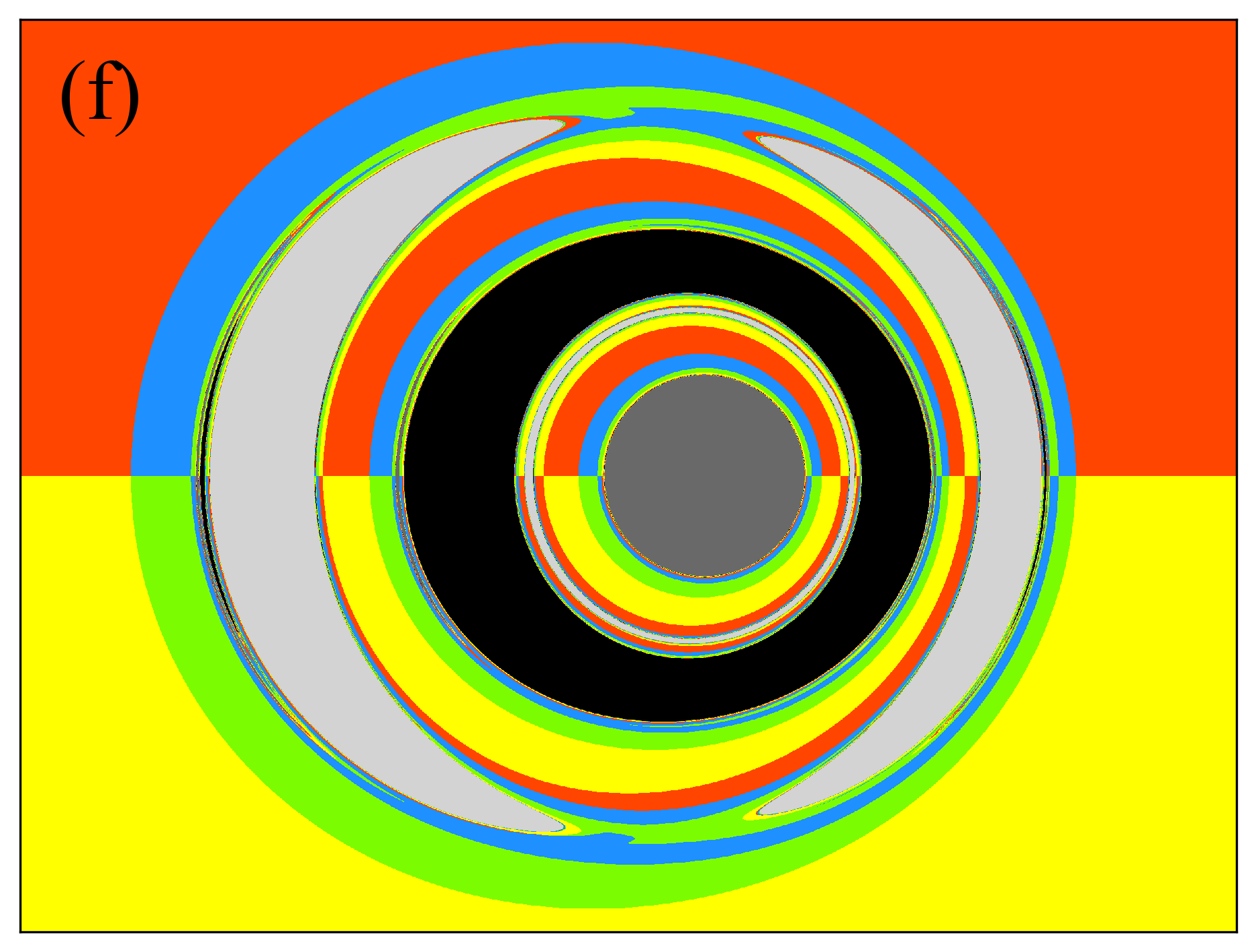}
\includegraphics[width=4cm]{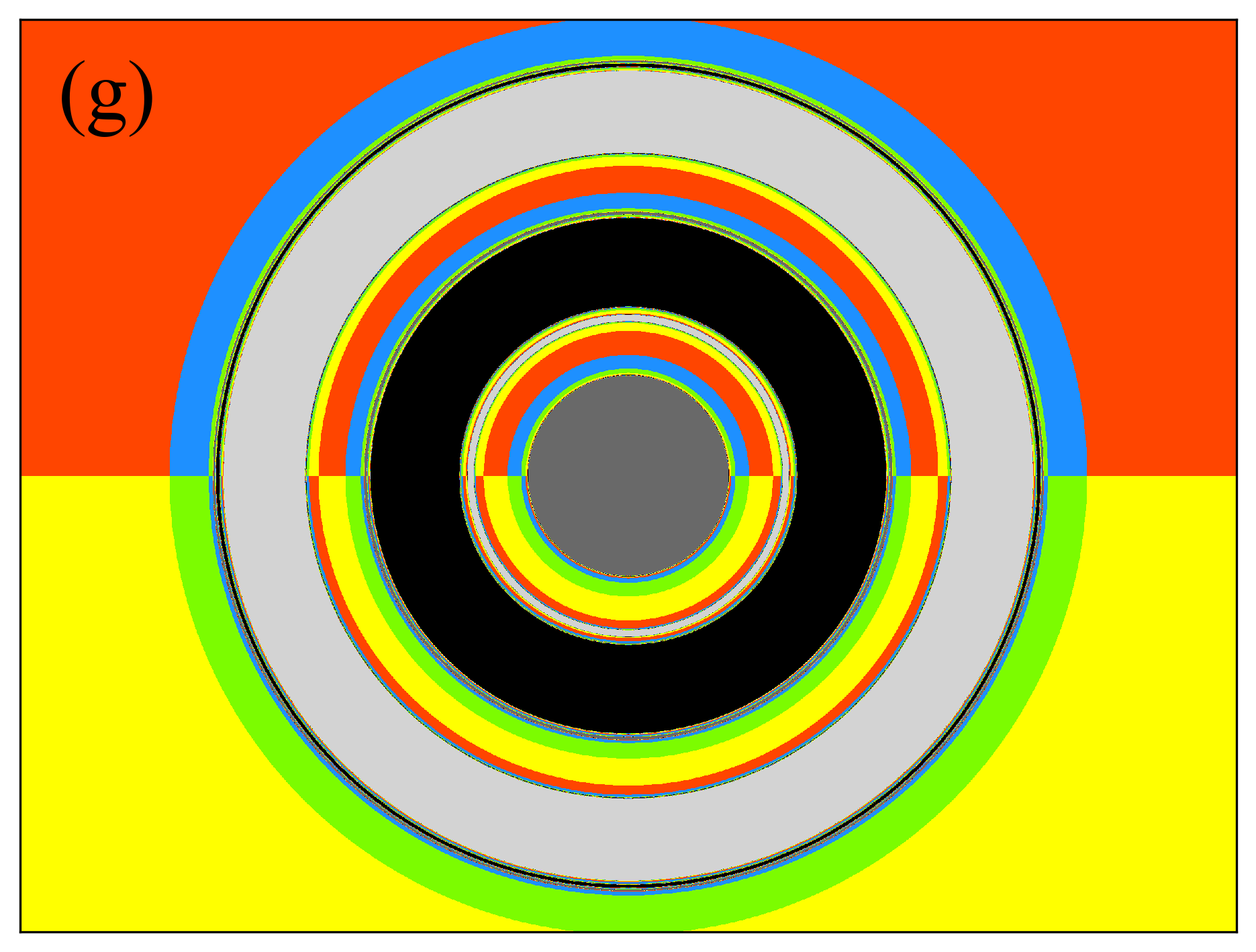}
\includegraphics[width=4cm]{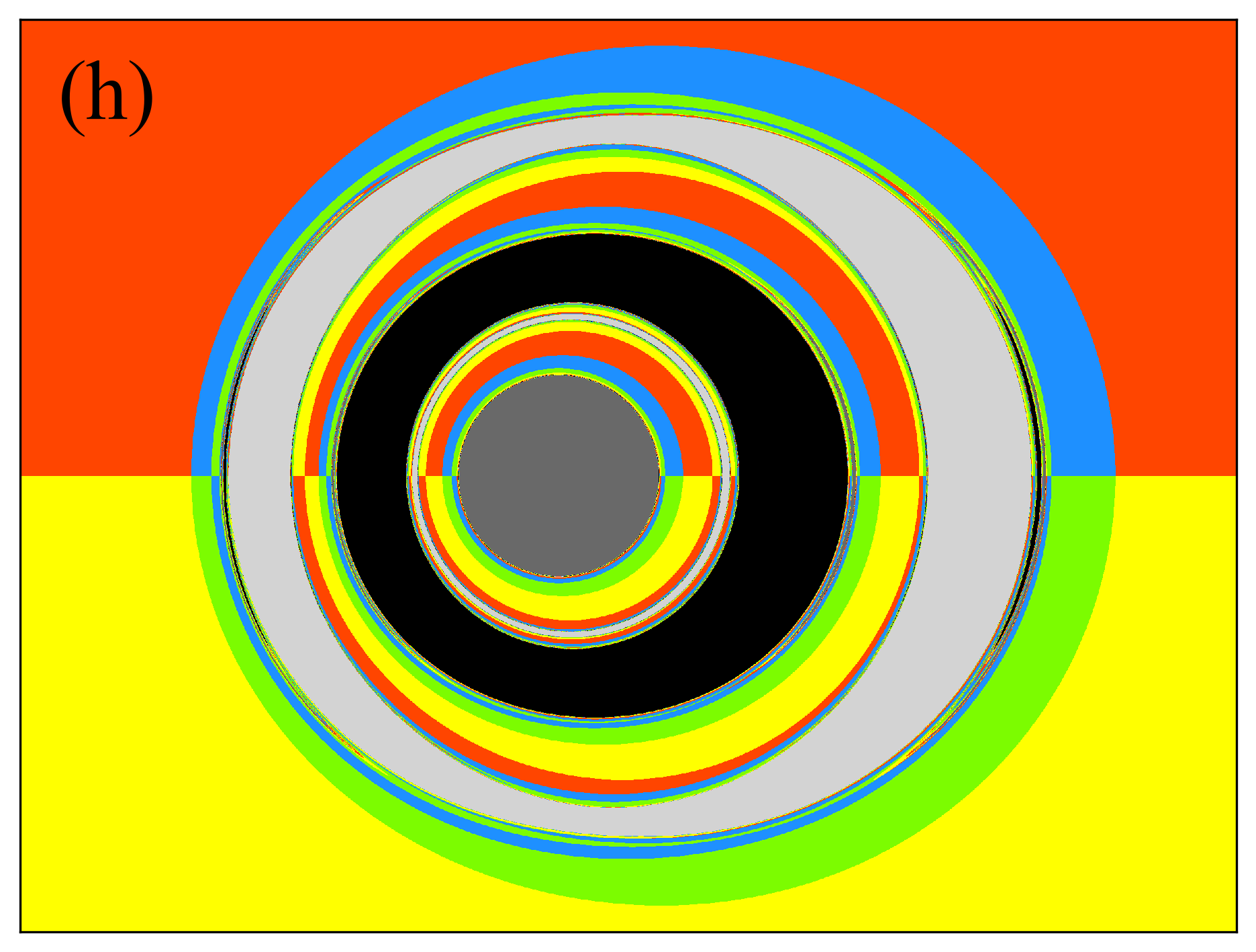}
\includegraphics[width=4cm]{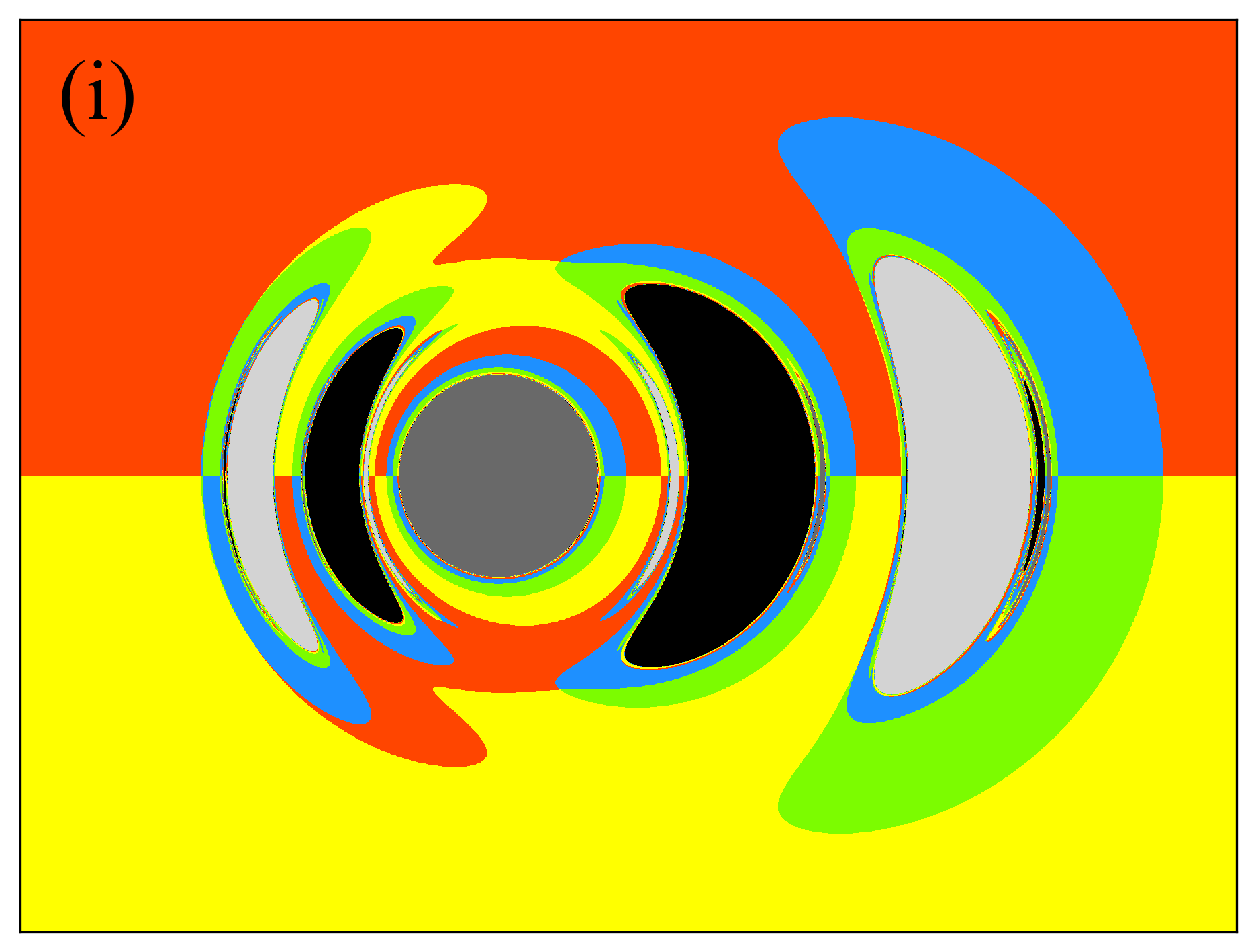}
\includegraphics[width=4cm]{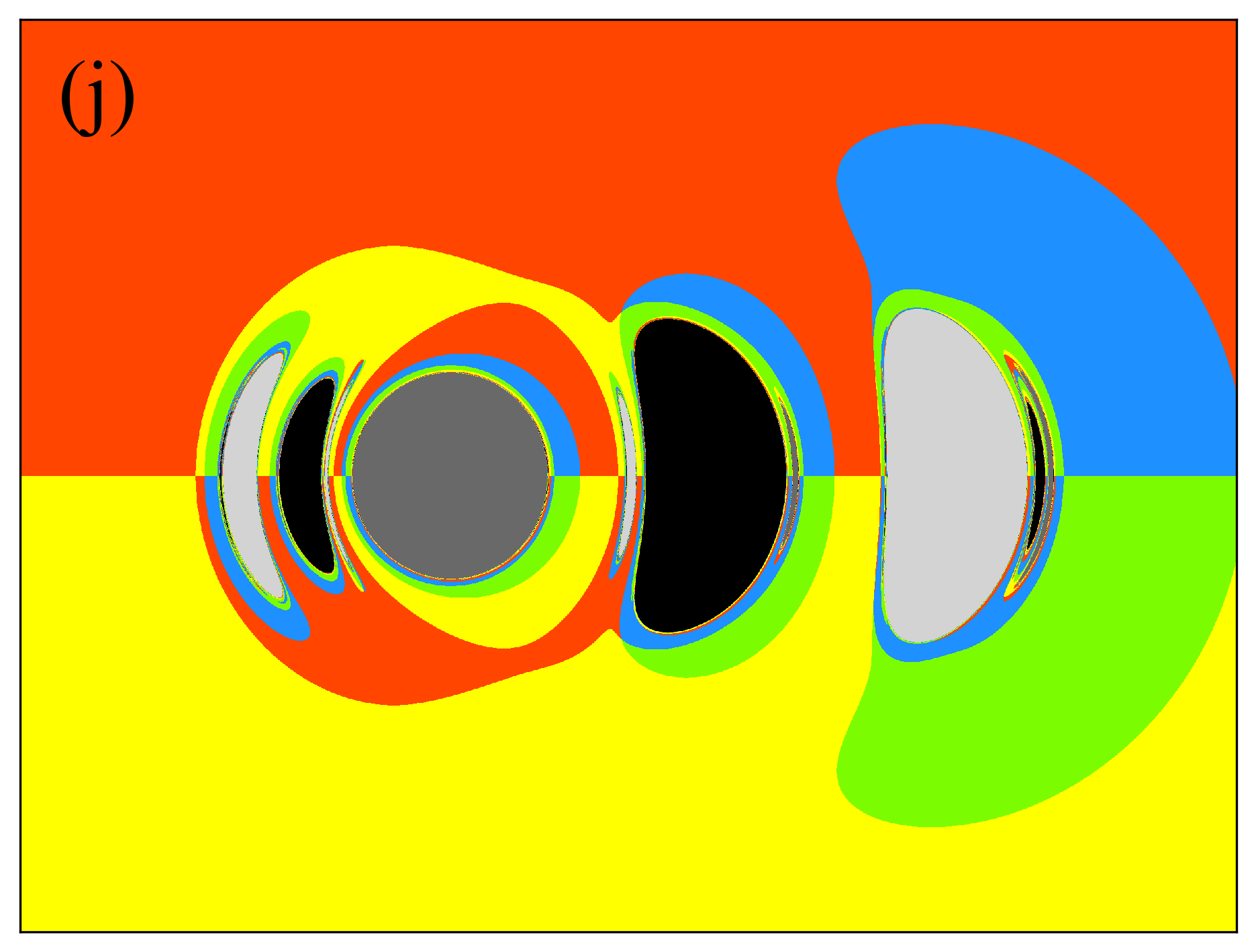}
\includegraphics[width=4cm]{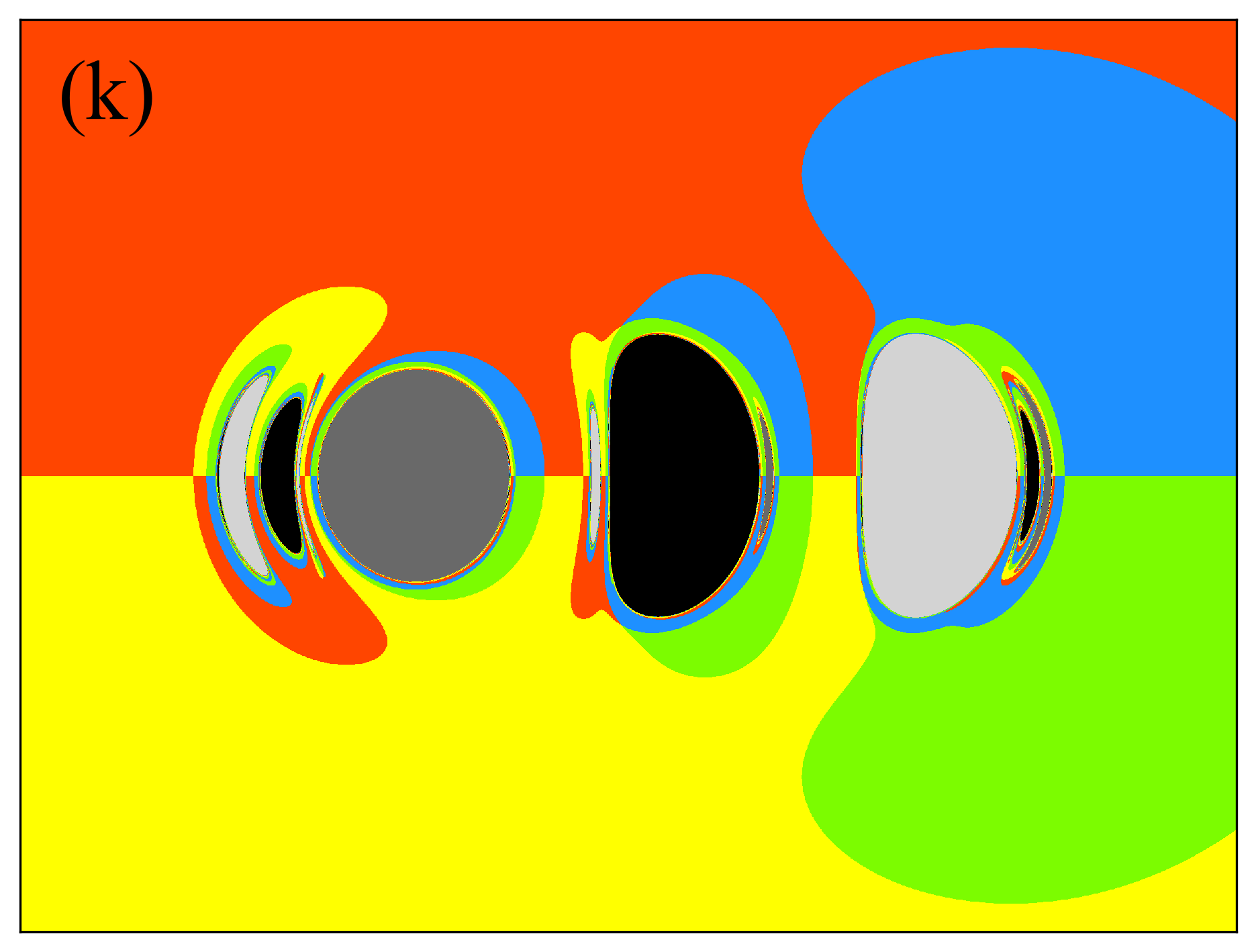}
\includegraphics[width=4cm]{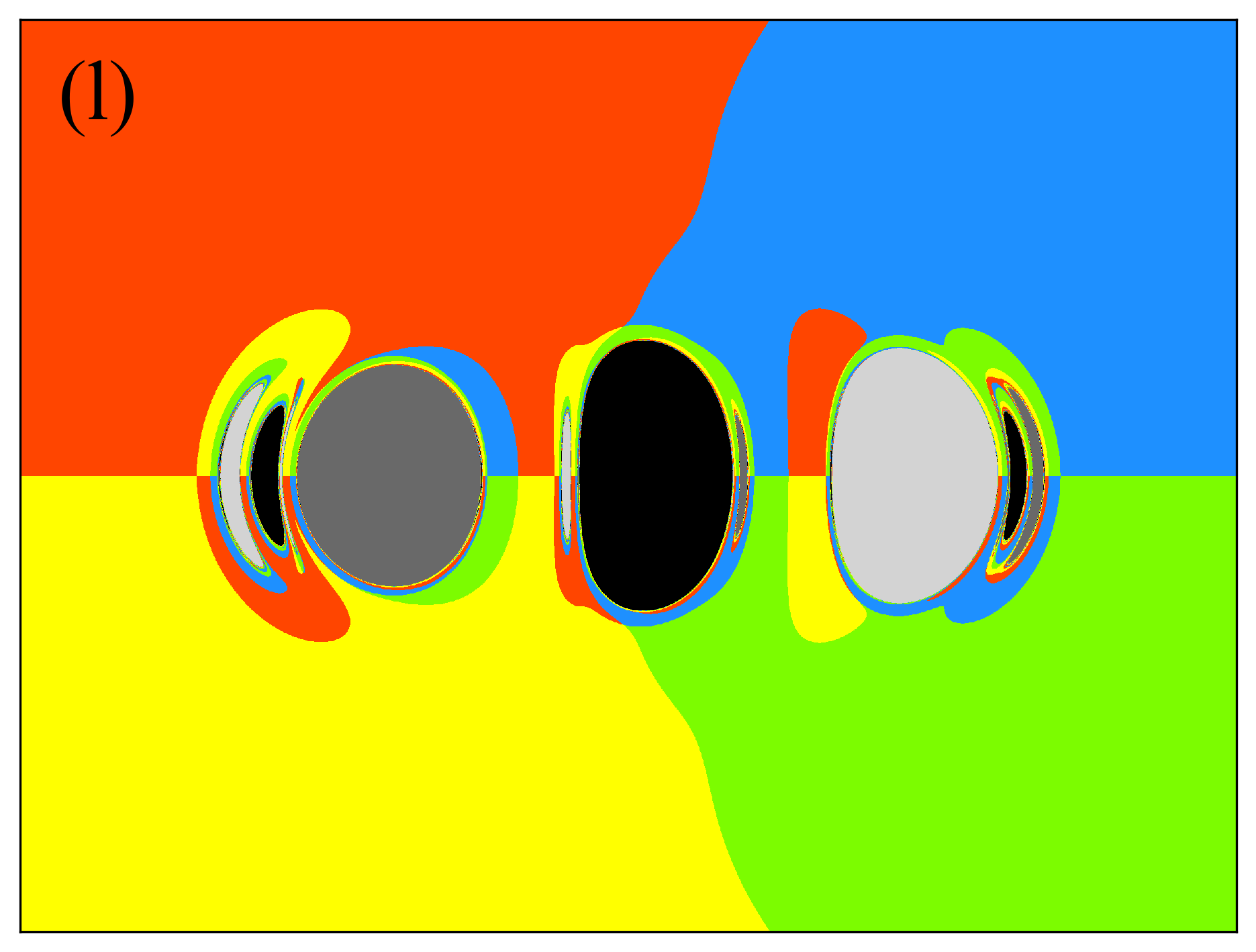}
\includegraphics[width=4cm]{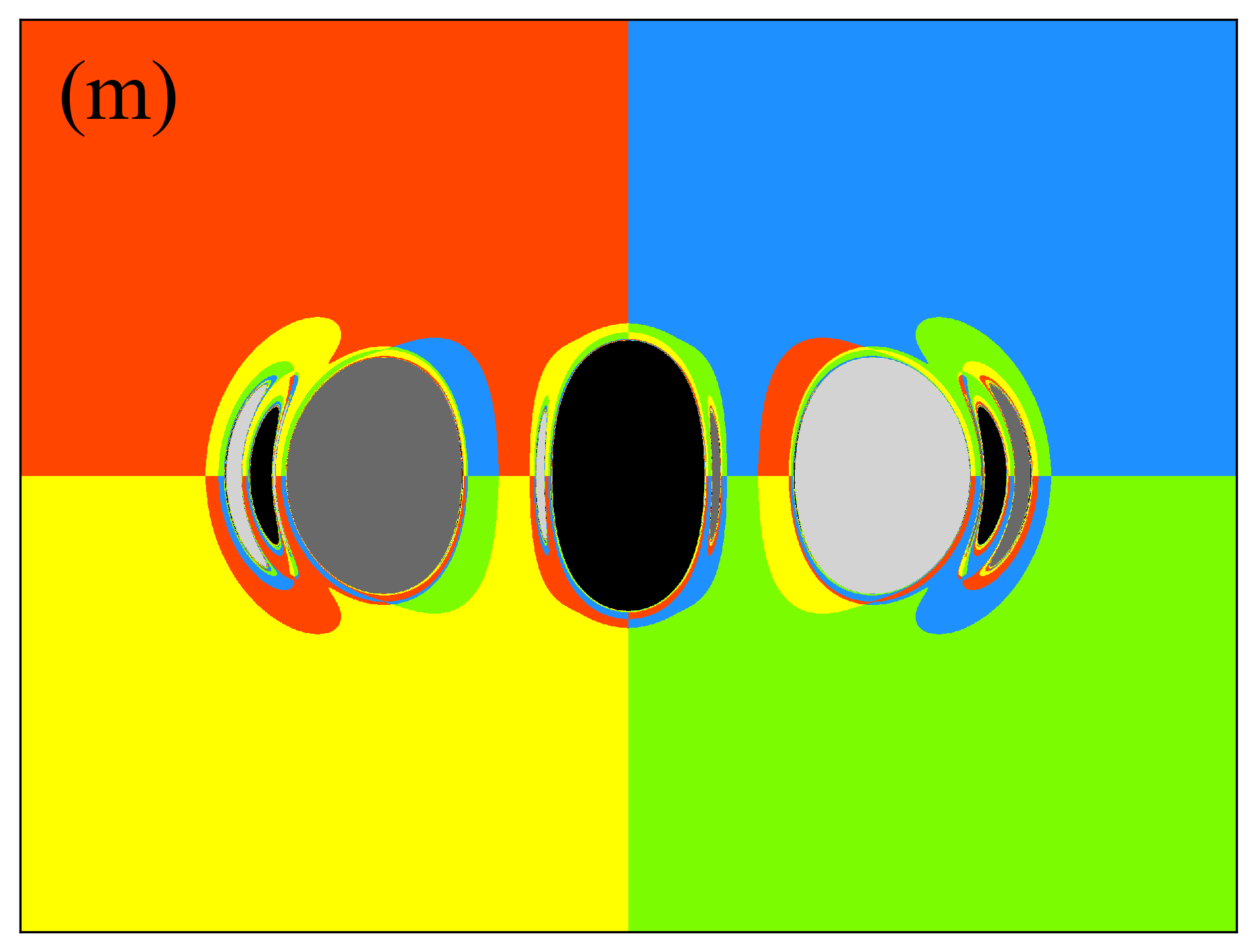}
\includegraphics[width=4cm]{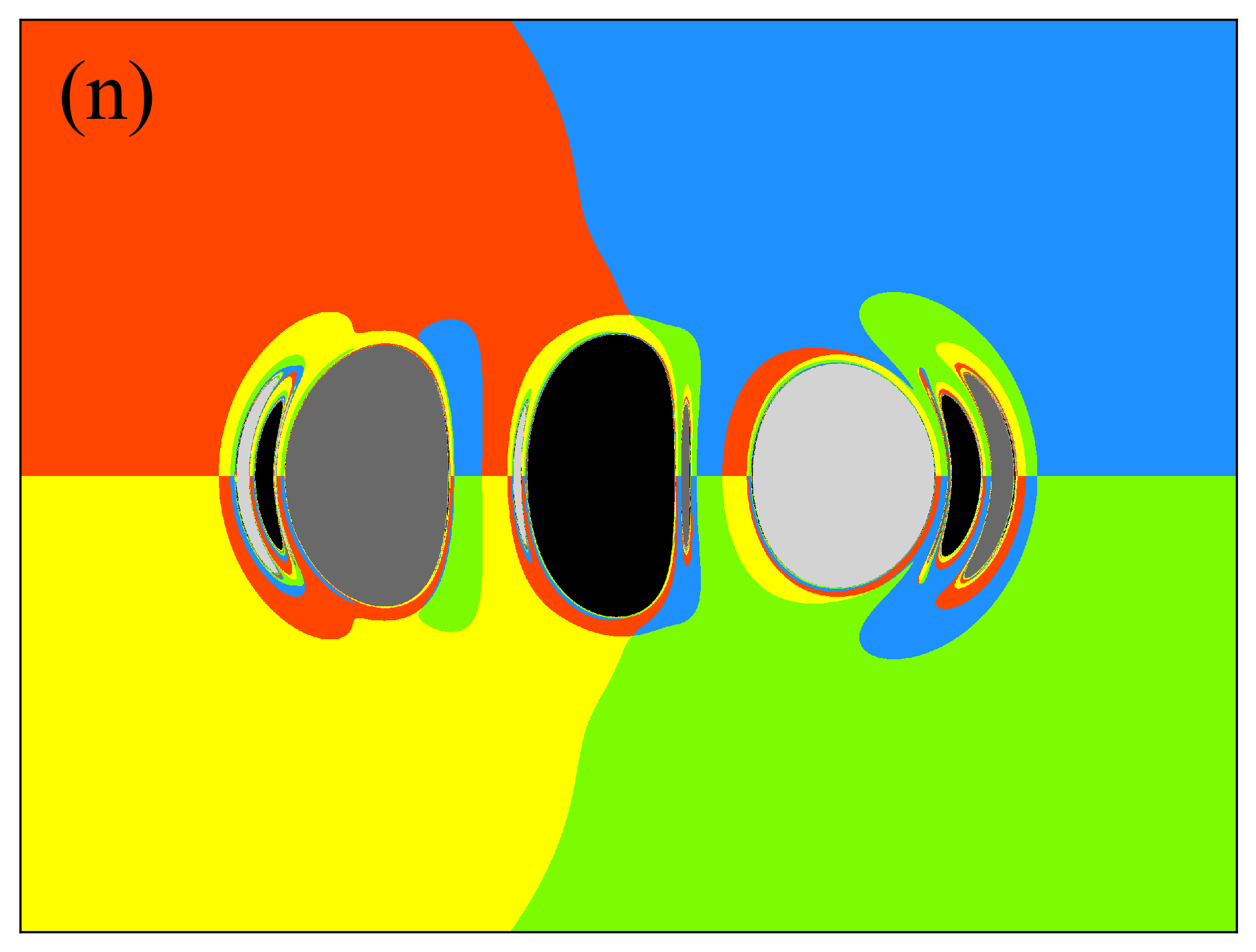}
\includegraphics[width=4cm]{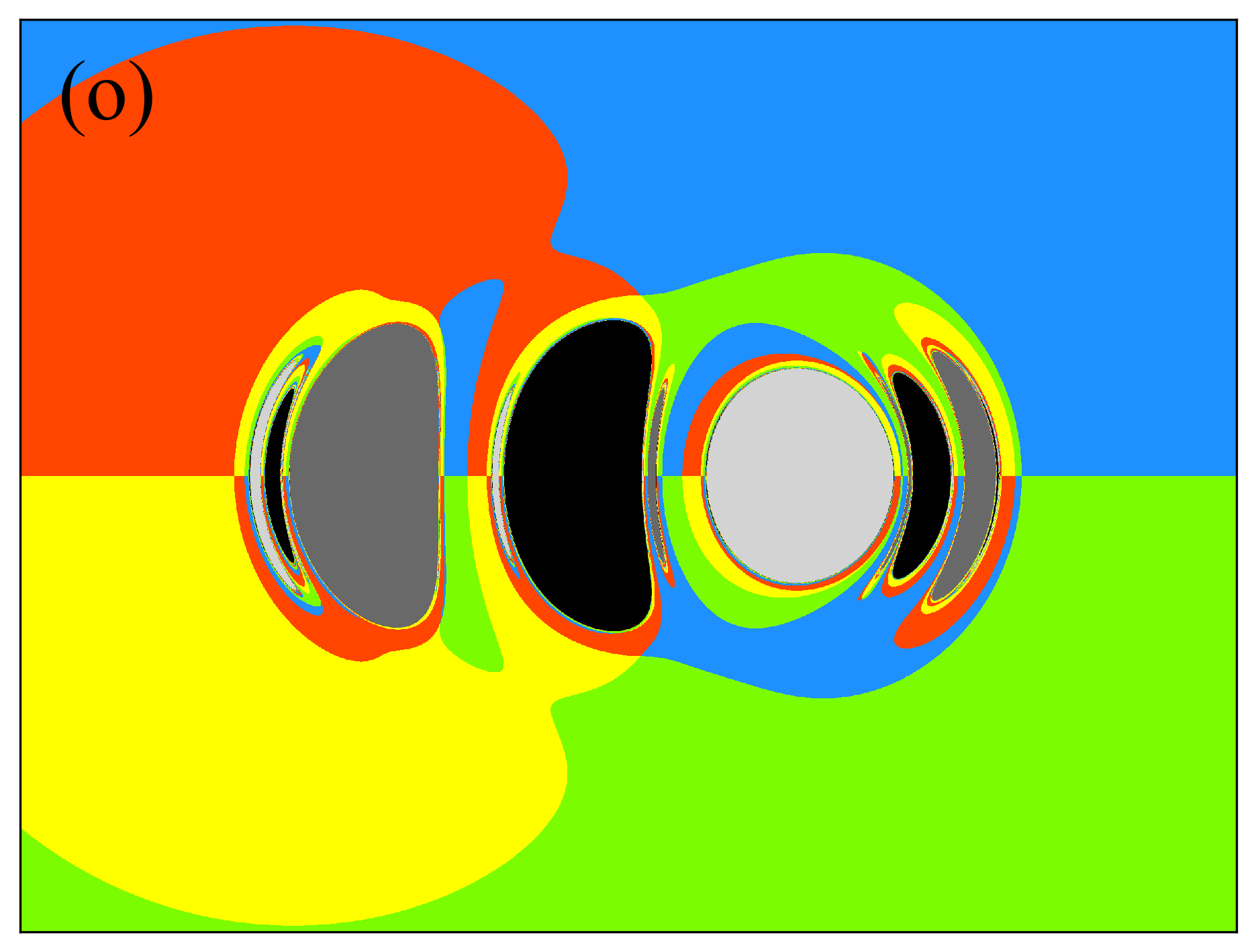}
\includegraphics[width=4cm]{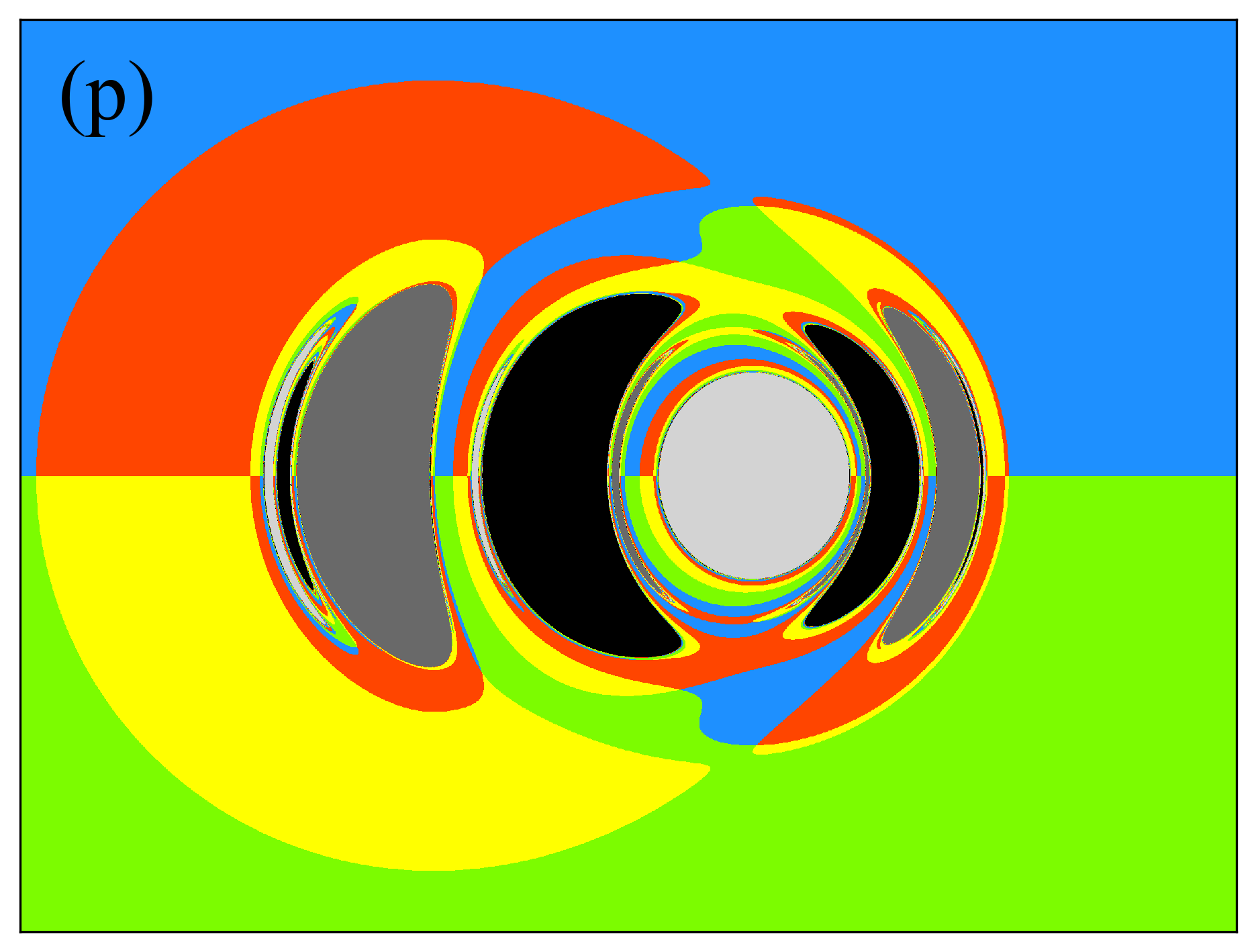}
\includegraphics[width=4cm]{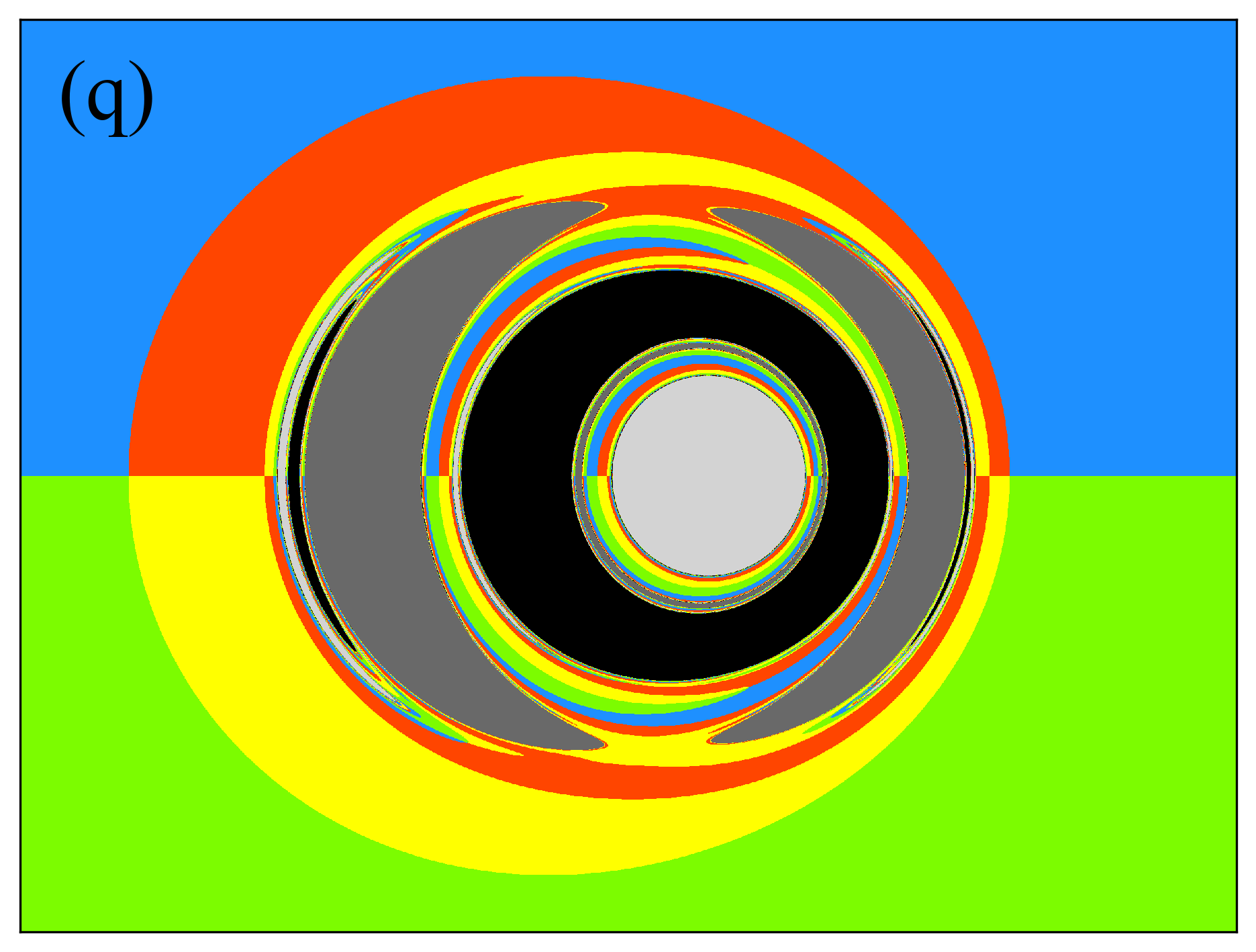}
\includegraphics[width=4cm]{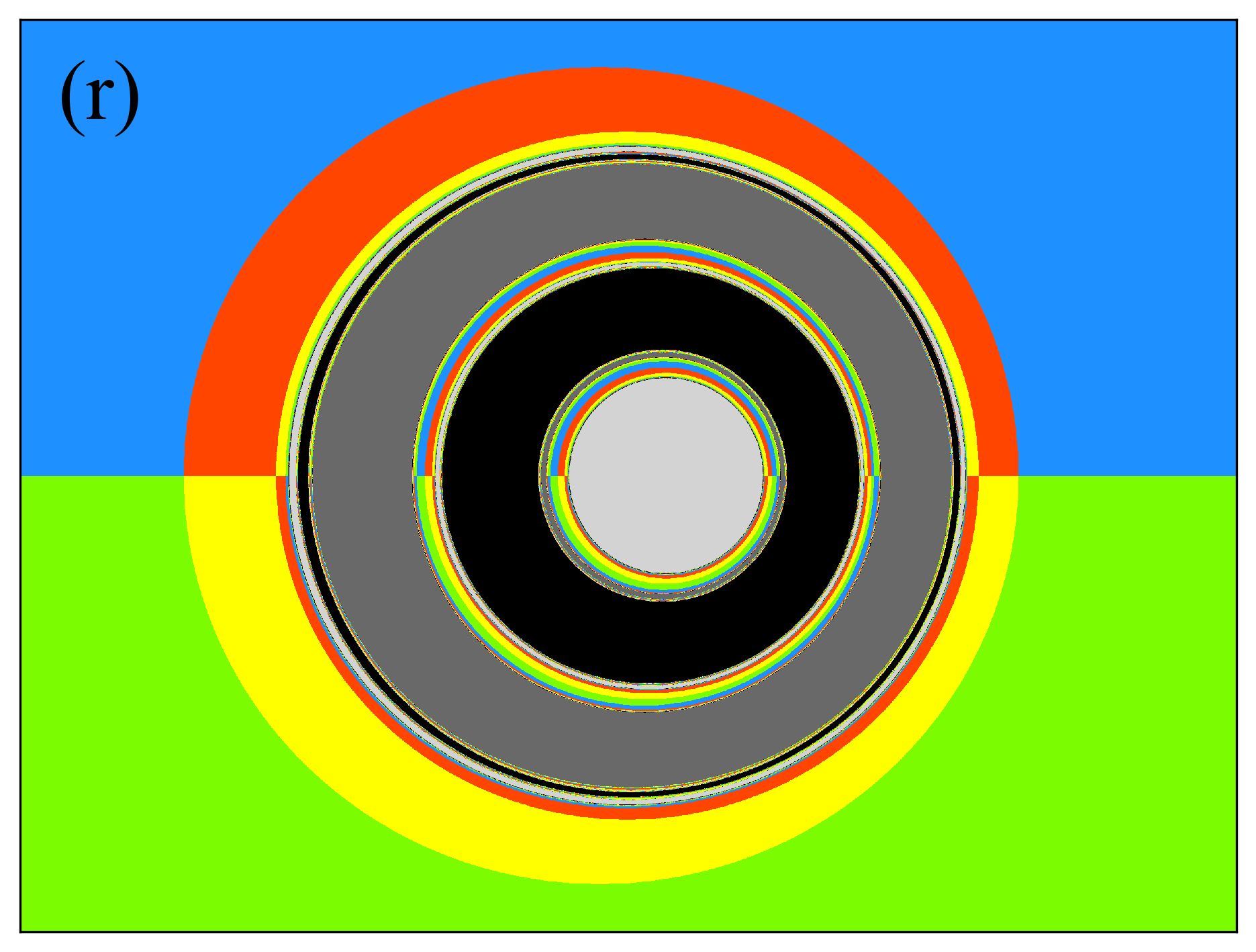}
\includegraphics[width=4cm]{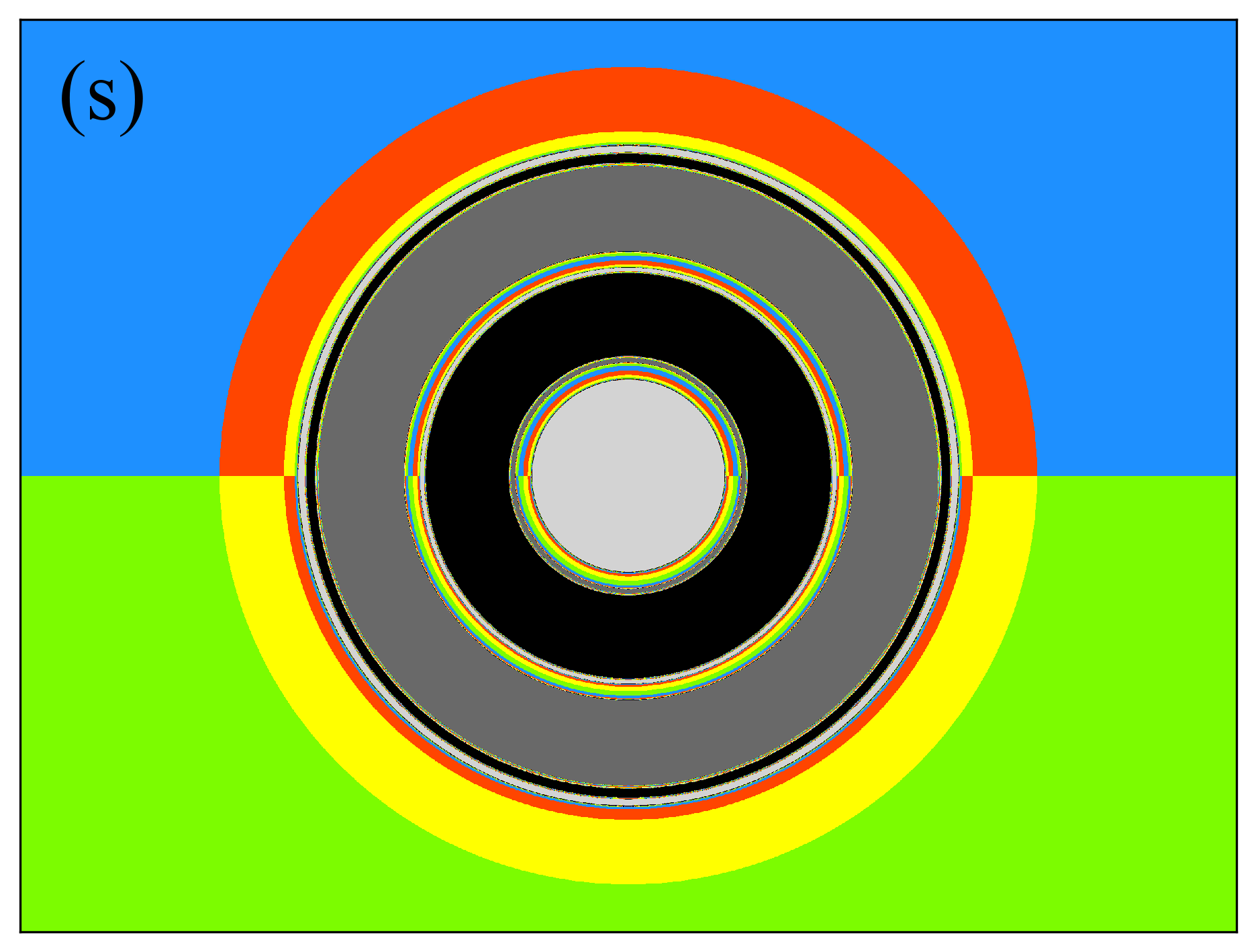}
\includegraphics[width=4cm]{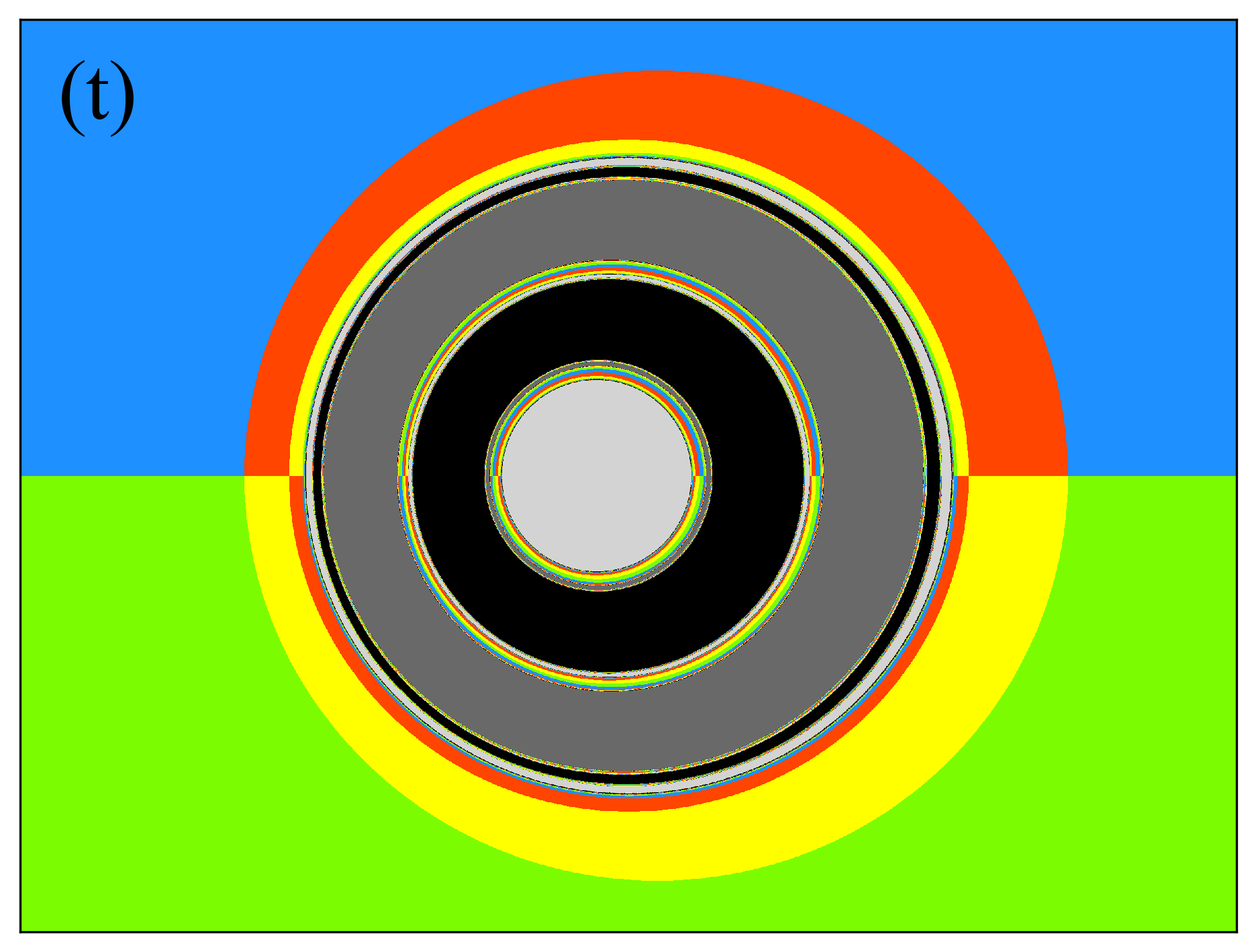}
\includegraphics[width=4cm]{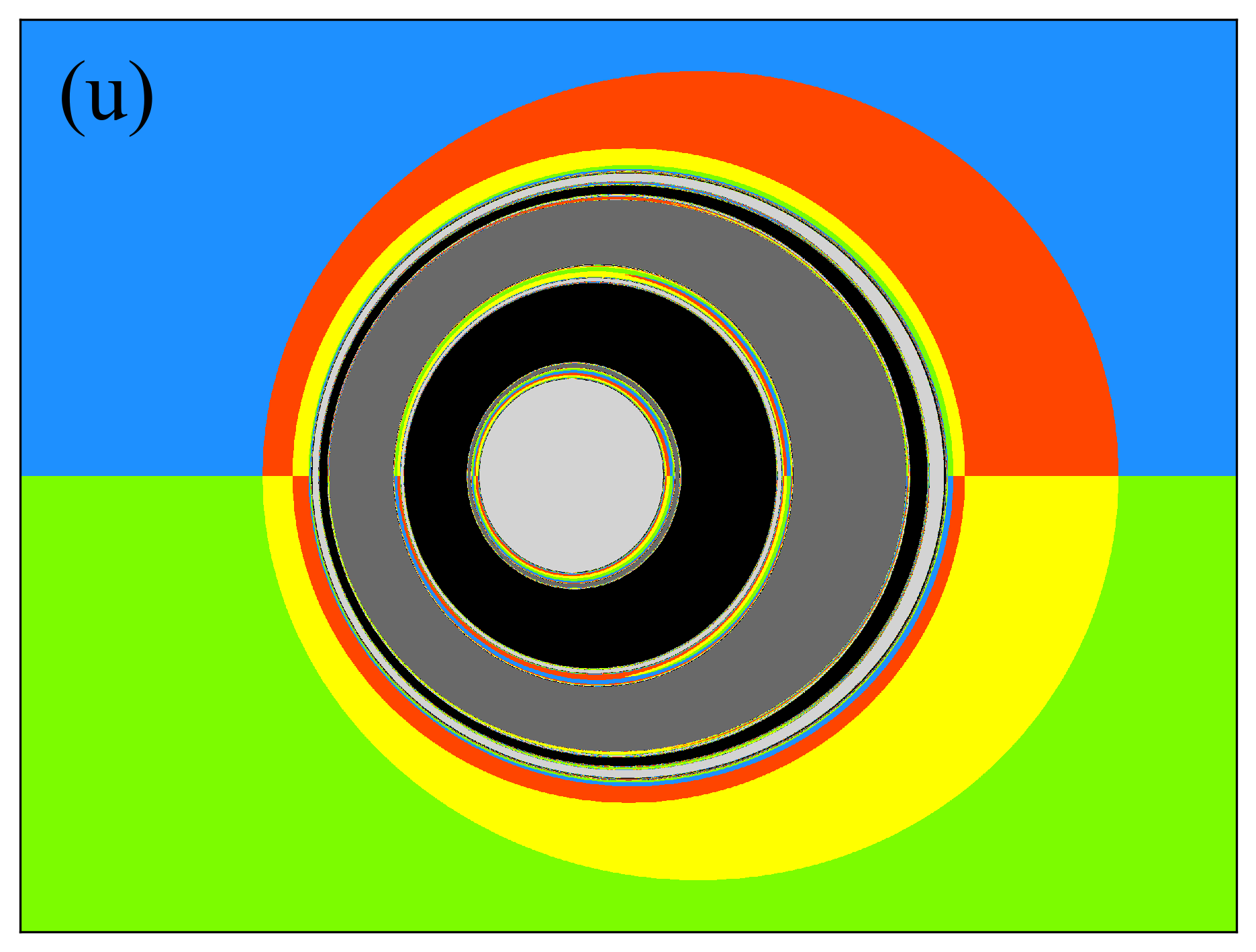}
\includegraphics[width=4cm]{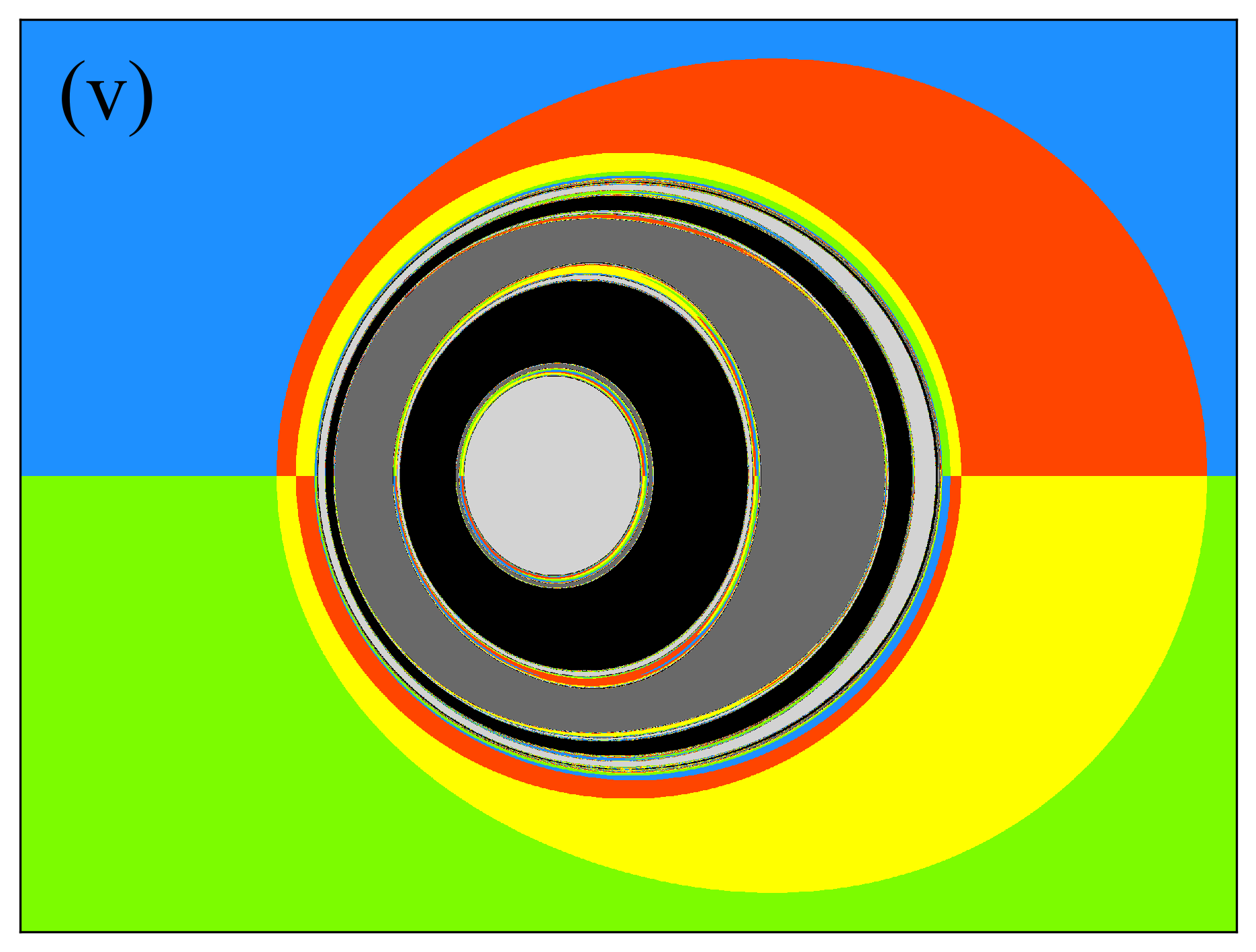}
\includegraphics[width=4cm]{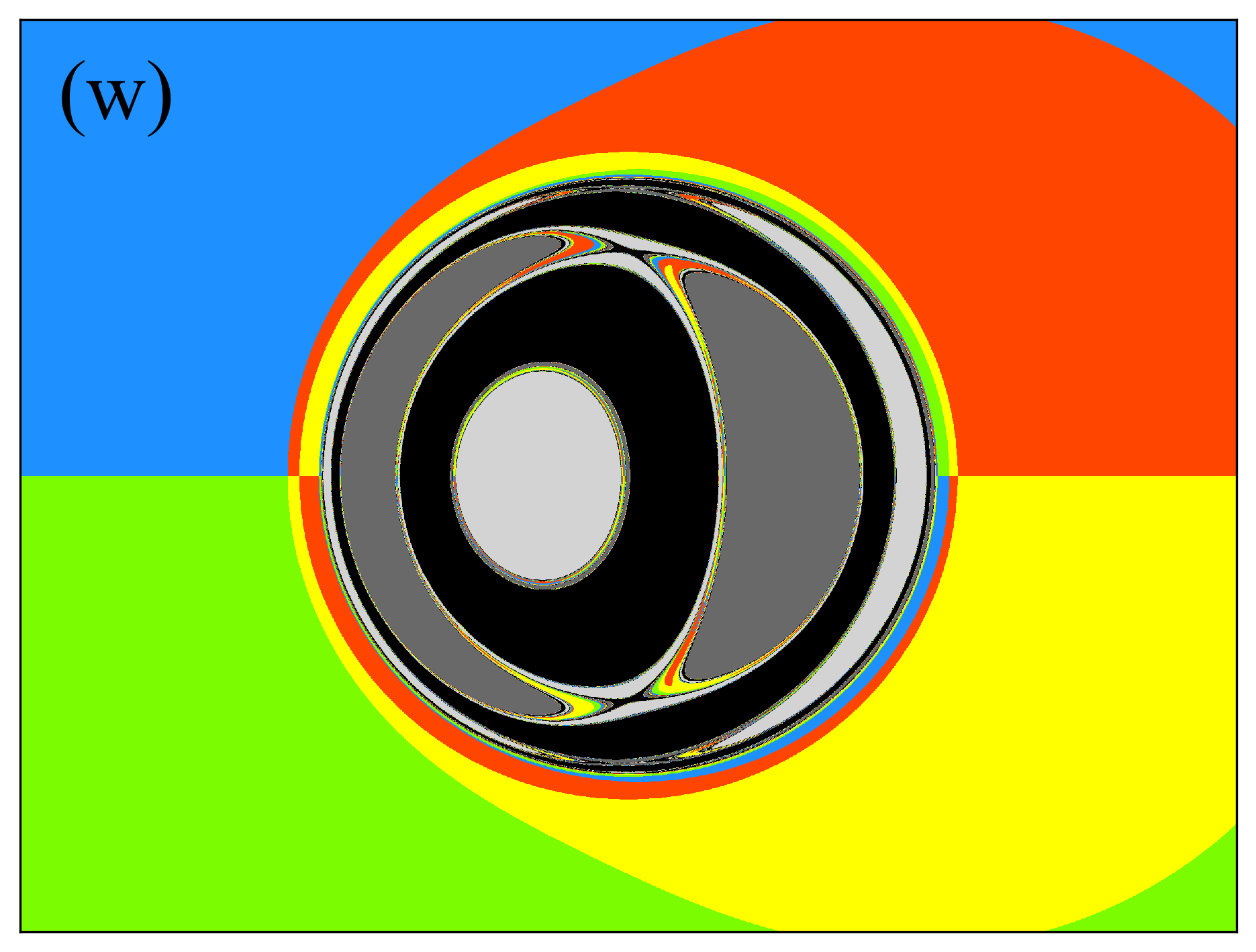}
\includegraphics[width=4cm]{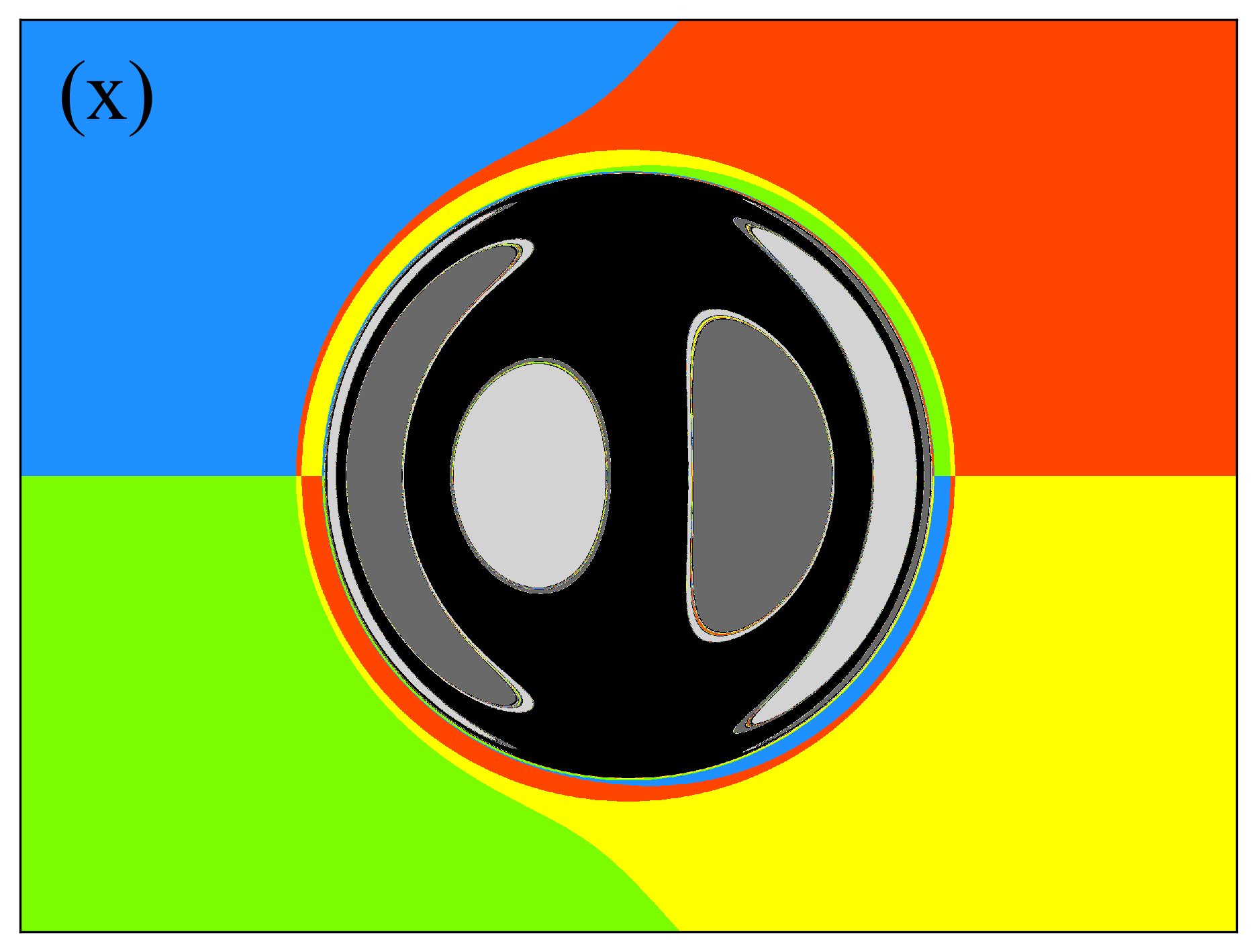}
\caption{Shadows of triple black holes across varying separations $l$ and azimuthal angles $\Phi$. From panel (a) to (x), the separation decreases from $l=4$ to $0.1$ in increments of $\delta l\approx-0.1695$, while $\Phi$ ranges from $0^{\circ}$ to $345^{\circ}$ with $\delta \Phi = 15^{\circ}$ intervals, all at a fixed observation inclination of $\Theta=90^{\circ}$. These results provide qualitative insights into shadow evolution during the inspiral and merger phases of triple black hole systems.}}\label{fig6}
\end{figure*}

\begin{figure*}%[tbph]
\center{
\includegraphics[width=5cm]{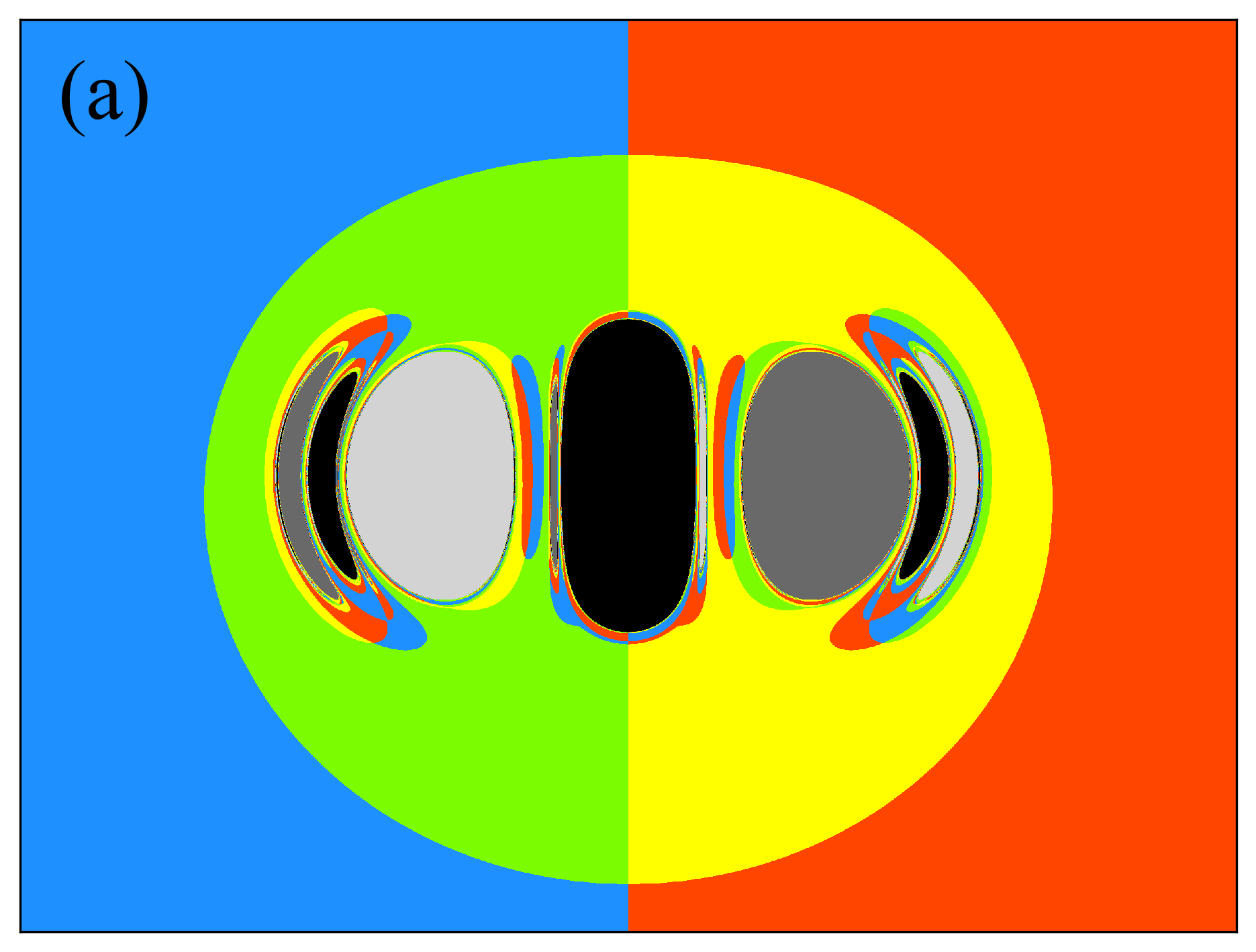}
\includegraphics[width=5cm]{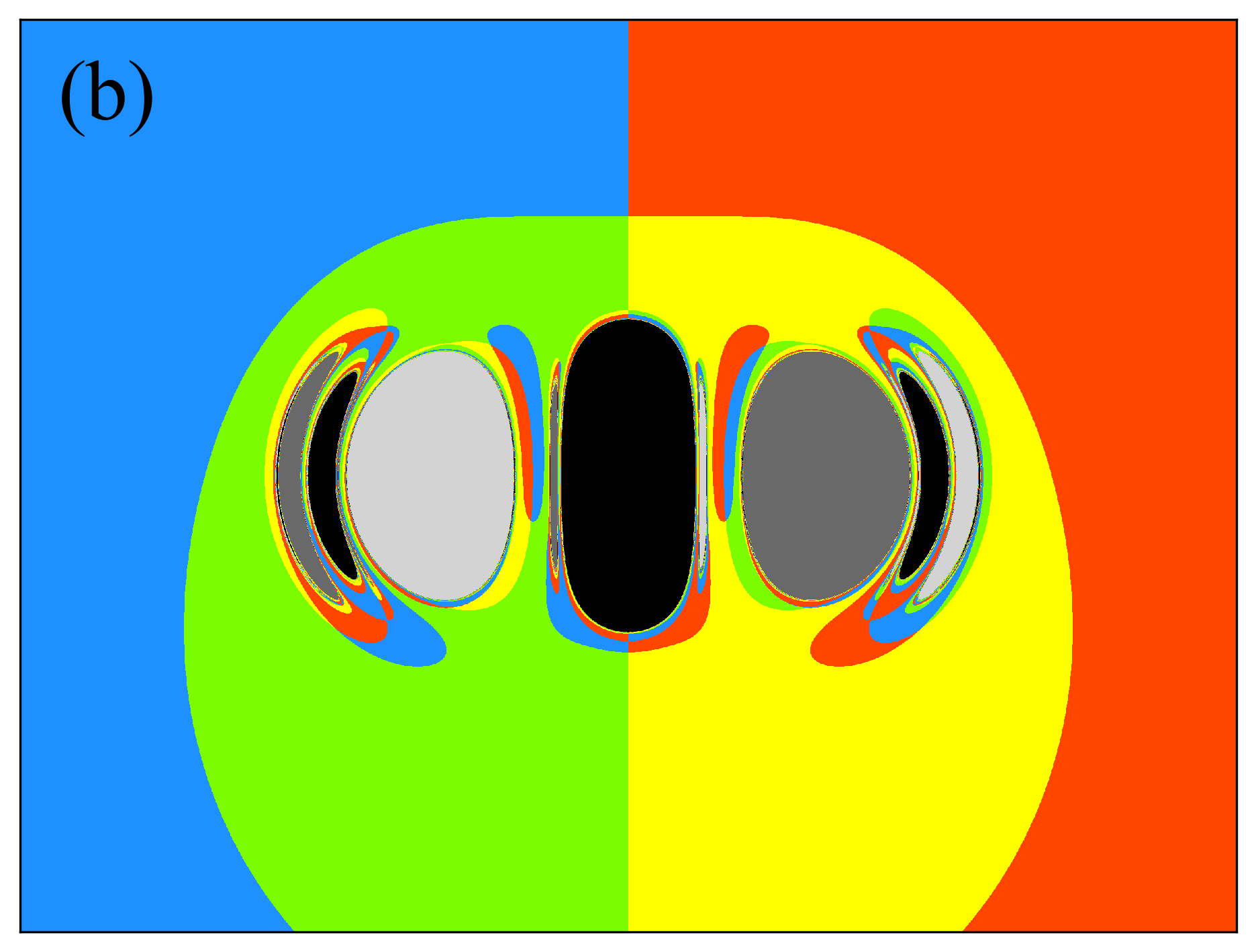}
\includegraphics[width=5cm]{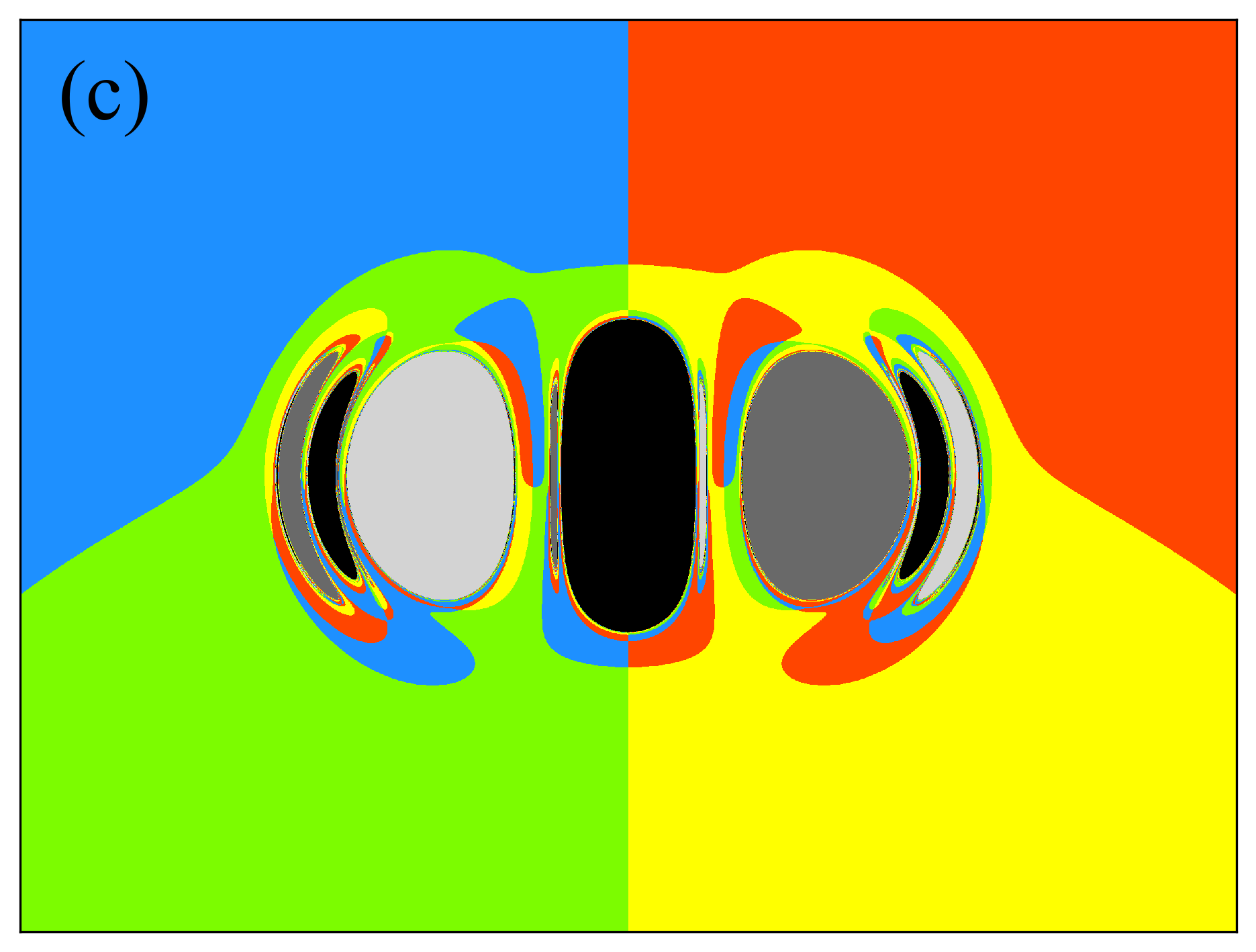}
\includegraphics[width=5cm]{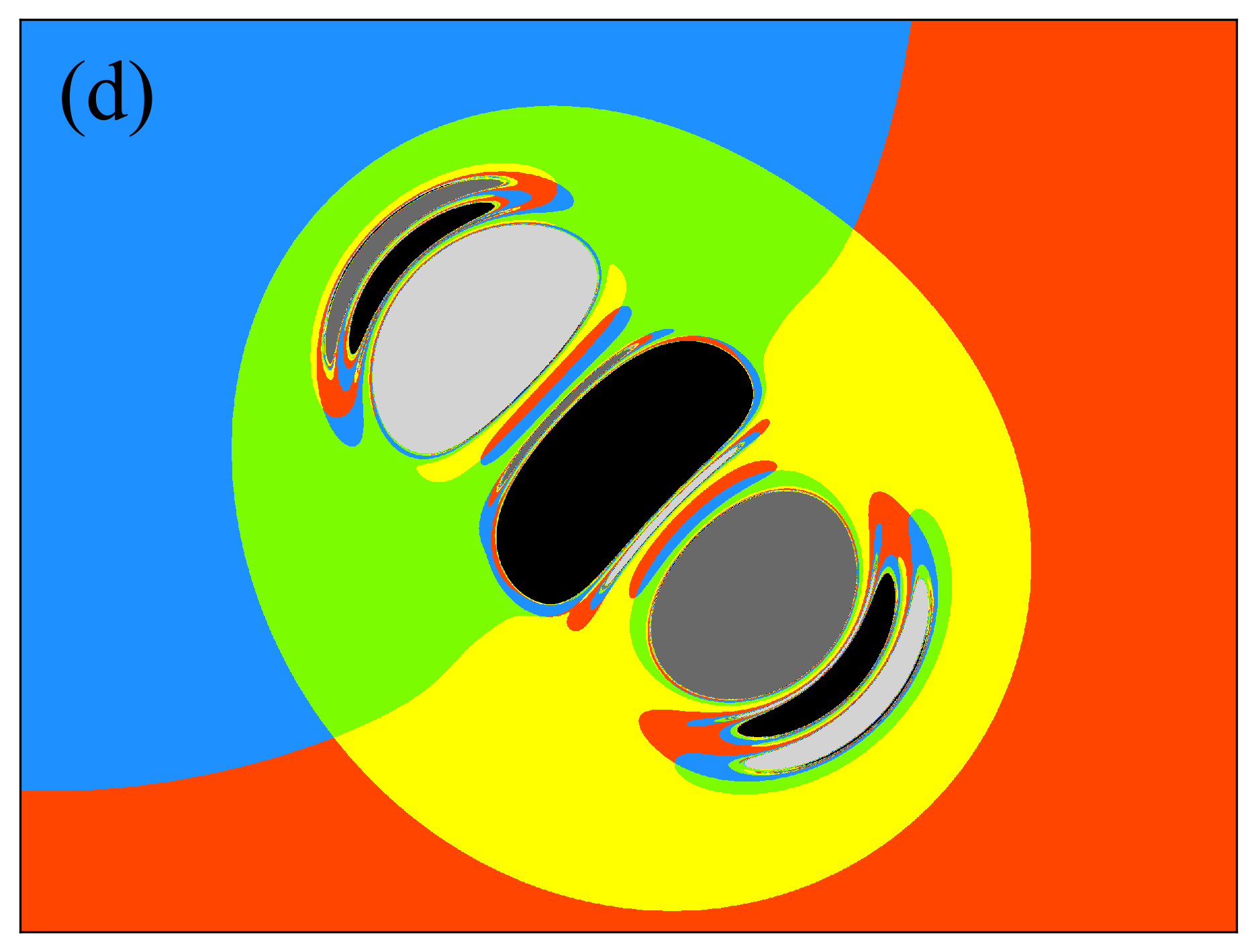}
\includegraphics[width=5cm]{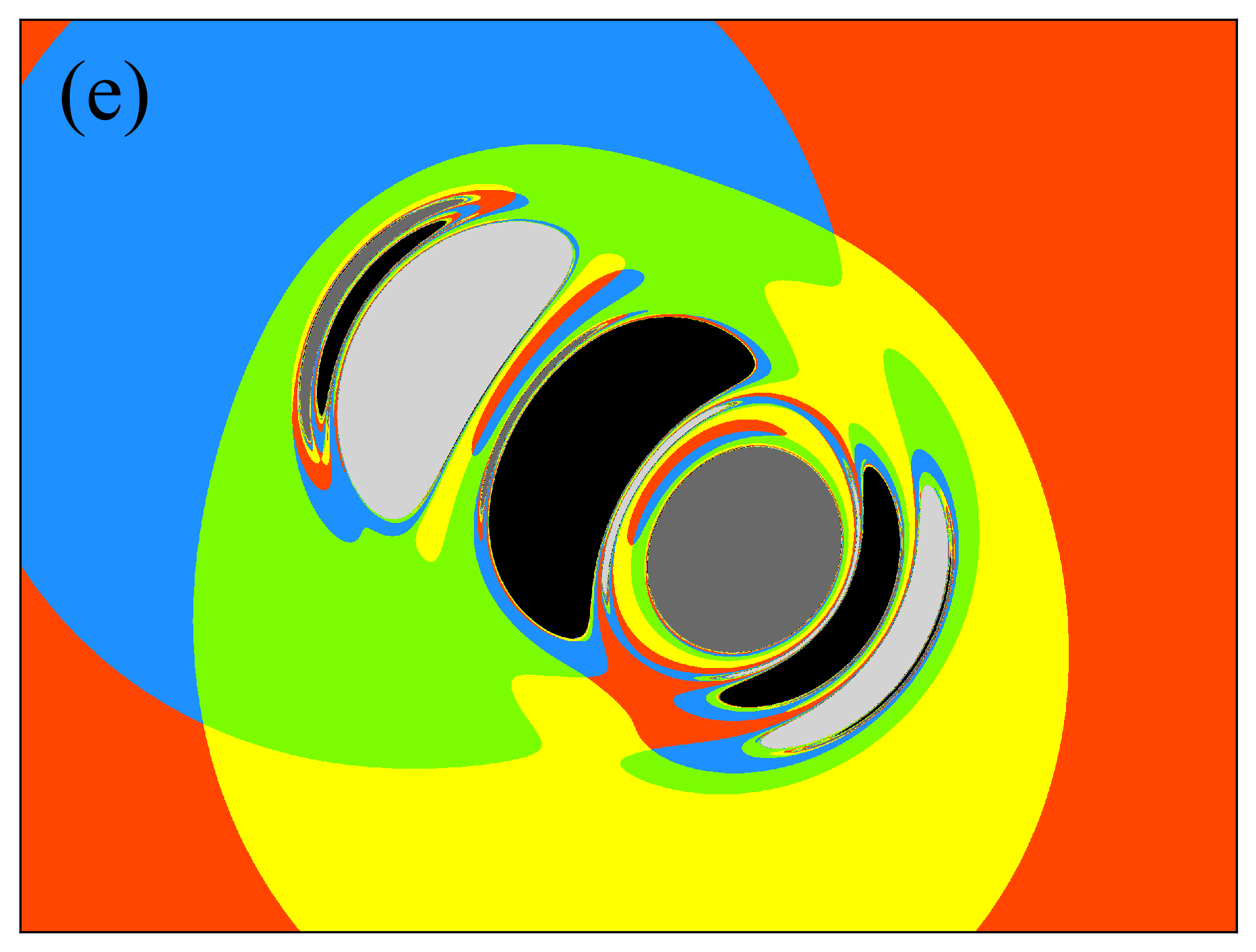}
\includegraphics[width=5cm]{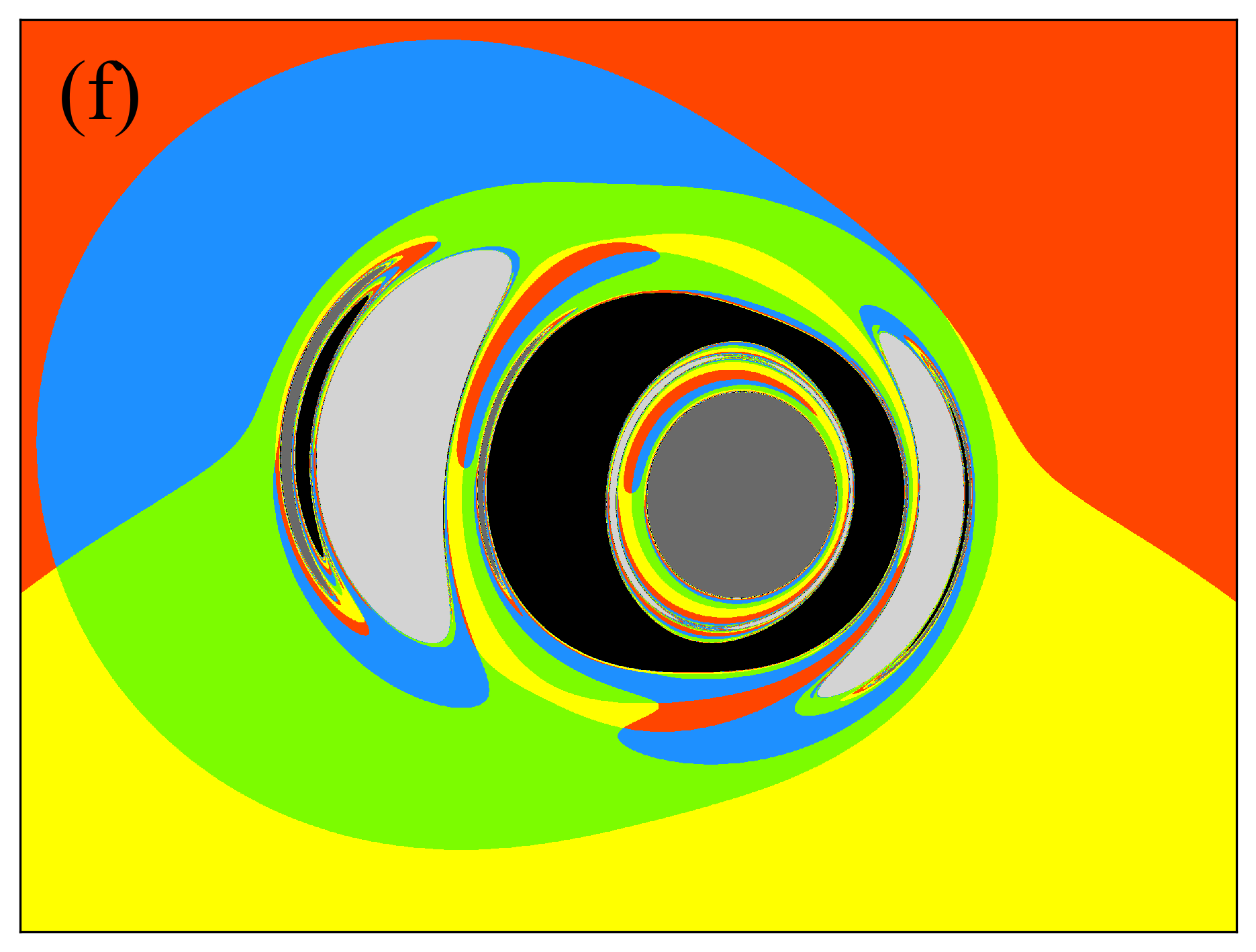}
\includegraphics[width=5cm]{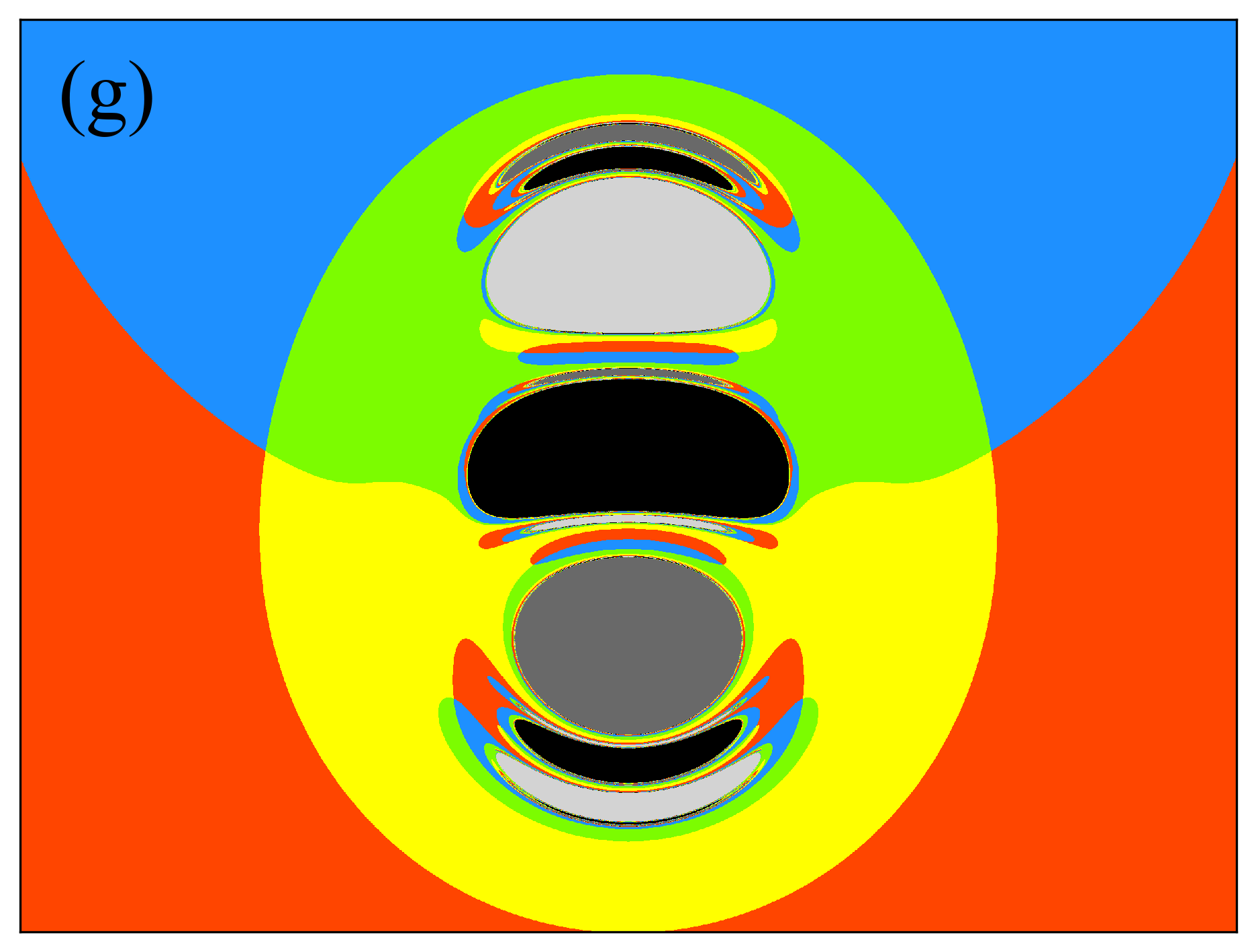}
\includegraphics[width=5cm]{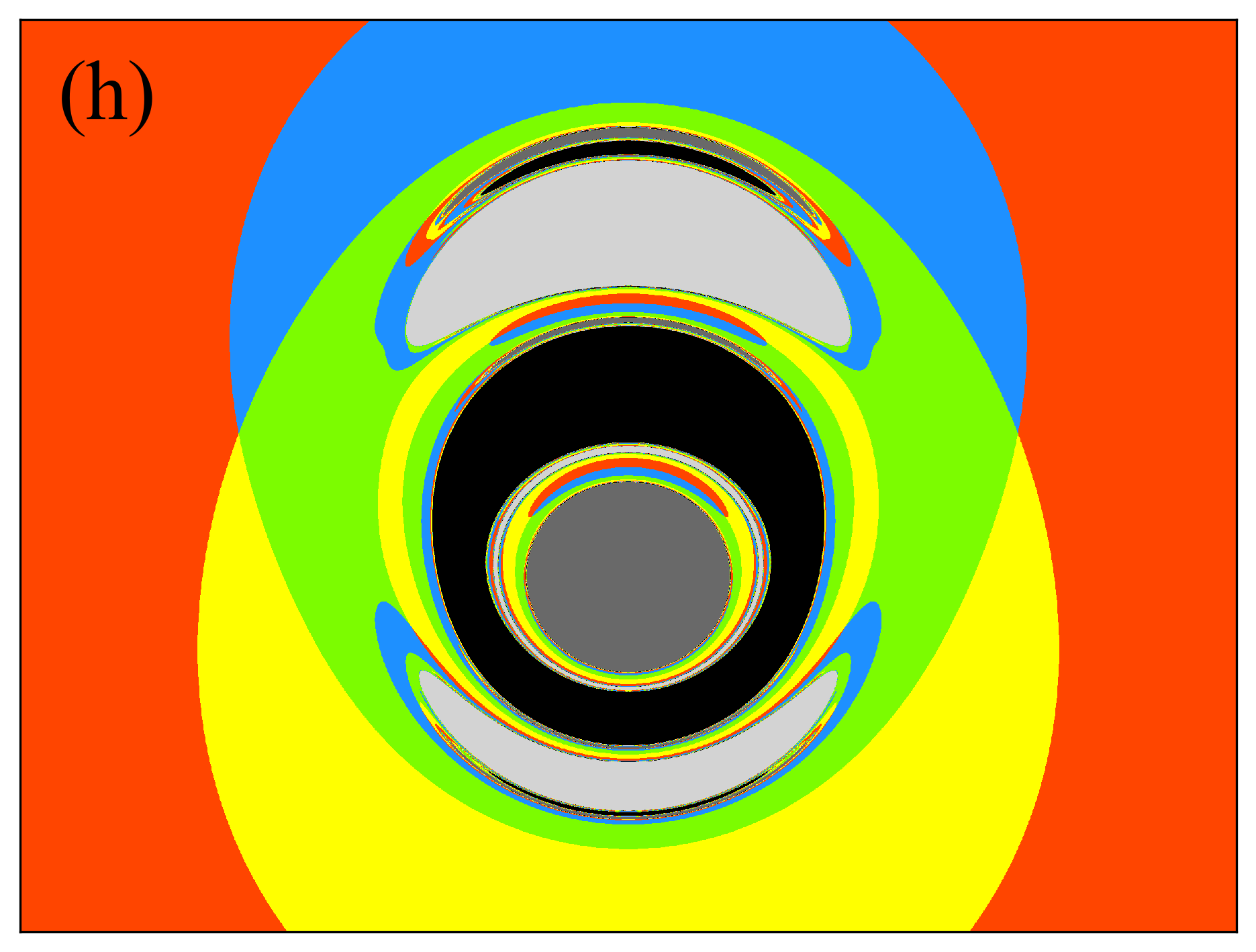}
\includegraphics[width=5cm]{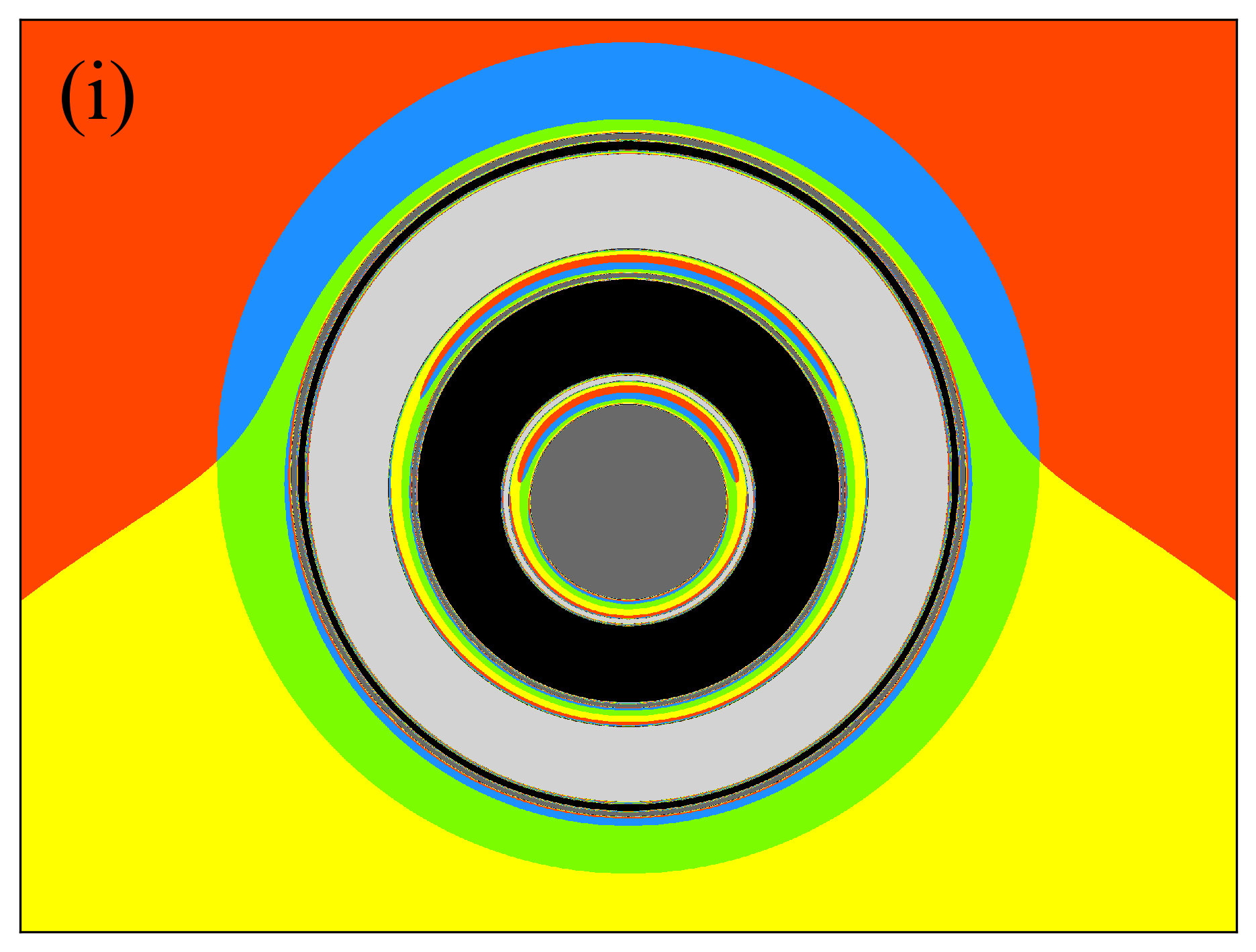}
\includegraphics[width=5cm]{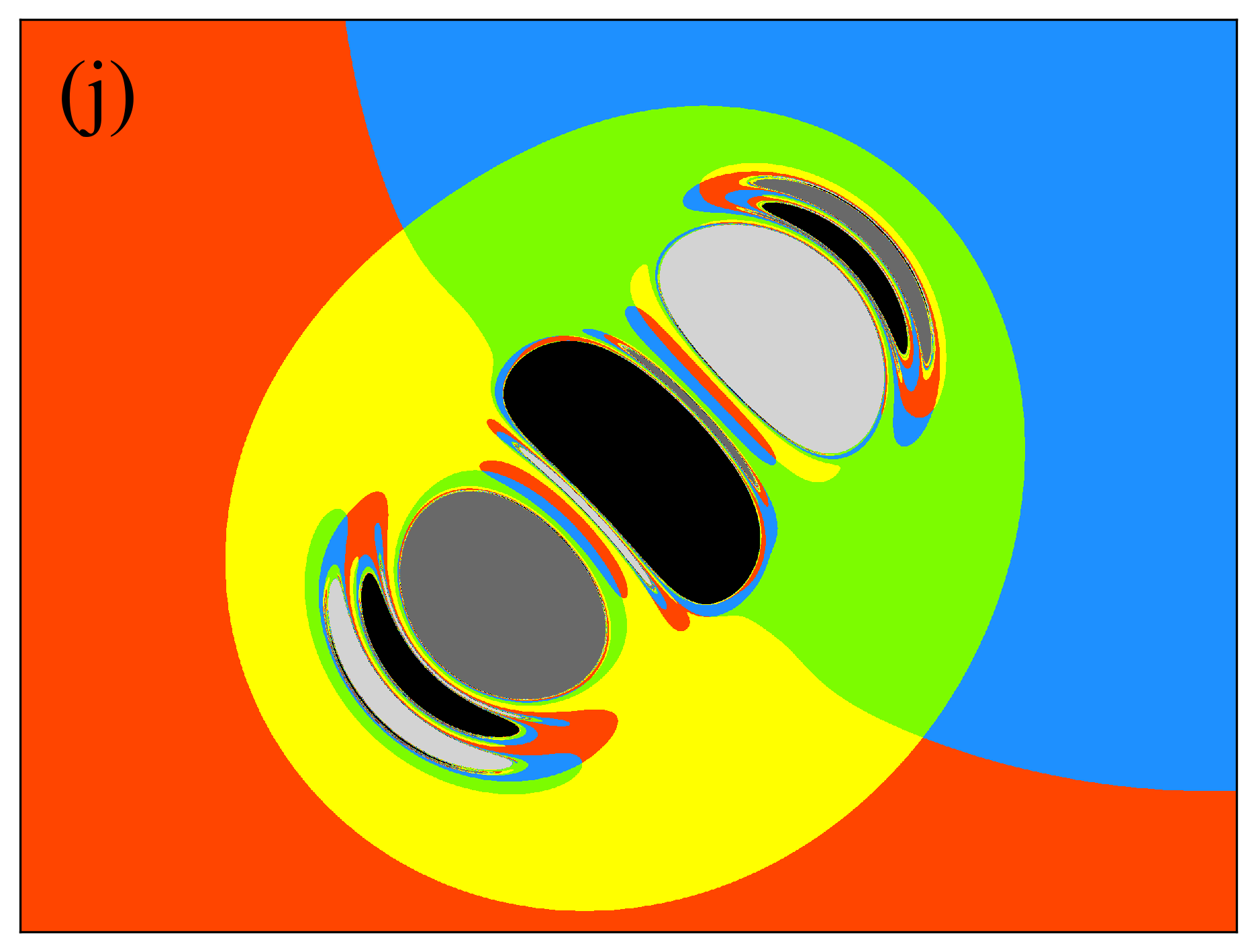}
\includegraphics[width=5cm]{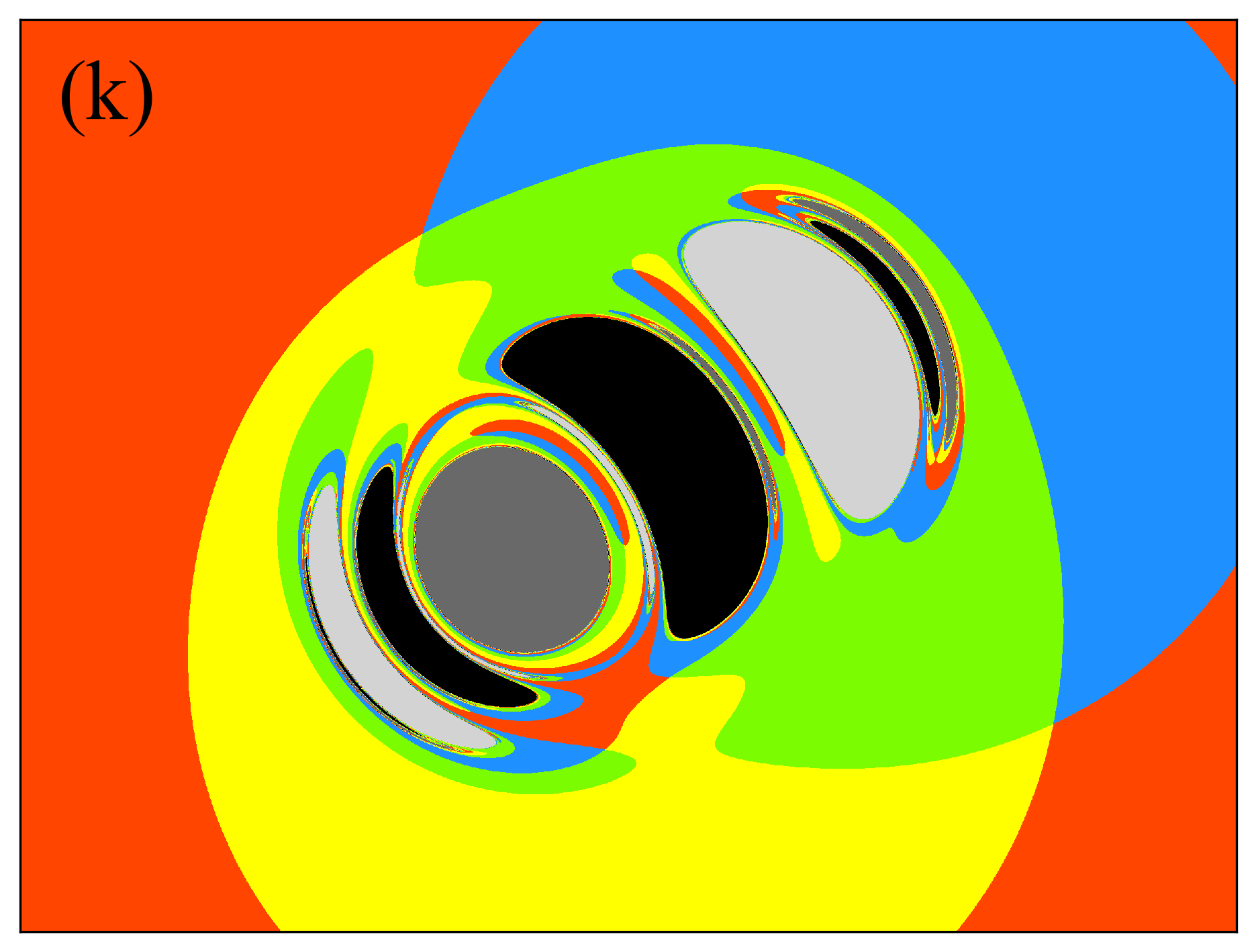}
\includegraphics[width=5cm]{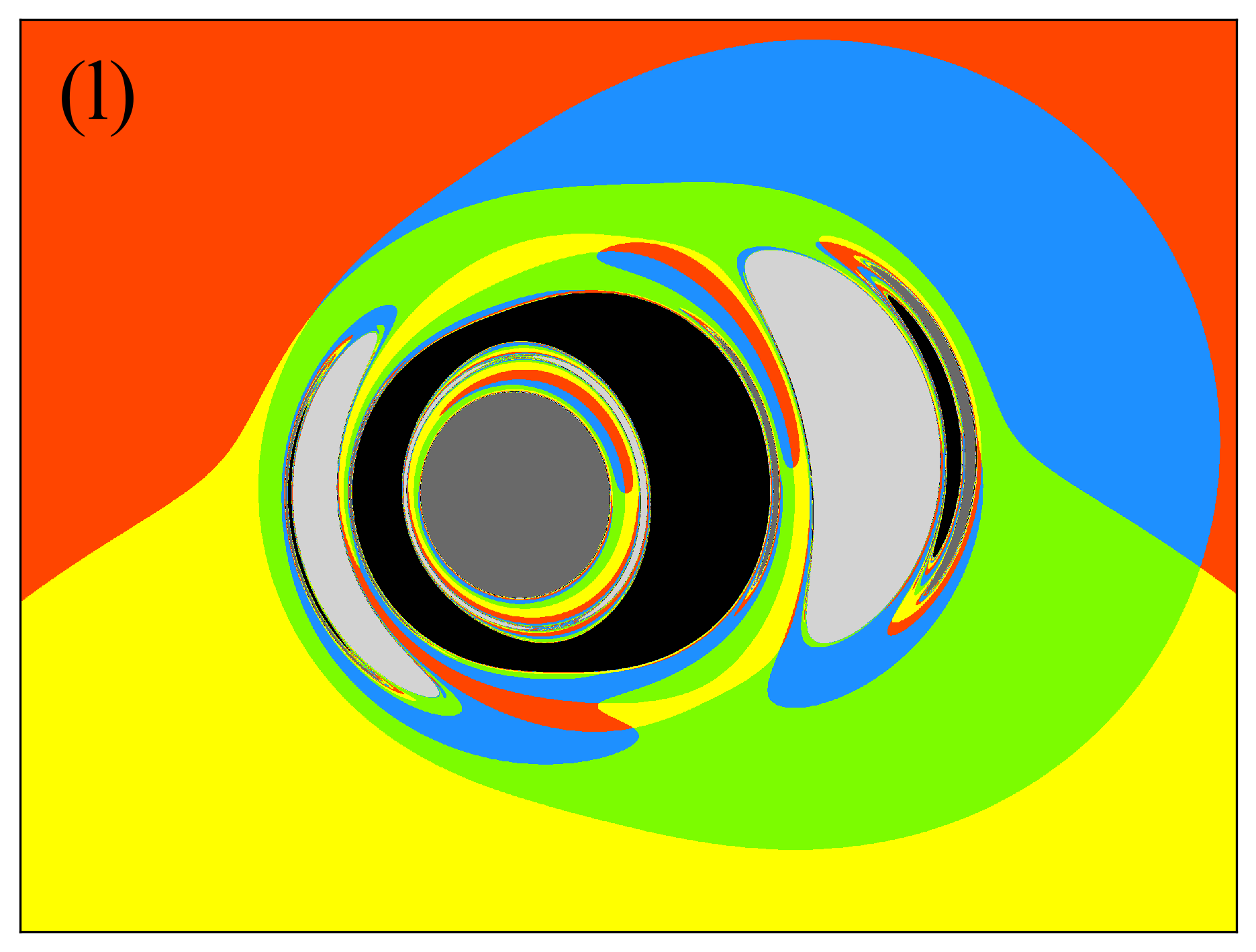}
\includegraphics[width=5cm]{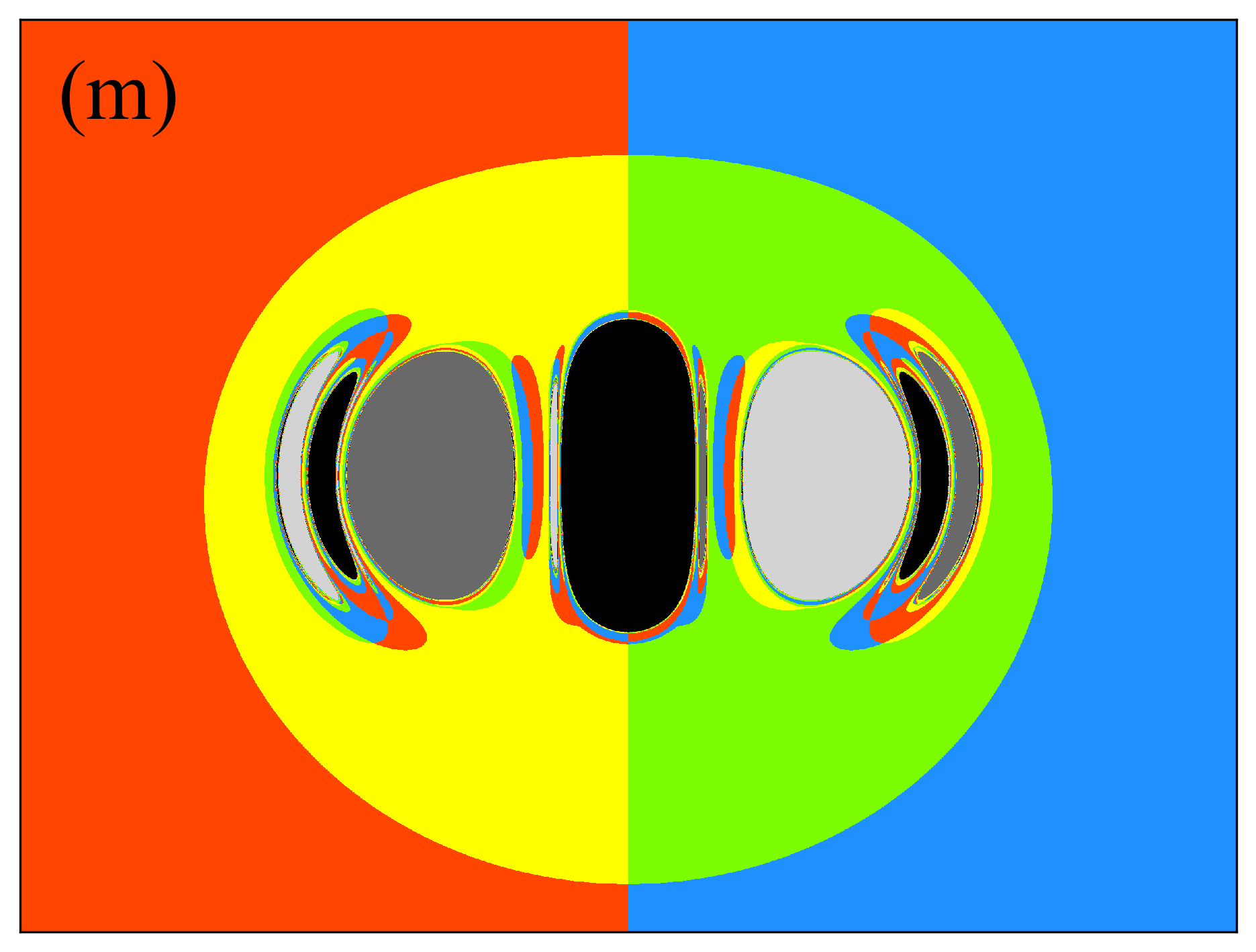}
\includegraphics[width=5cm]{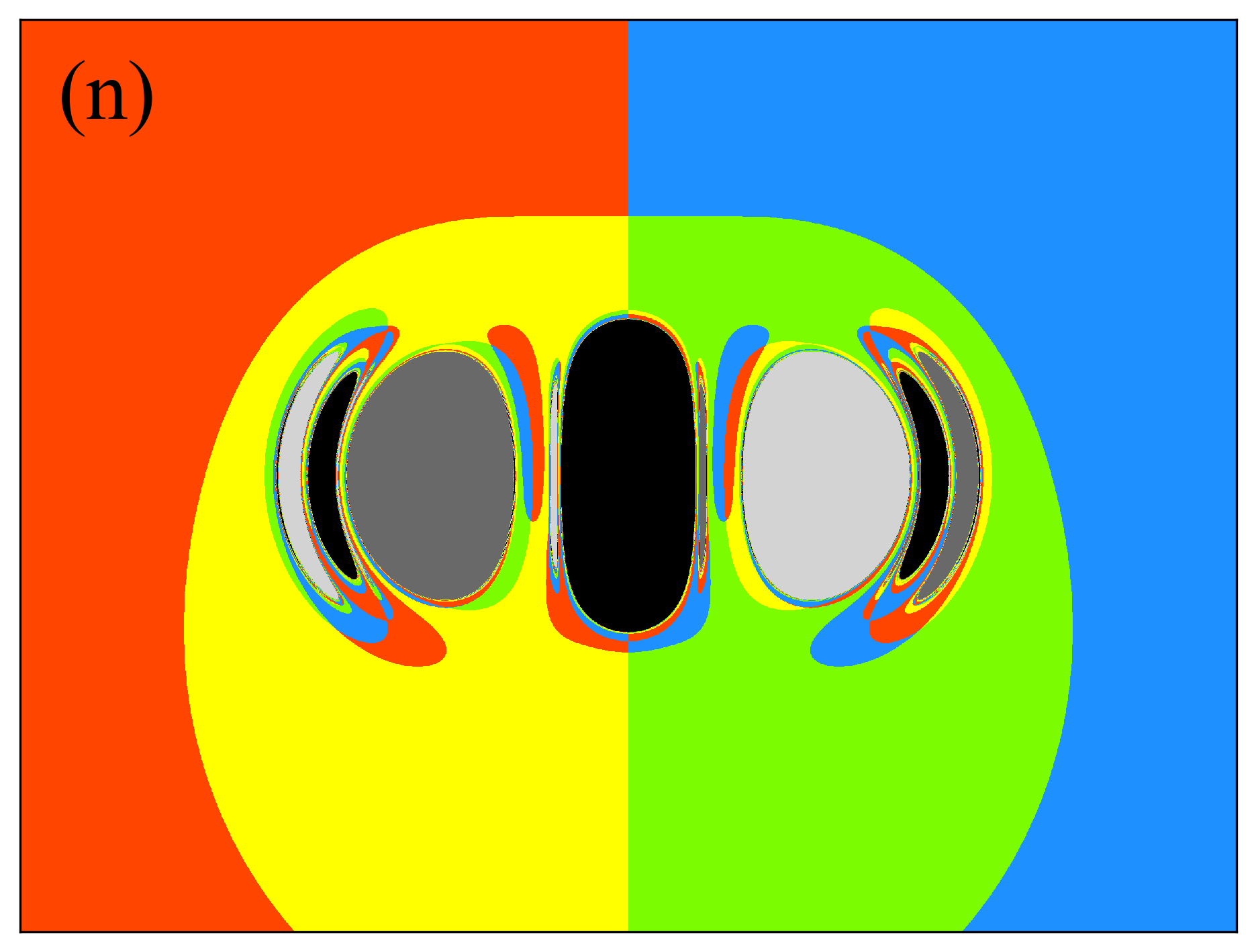}
\includegraphics[width=5cm]{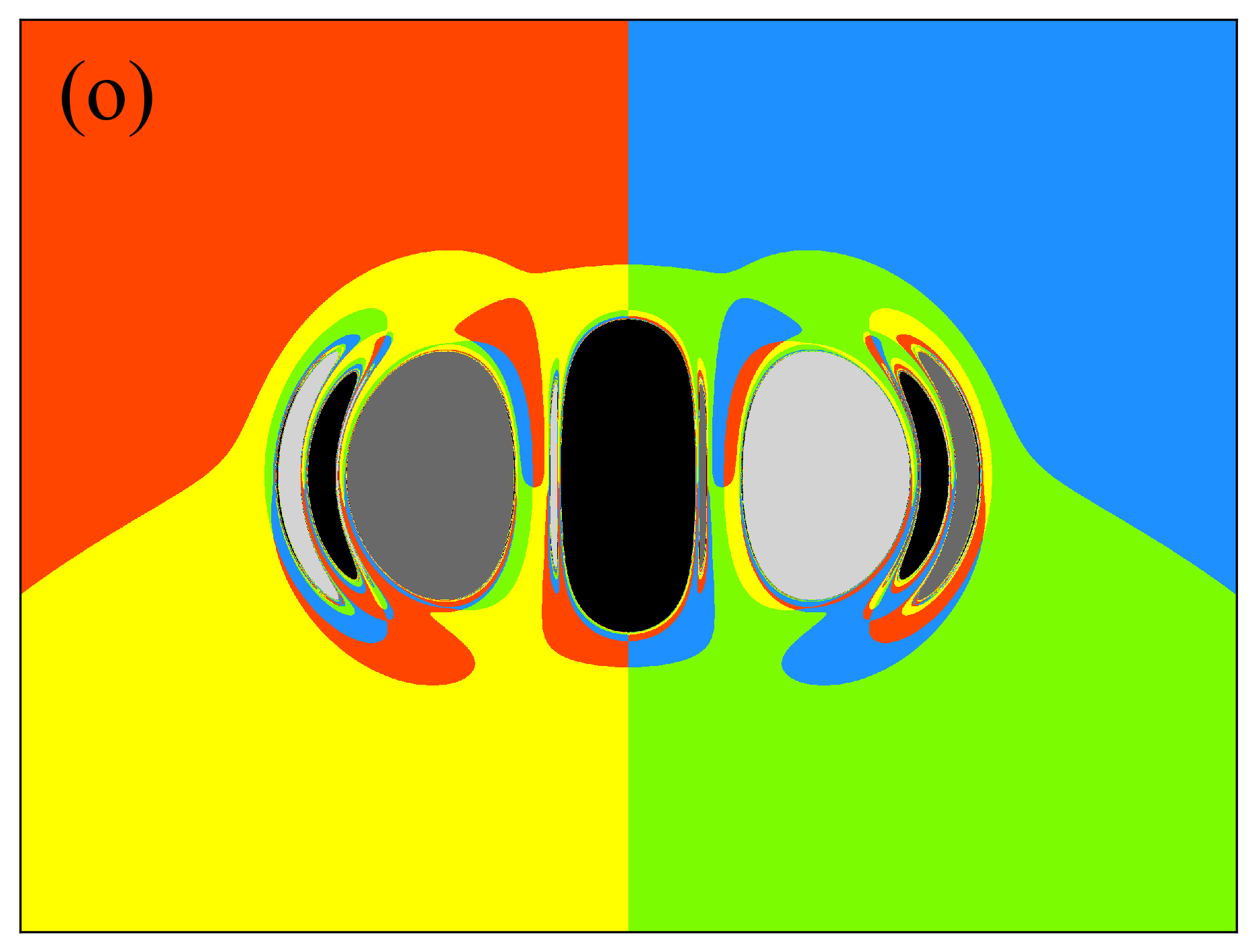}
\caption{Shadow morphology of triple black holes at fixed separation $l=1.2$ for varying observation inclinations $\Theta$ and azimuthal angles $\Phi$. From left to right: $\Theta=17^{\circ}$, $\Theta=50^{\circ}$, and $\Theta=80^{\circ}$. From top to bottom: $\Phi=0^{\circ}$, $\Phi=45^{\circ}$, $\Phi=90^{\circ}$, $\Phi=135^{\circ}$, and $\Phi=180^{\circ}$. The shadow contours exhibit $\Theta$-independent features in the cases of $\Phi=0^{\circ}$ and $\Phi=180^{\circ}$, whereas under other configurations, the inclination angle significantly rotate the black hole shadows and modulate the strength of gravitational lensing effects.}}\label{fig7}
\end{figure*}
Next, we investigate how the observation inclination $\Theta$ affects the images. Fig. 7 illustrates the dependence of the triple black hole shadow on the inclination angle $\Theta$, while maintaining fixed black hole separation $l$ and systematically varying the observer's azimuthal angle $\Phi$. The analysis reveals that the effects induced by $\Theta$ can be categorized into two distinct types. First type (panels (a)-(c) and (m)-(o)): Varying the inclination angle $\Theta$ does not alter the shadow contours of any of the three black holes---neither the primary nor secondary shadows exhibit morphological changes. Nevertheless, we can still identify changes in the background source distribution---that is, when accounting for source emission, luminosity variations with viewing angle are indeed expected. Second type (panels (d)-(l)): Increasing the inclination angle induces significant rotation and deformation of the black hole shadows. This occurs because in these configurations, the triple black hole system is no longer symmetric about the line of sight, and varying the viewing angle consequently modifies the gravitational lensing effects.
\subsection{Triangular configuration}
In the following numerical simulation, we set the observation plane with $x^{\prime} \in [-8,8]$ $M_{s}$ and $y^{\prime} \in [-8,8]$ $M_{s}$ at a resolution of $2000 \times 2000$ pixels. Fig. 8 demonstrates the influence of both the inclination angle $\Theta$ and side length $l$ on the shadows when the three black holes are positioned at the vertices of an equilateral triangle. When $l$ is small (first row), we observe that the shadows of the three black holes are closely interconnected, with their positions and shapes strongly dependent on the inclination angle. Notably, despite the extreme proximity of the black holes, their shadows do not merge into a seamless structure as seen in the colinear case. Instead, distinct small-scale bright spots become clearly visible. As $l$ increases, we observe a growth in the size of the bright spots. Simultaneously, the shadows of the three black holes progressively migrate toward the field boundary while undergoing significant morphological changes. Eyebrow-like secondary shadows emerge with larger $l$, though their spatial distribution depends critically on the inclination angle $\Theta$. For instance, at $\Theta=45^{\circ}$ or $\Theta=90^{\circ}$, the secondary shadows of $m_{1}$ and $m_{3}$ flank the quasi-circular shadow of $m_{2}$, as exemplified in panels (e), (f), (h), and (i). Notably, these features are entirely absent in the $\Theta=0^{\circ}$ case. As the black holes separate further, their shadows progressively separate. Remarkably, regardless of the observer's inclination angle, each black hole consistently exhibits at least two clearly identifiable eyebrow-shaped secondary shadows.
\begin{figure*}%[tbph]
\center{
\includegraphics[width=4.5cm]{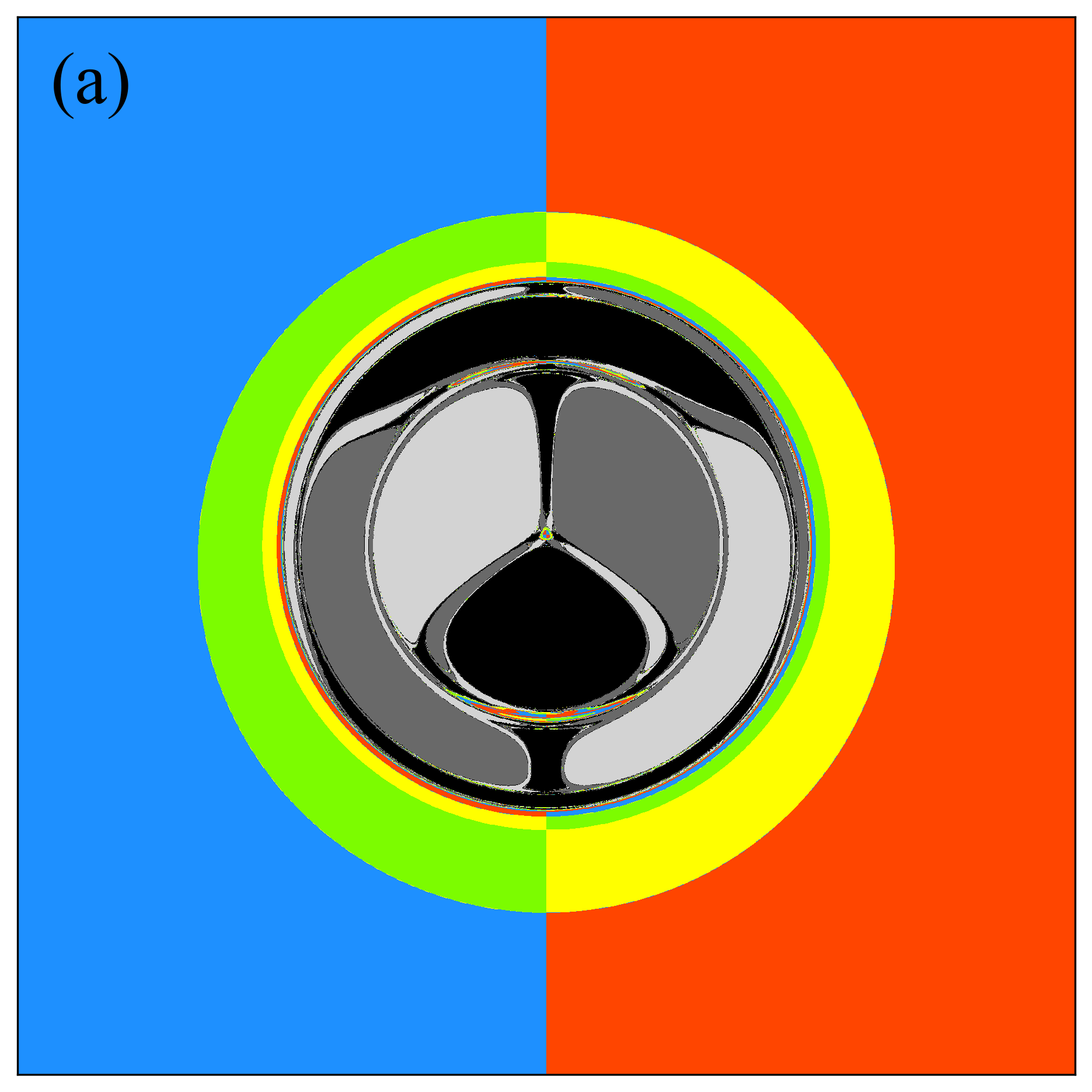}
\includegraphics[width=4.5cm]{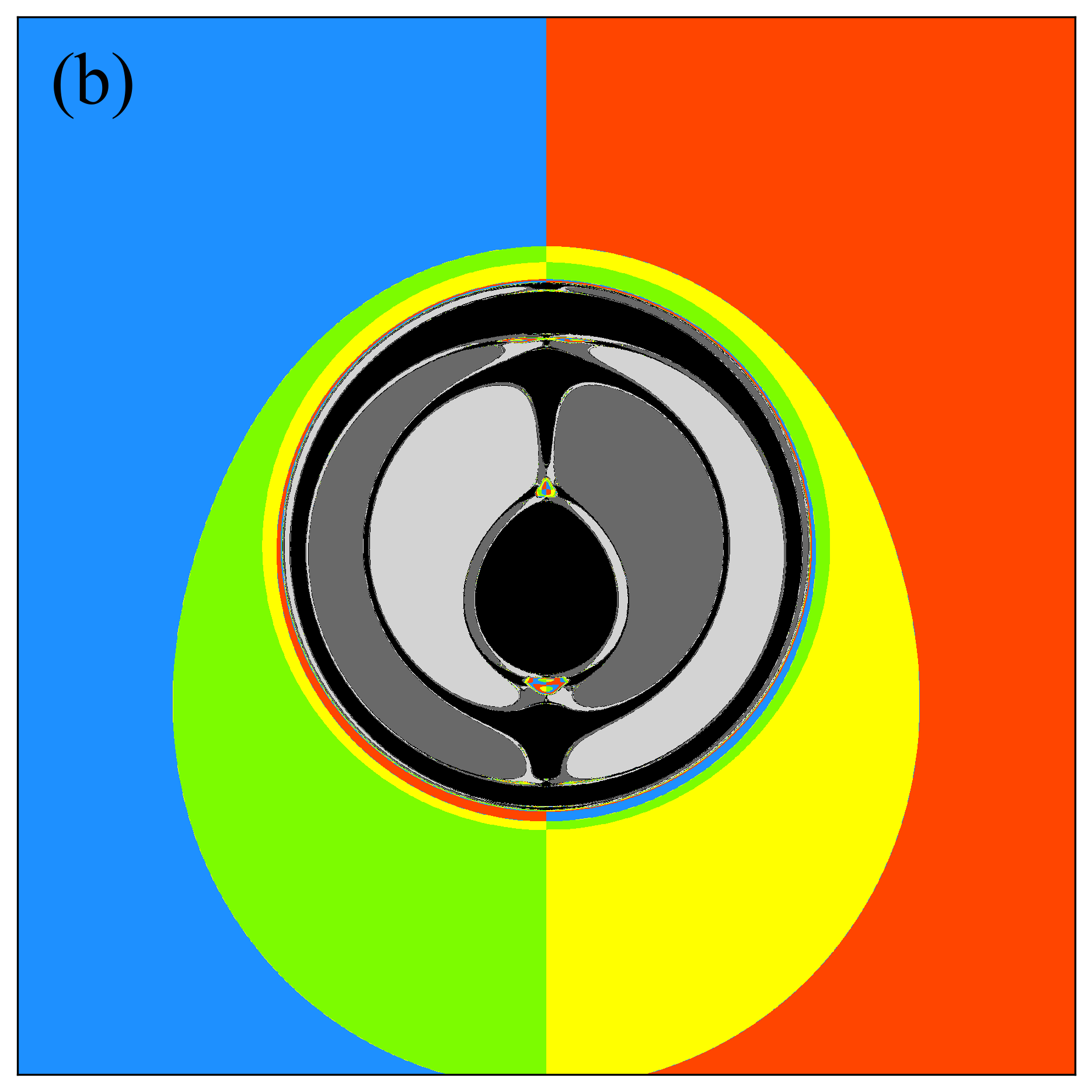}
\includegraphics[width=4.5cm]{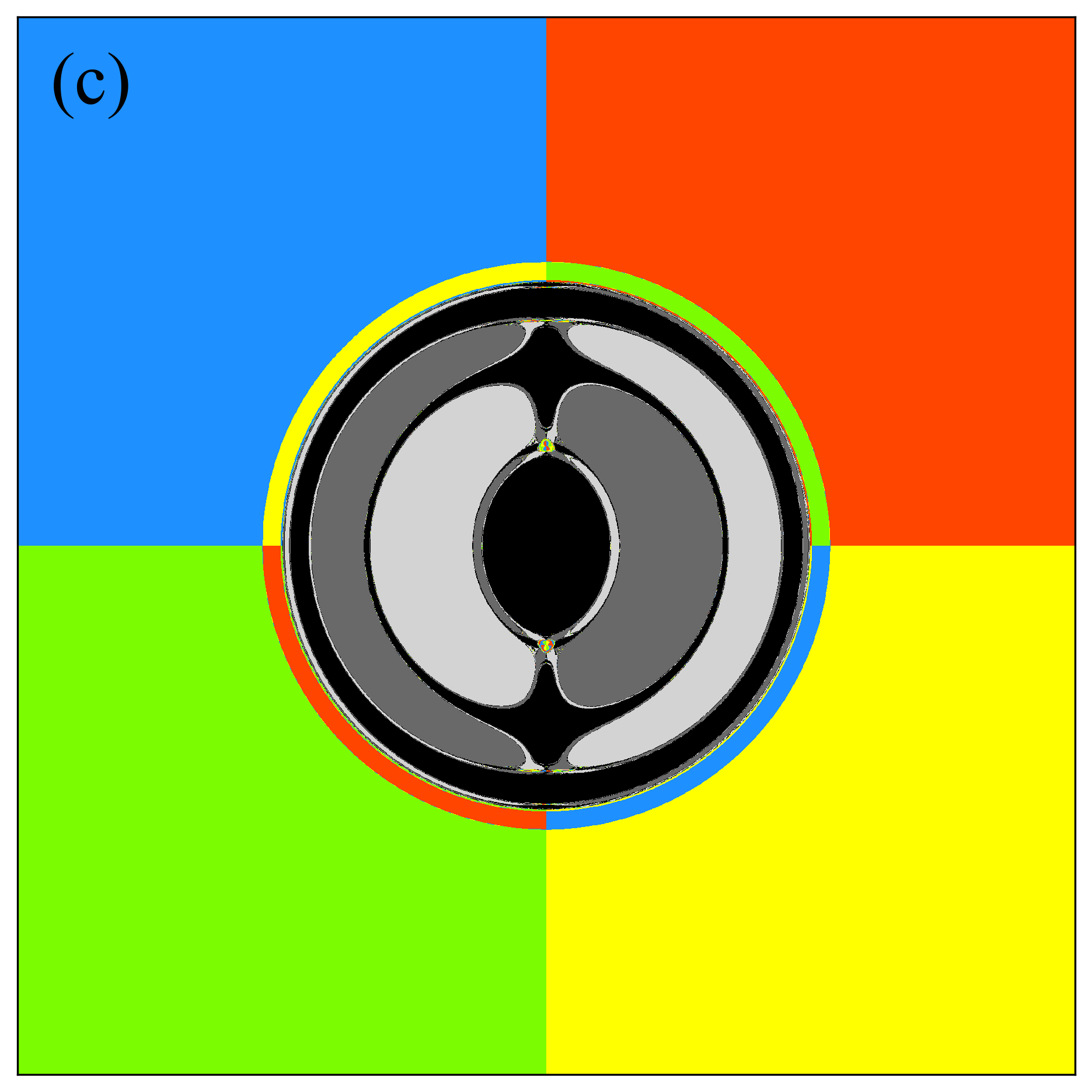}
\includegraphics[width=4.5cm]{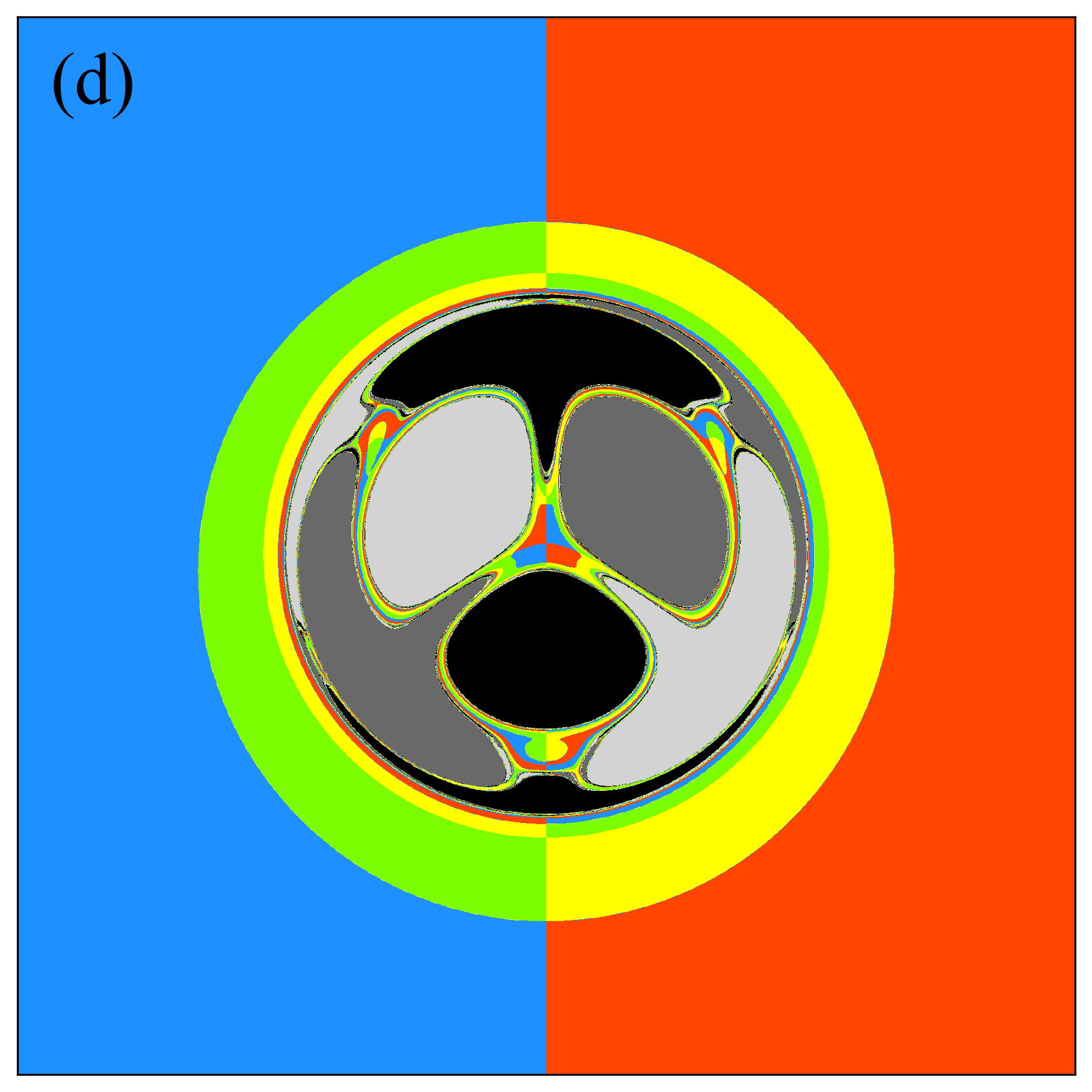}
\includegraphics[width=4.5cm]{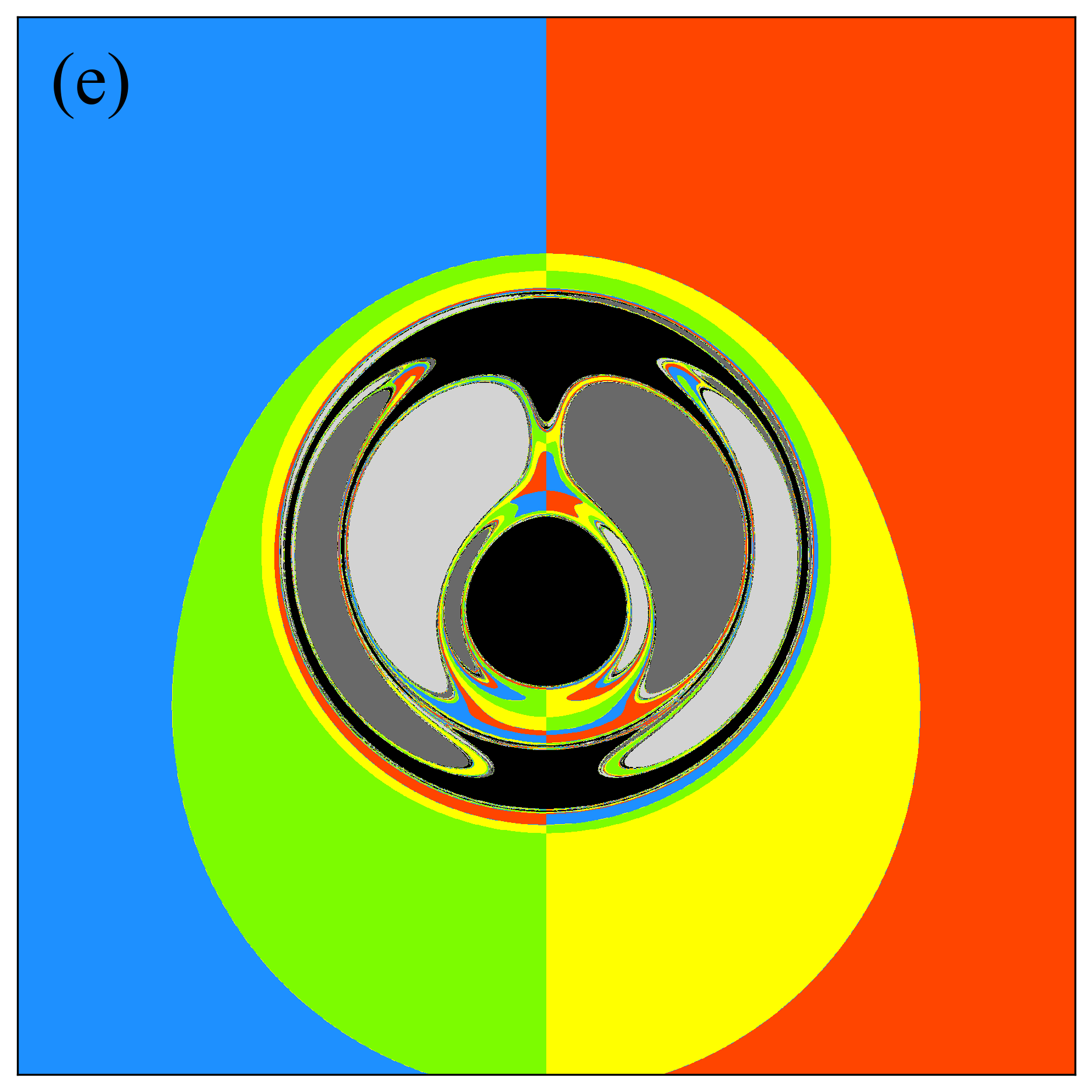}
\includegraphics[width=4.5cm]{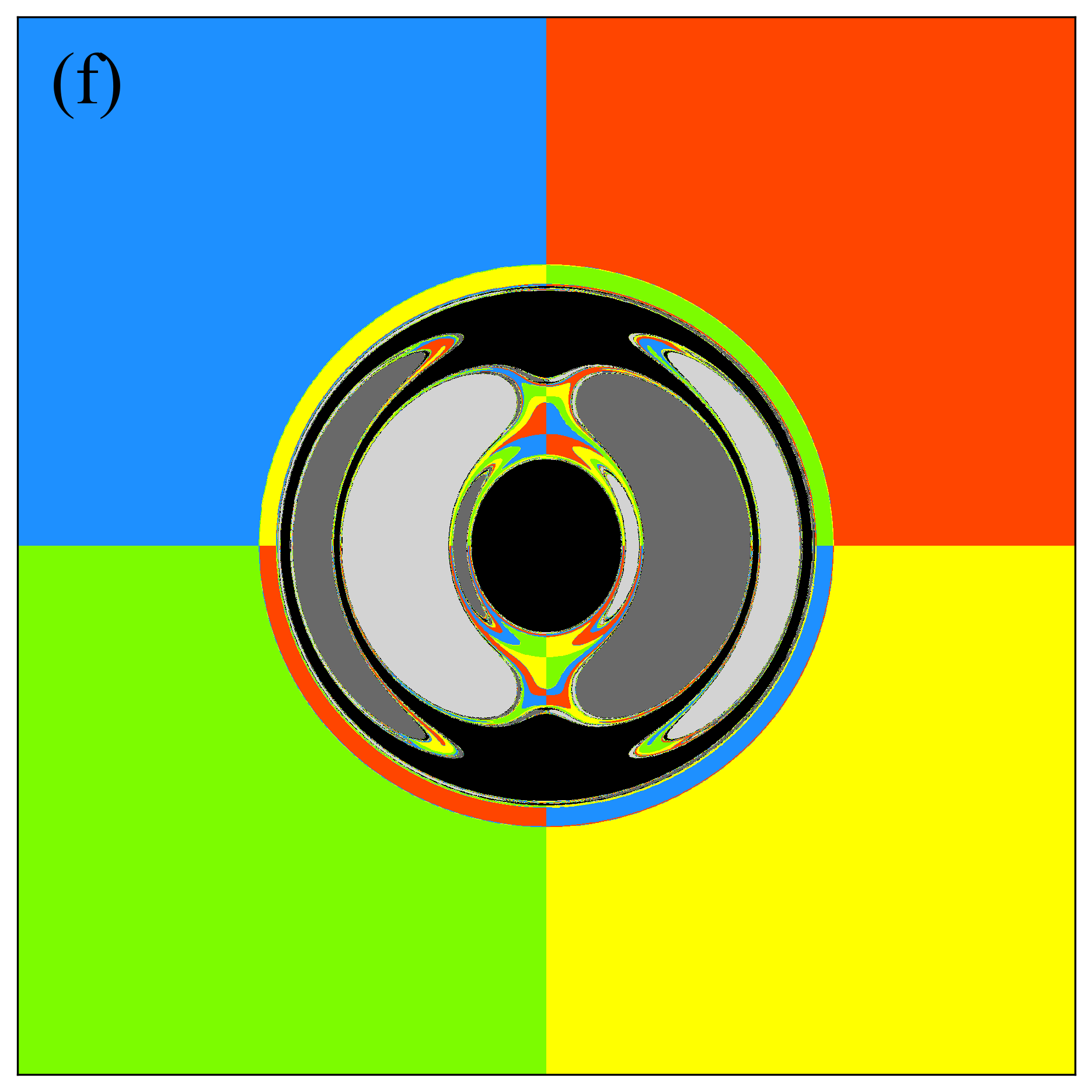}
\includegraphics[width=4.5cm]{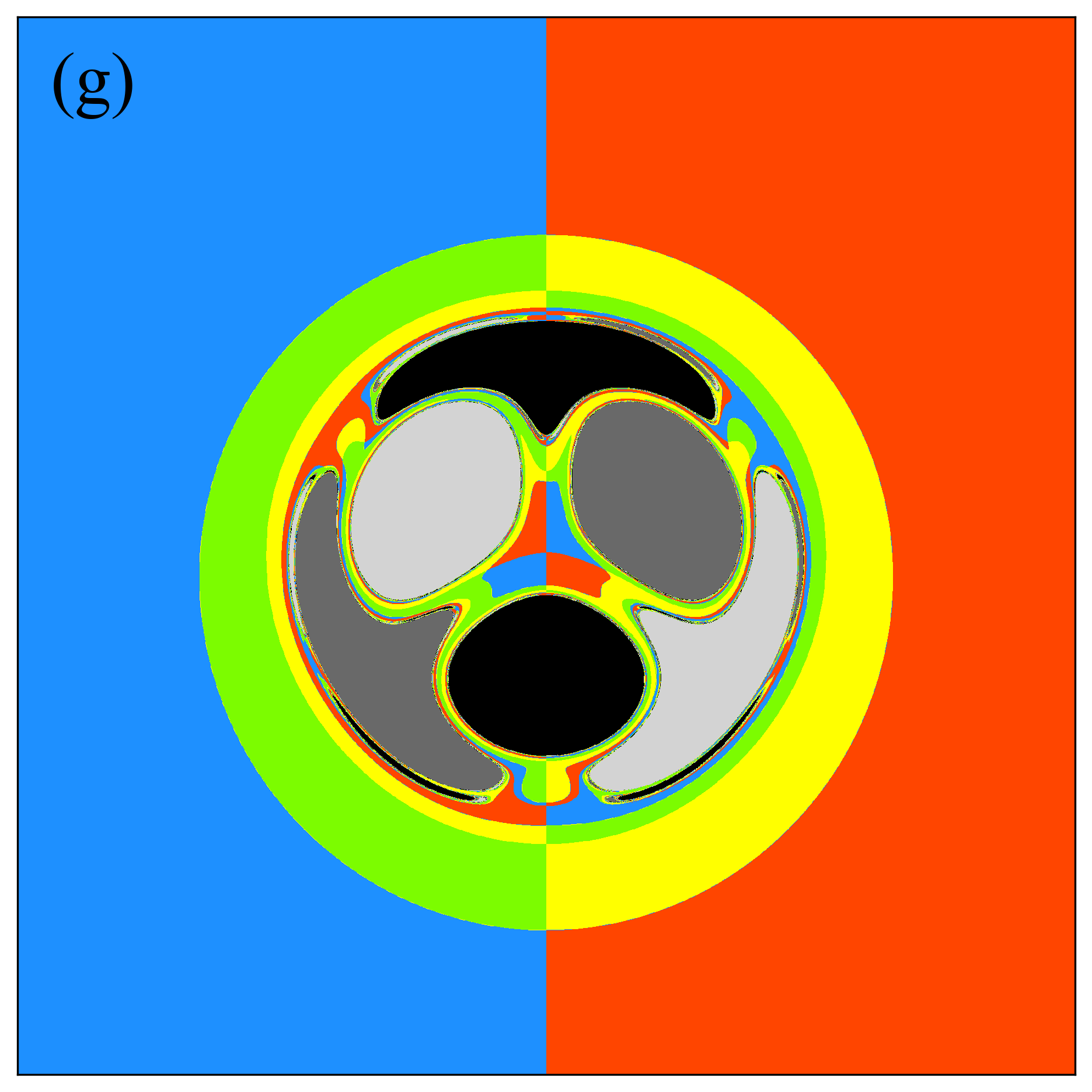}
\includegraphics[width=4.5cm]{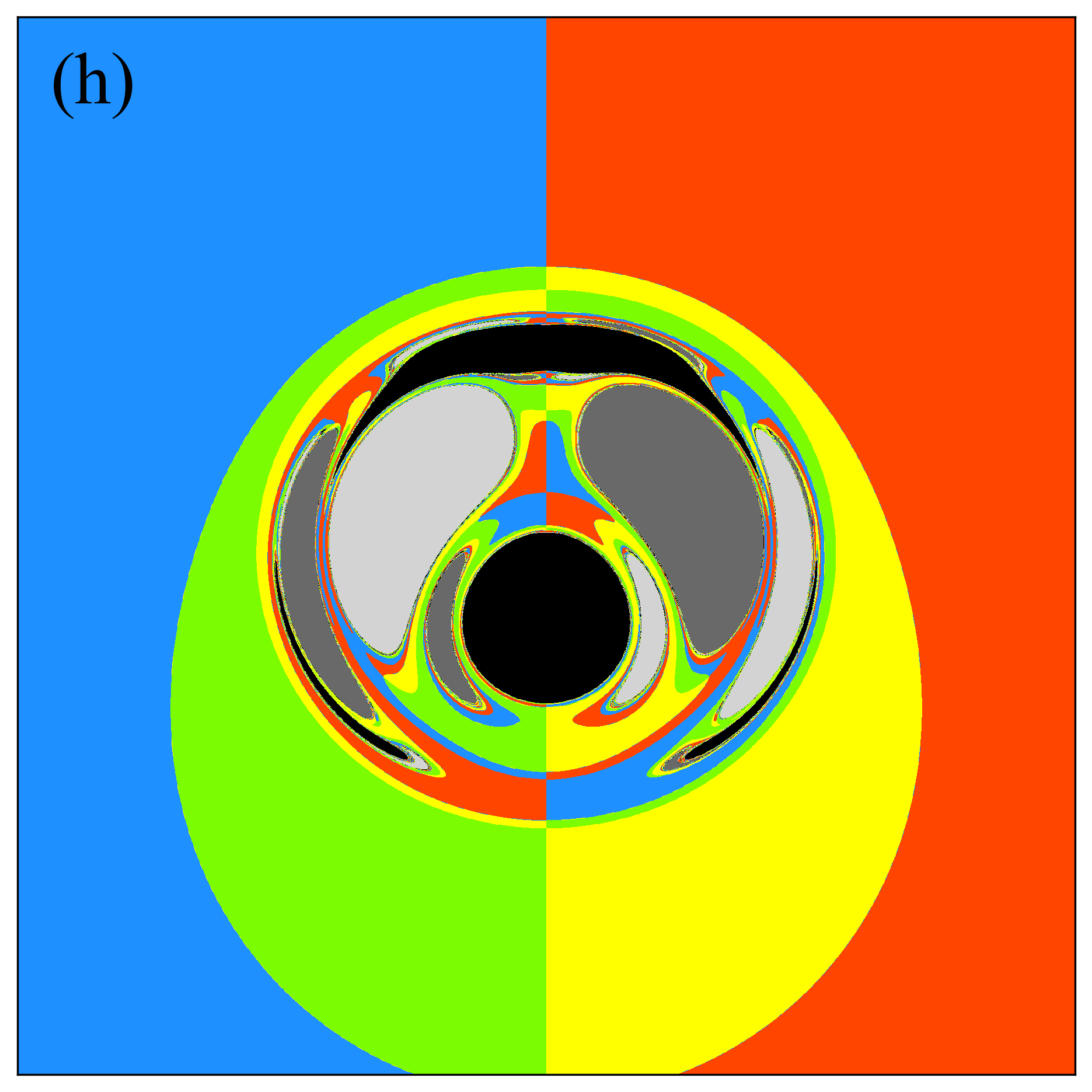}
\includegraphics[width=4.5cm]{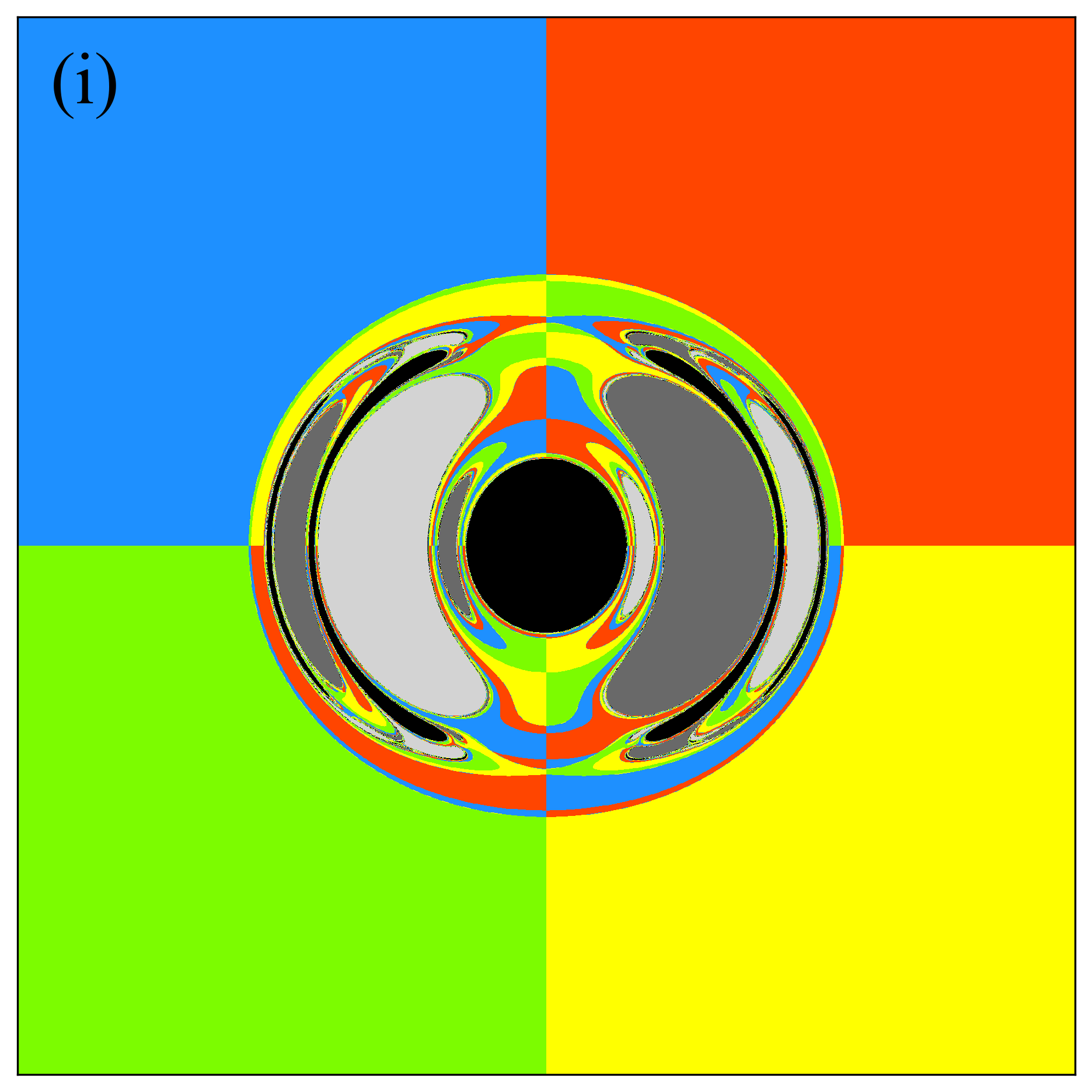}
\includegraphics[width=4.5cm]{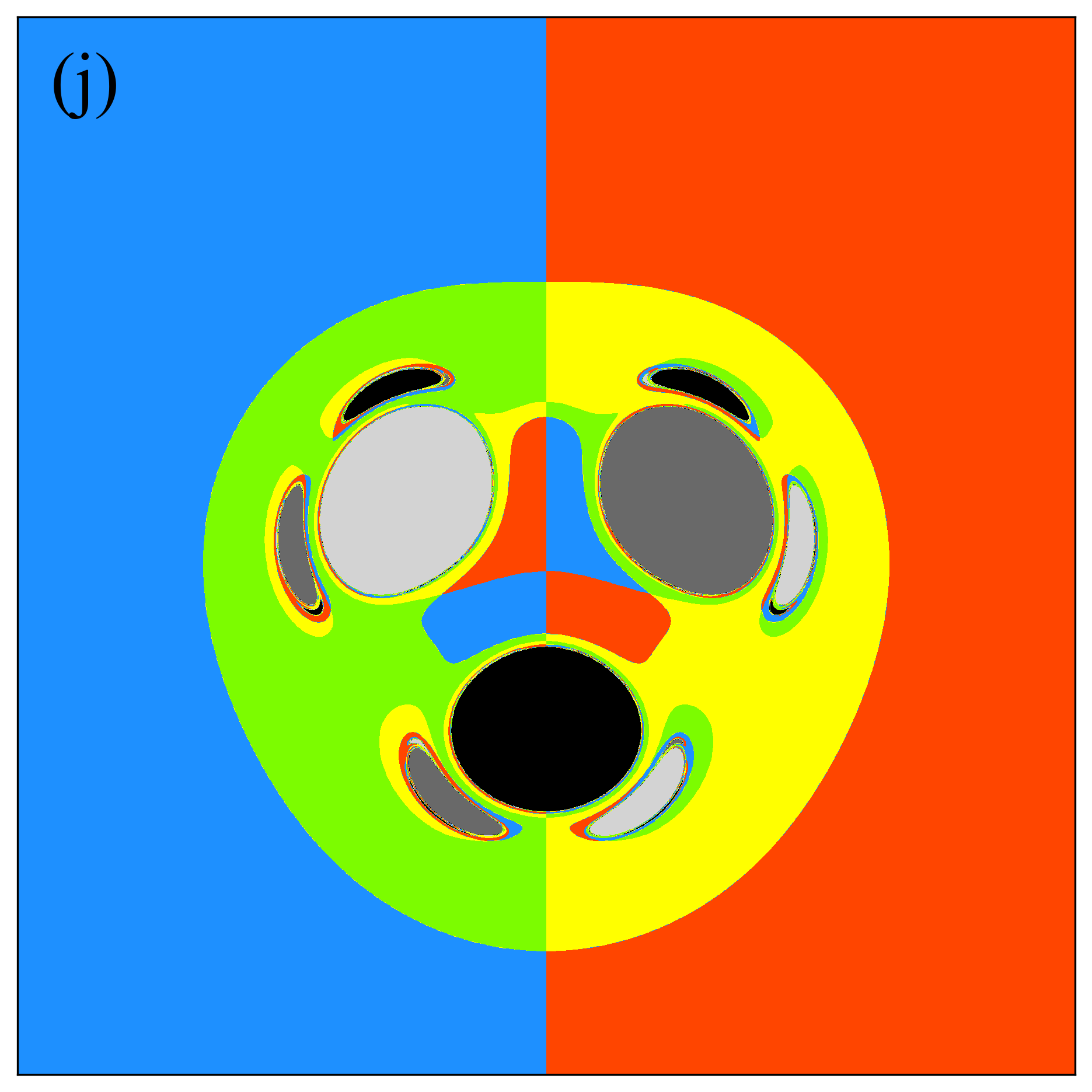}
\includegraphics[width=4.5cm]{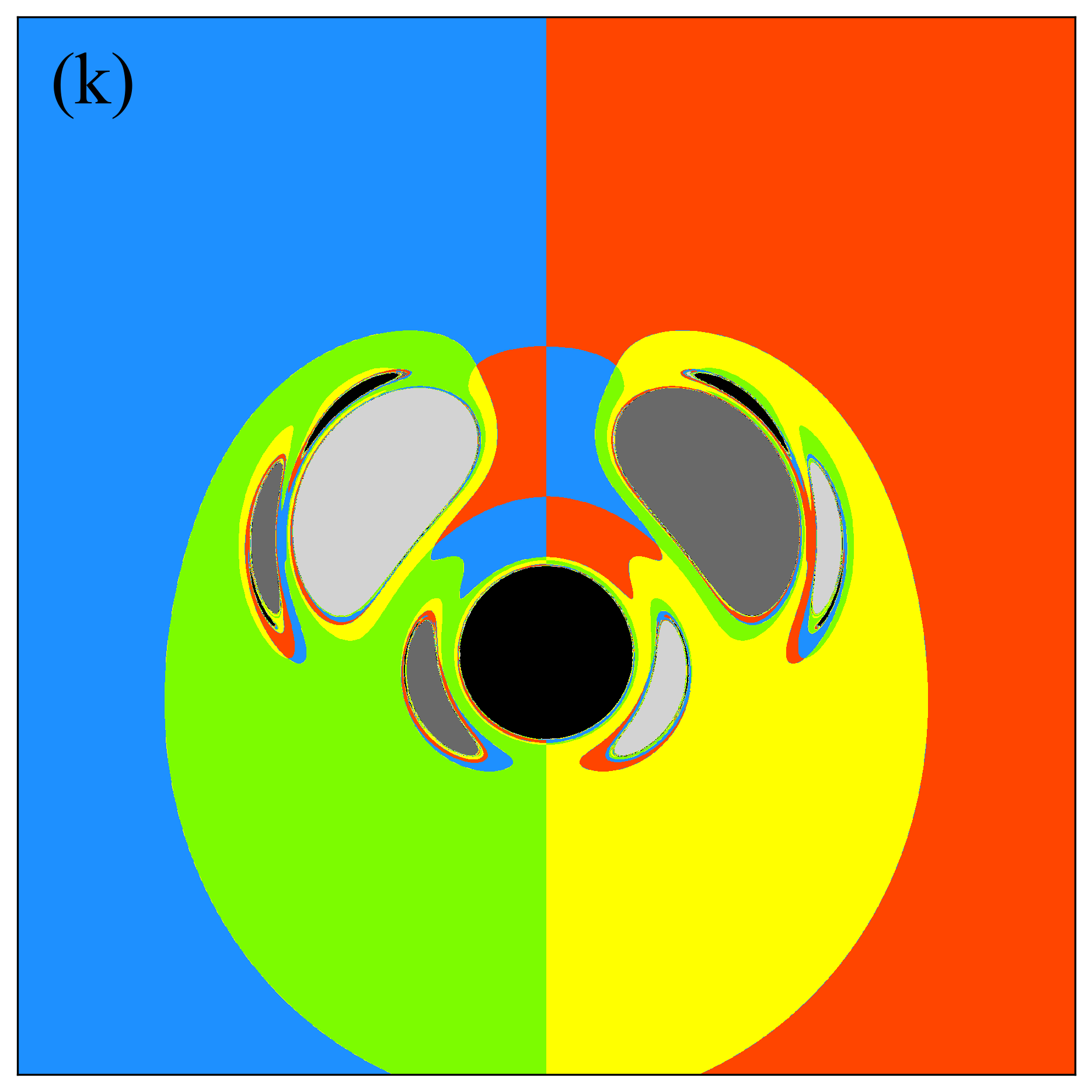}
\includegraphics[width=4.5cm]{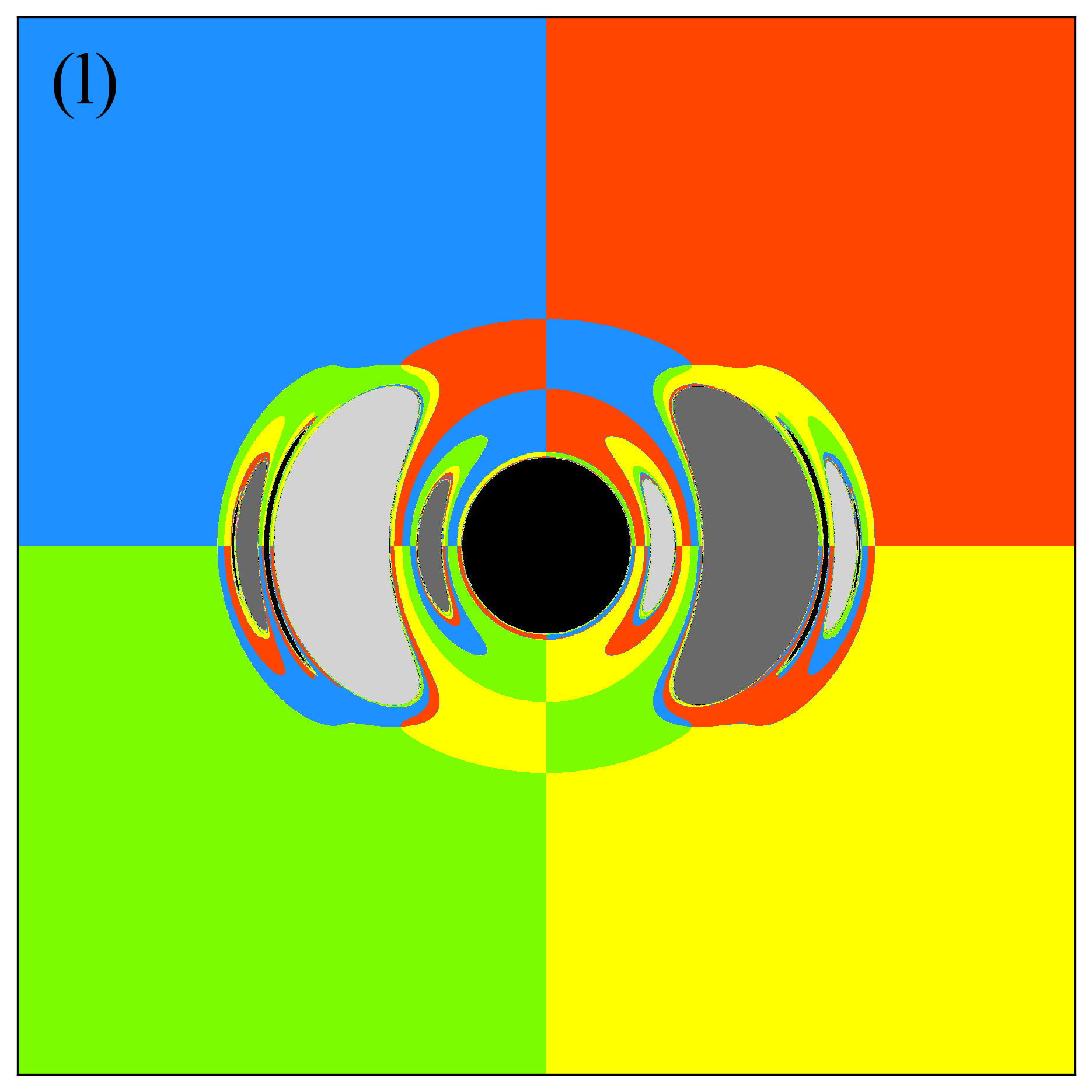}
\includegraphics[width=4.5cm]{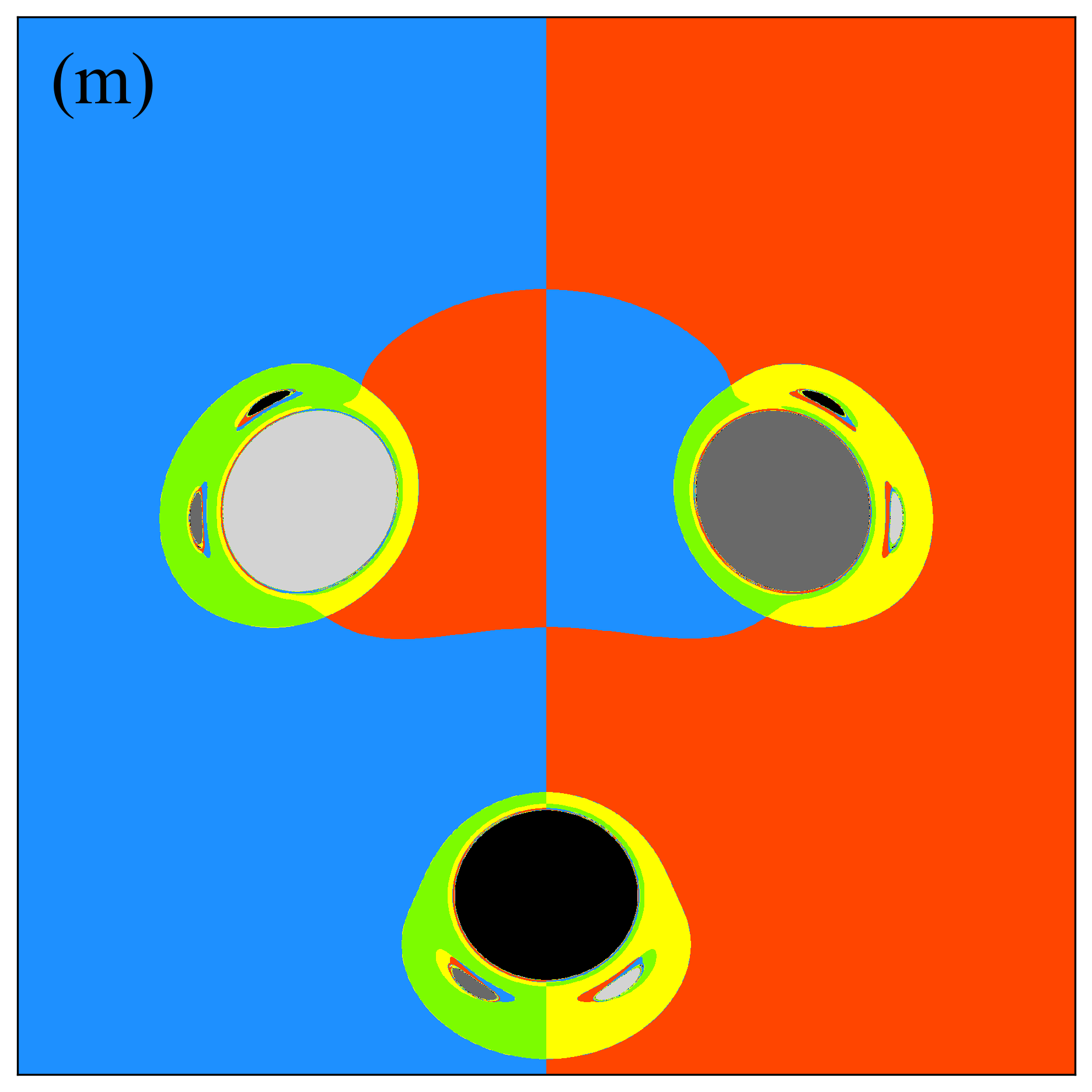}
\includegraphics[width=4.5cm]{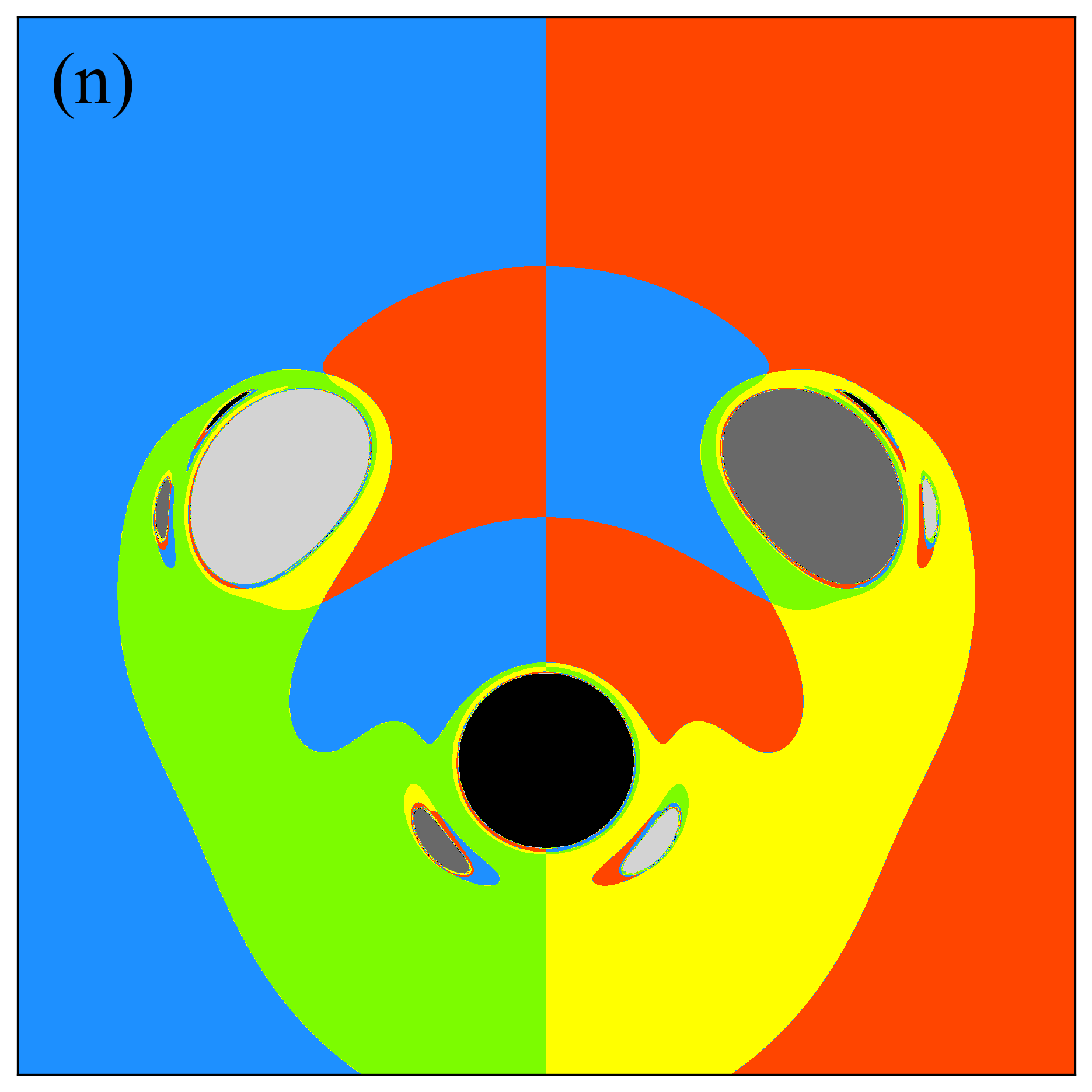}
\includegraphics[width=4.5cm]{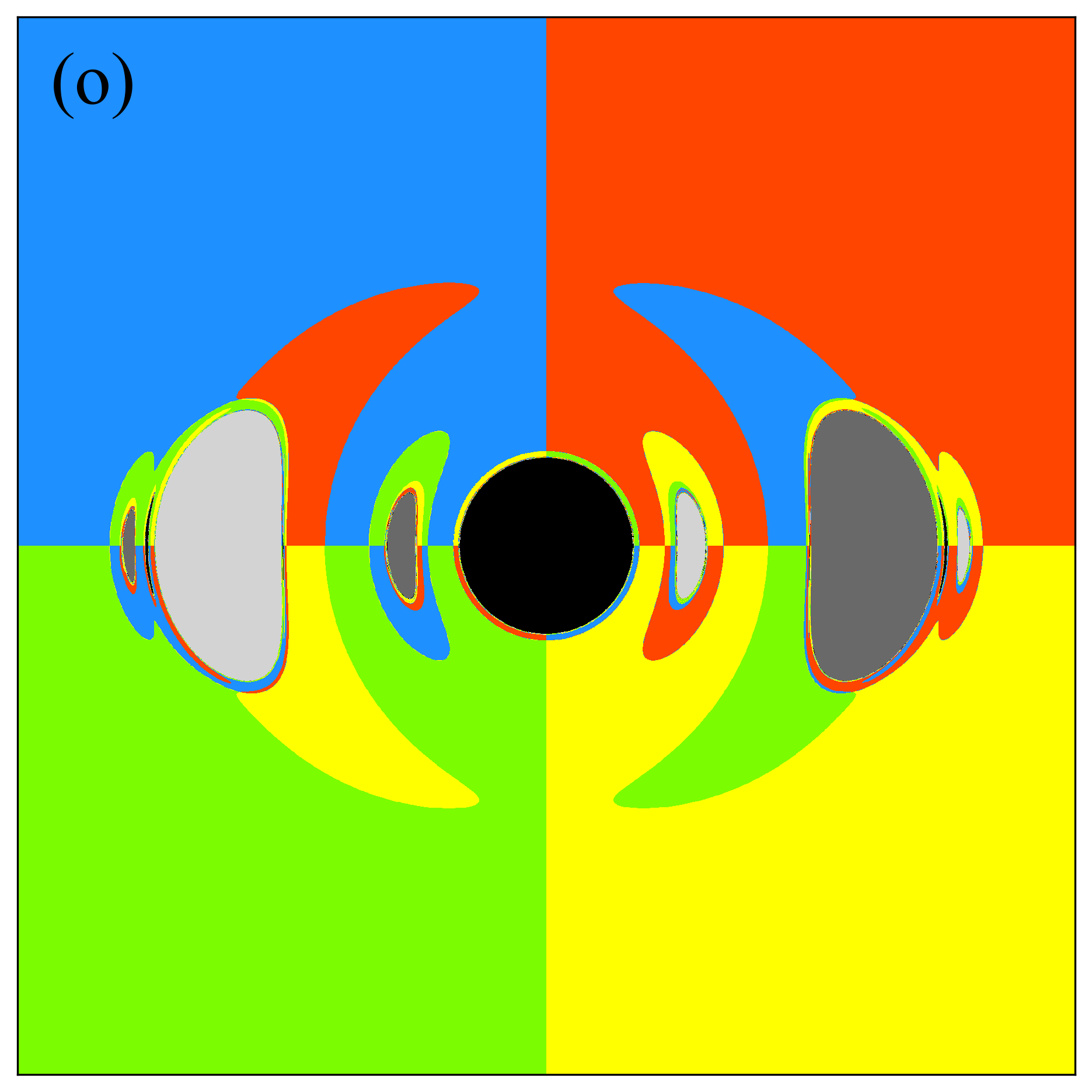}
\caption{Three black hole shadows with different parameters in the case of equilateral triangle configuration. From left to right, the observation angles are $0^{\circ}$, $45^{\circ}$, and $90^{\circ}$; from top to bottom, the separation $l$ take values of $0.02$, $0.5$, $1$, $2$, and $5$. Here, we have the observation azimuth of $\Phi = 0^{\circ}$.}}\label{fig8}
\end{figure*}

We changed the observation azimuth to $45^{\circ}$ and repeated the numerical simulations as shown in Fig. 8, with the results presented in Fig. 9. It is evident that as $l$ increases, the trends in the changes of the three black hole shadows are similar to those in Fig. 8, i.e., the shadows deform while also moving toward the screen boundary, accompanied by the formation of multiple eyebrow-like shadows. The deformation and movement directions of the shadows are related to the observation angle. Furthermore, we observed some unique features: First, when the observation inclination is $0^{\circ}$ or $45^{\circ}$, the symmetry of the image about the $y^{\prime}$-axis is broken. Specifically, $m_{3}$ exhibits irregular shadows, as shown in panels (h) and (k). Second, we identified shadow rings caused by gravitational lensing, as seen in panels (l) and (o). This is due to $m_{3}$ appearing behind $m_{2}$.
\begin{figure*}%[tbph]
\center{
\includegraphics[width=4.5cm]{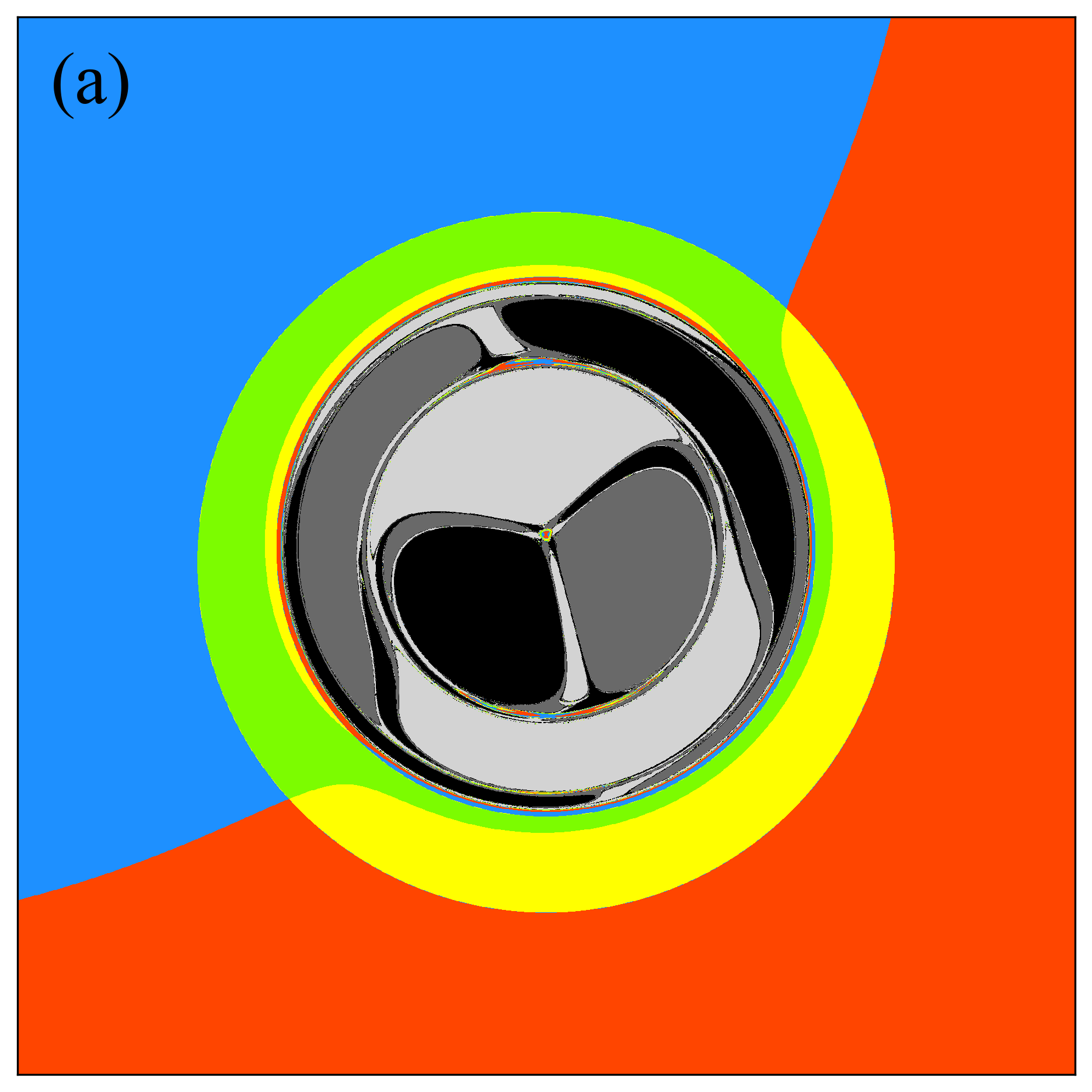}
\includegraphics[width=4.5cm]{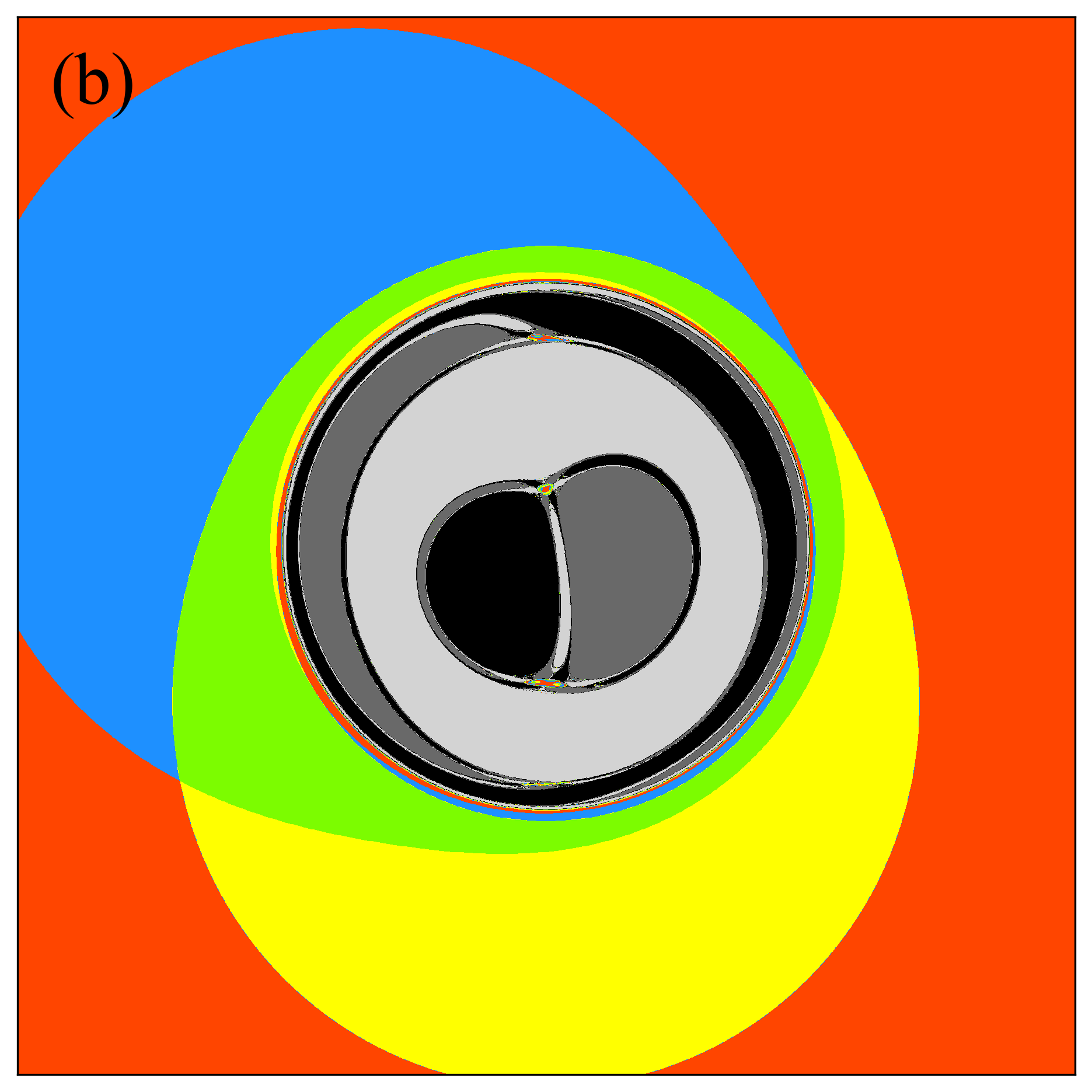}
\includegraphics[width=4.5cm]{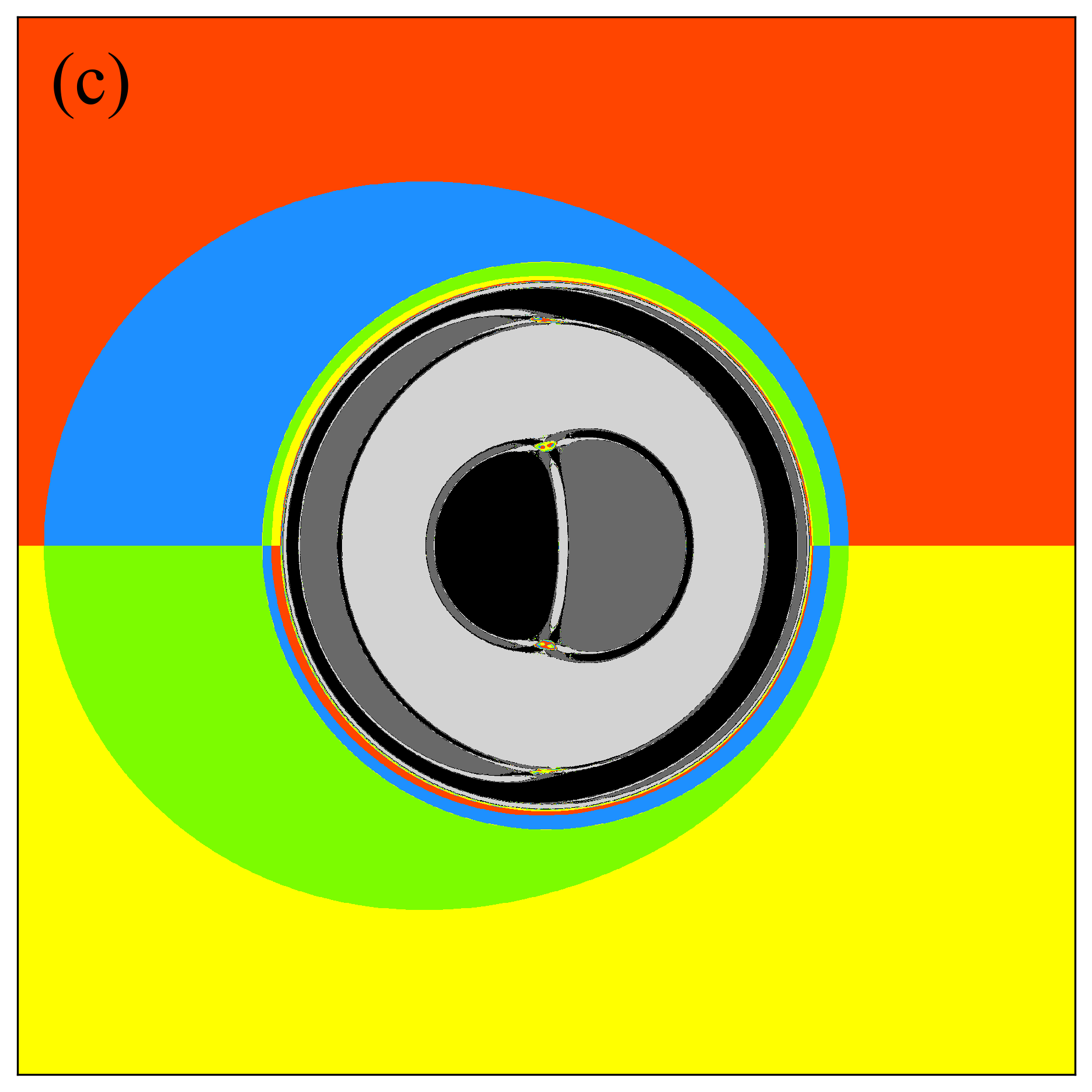}
\includegraphics[width=4.5cm]{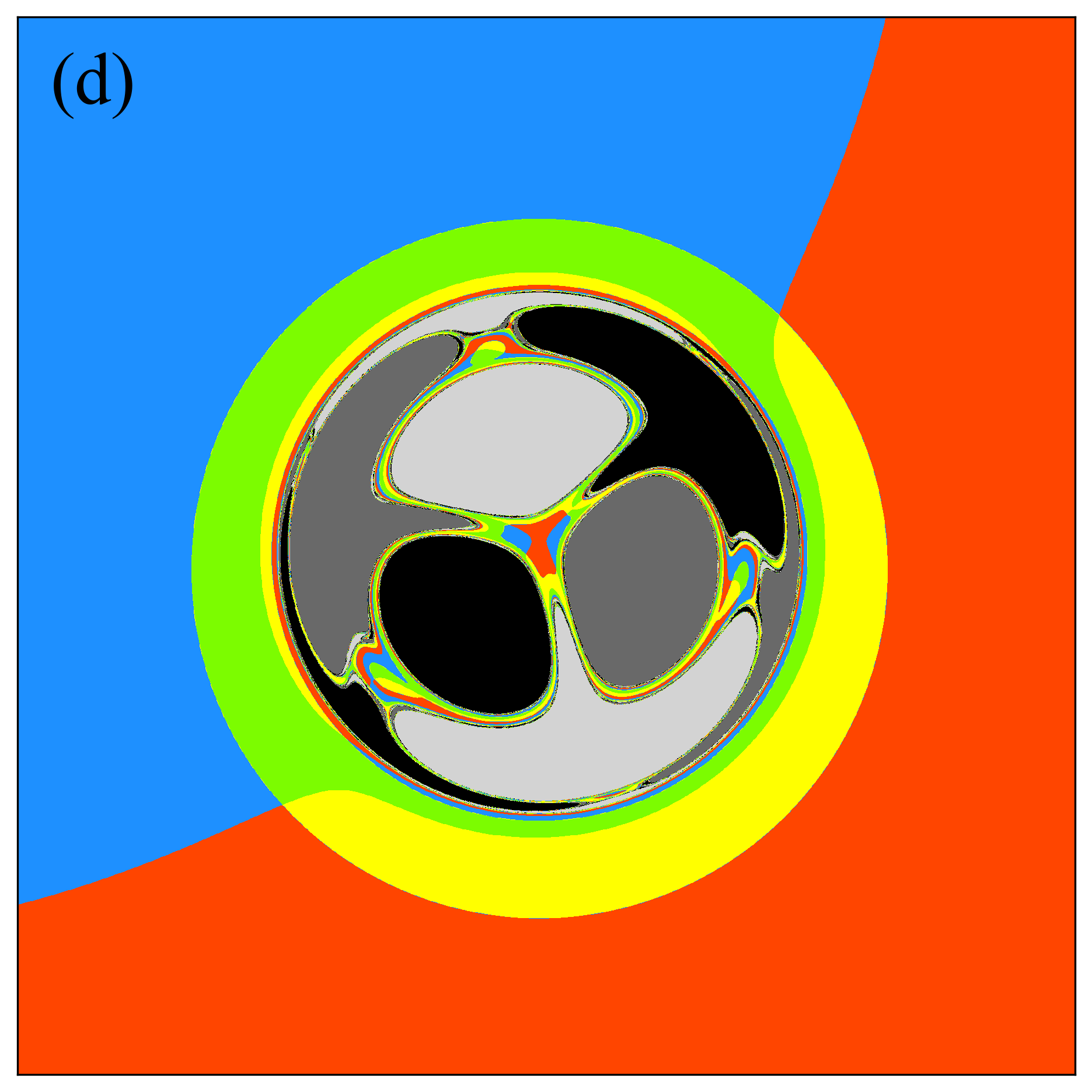}
\includegraphics[width=4.5cm]{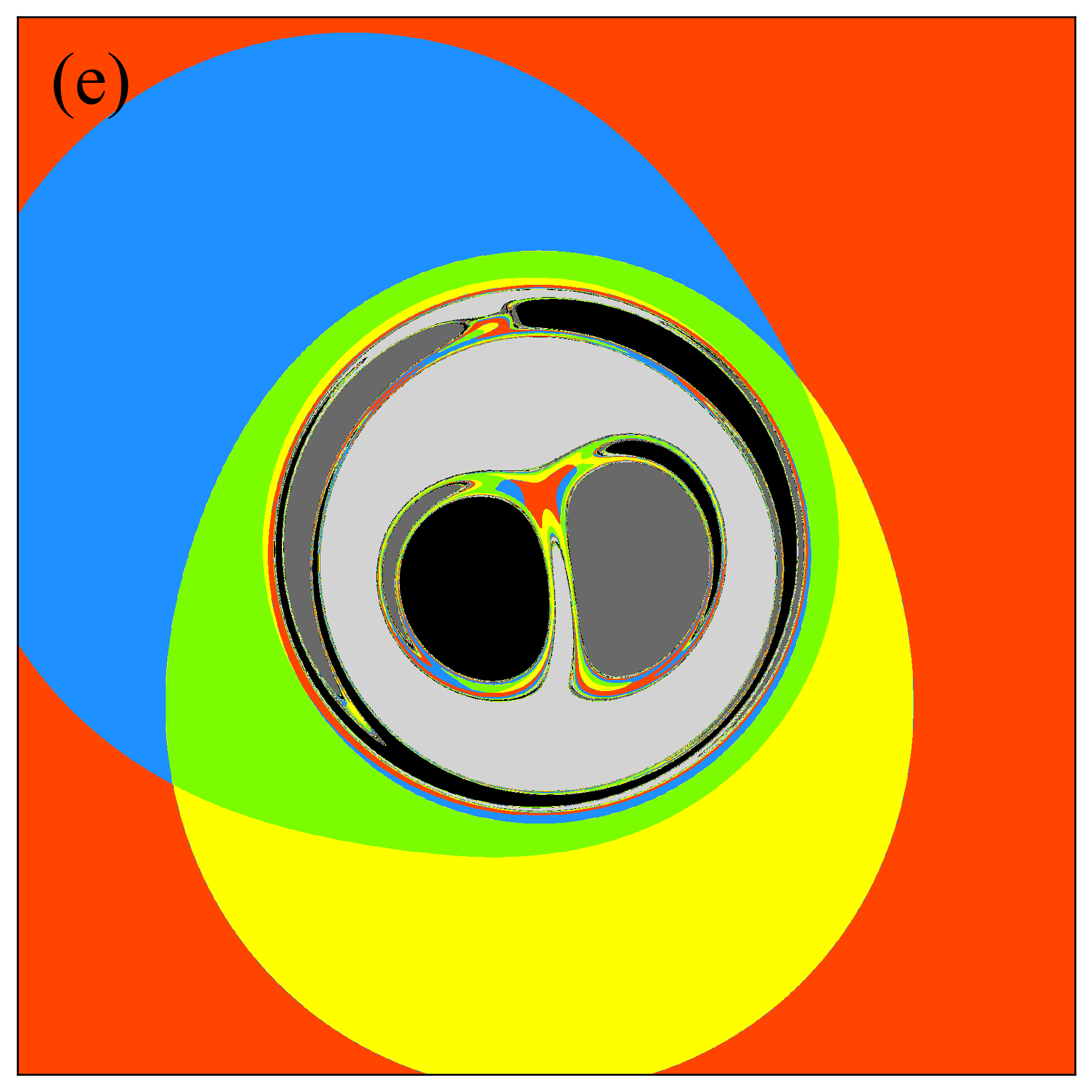}
\includegraphics[width=4.5cm]{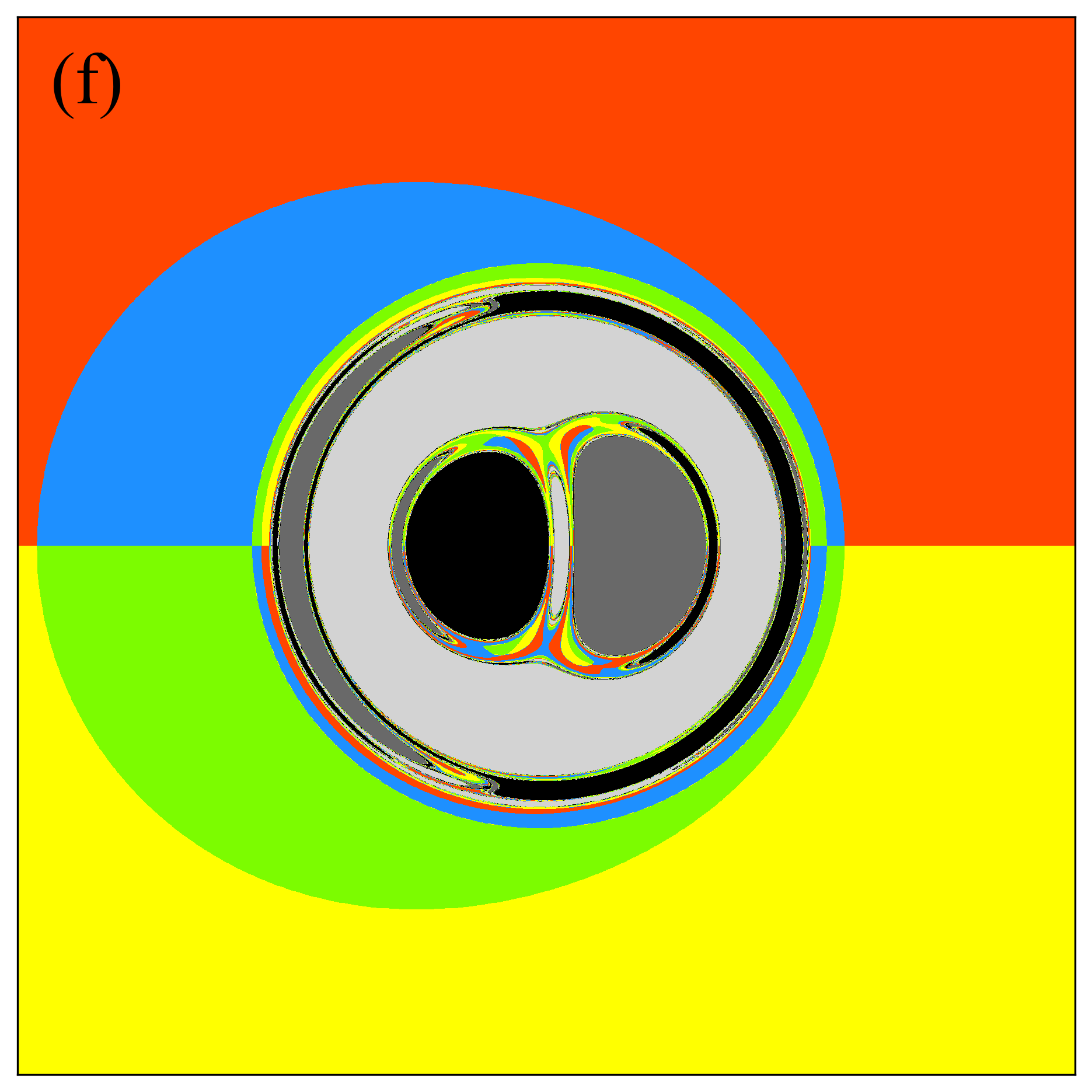}
\includegraphics[width=4.5cm]{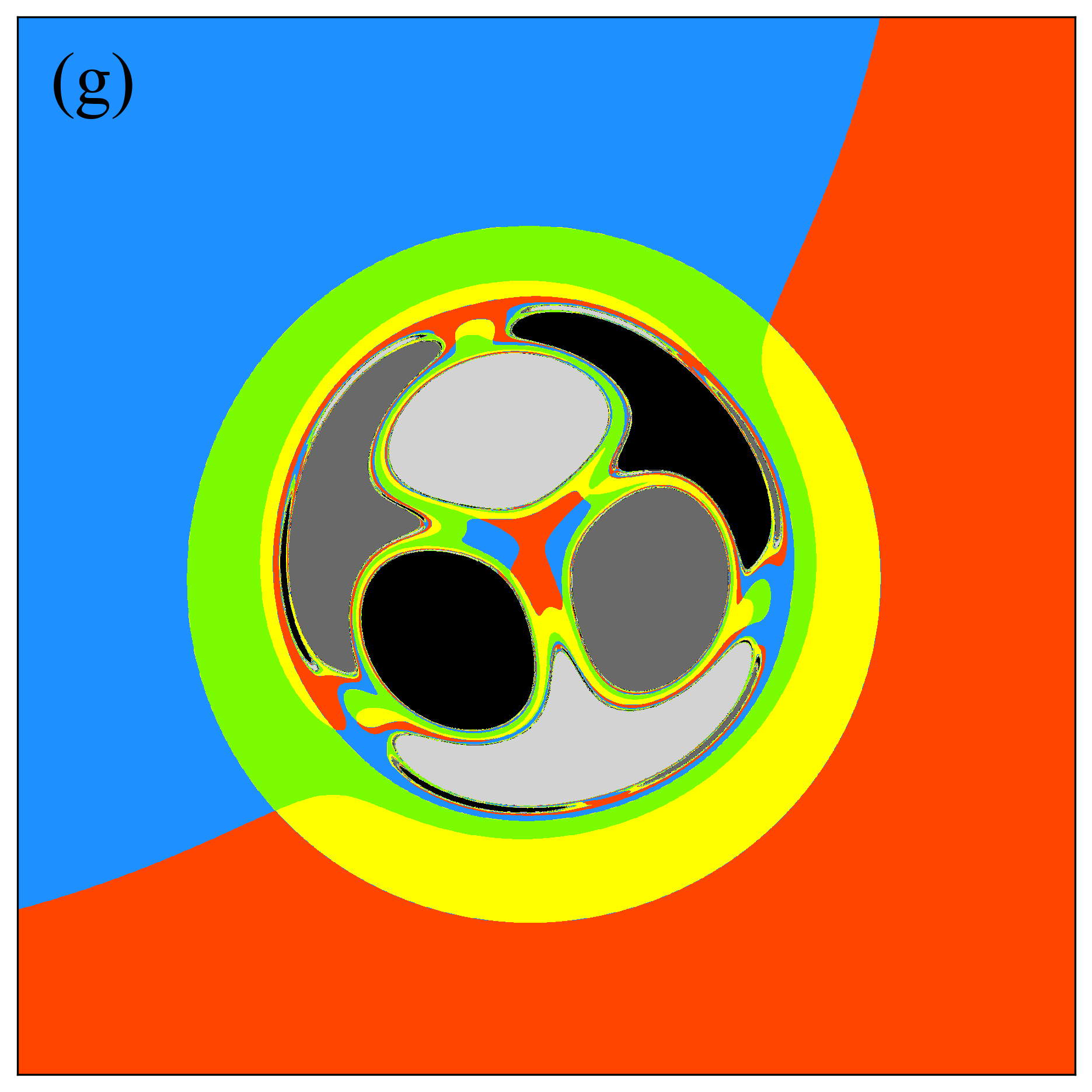}
\includegraphics[width=4.5cm]{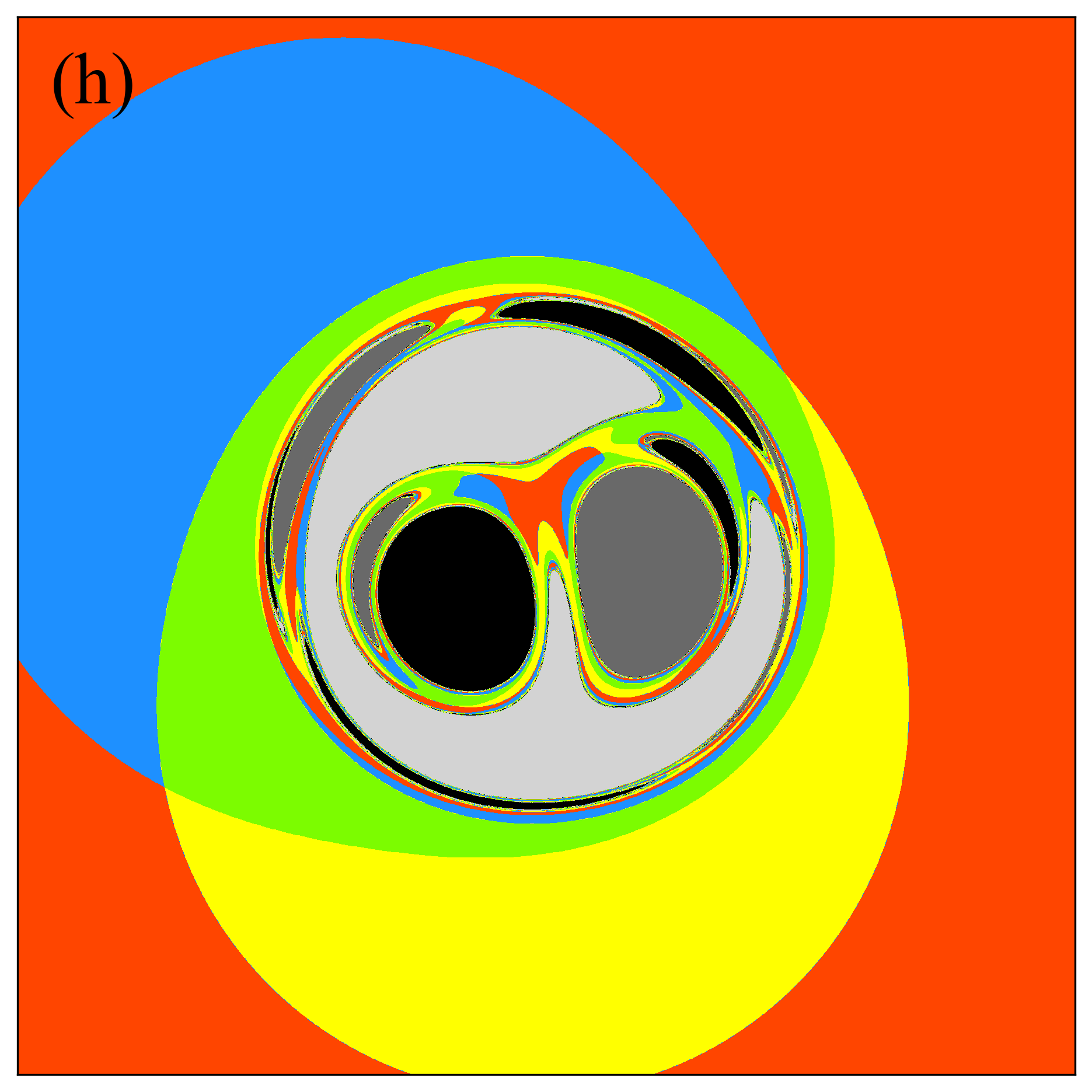}
\includegraphics[width=4.5cm]{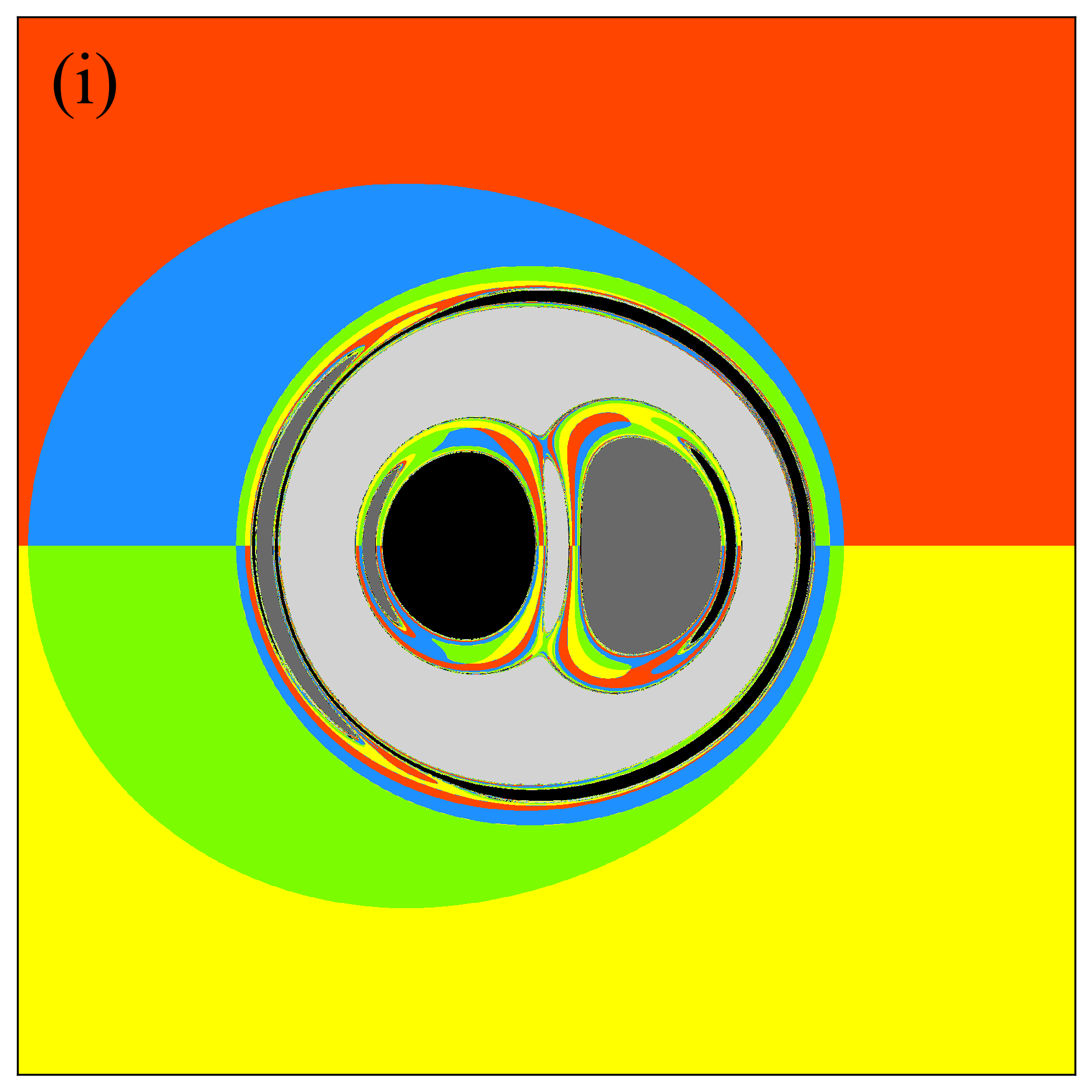}
\includegraphics[width=4.5cm]{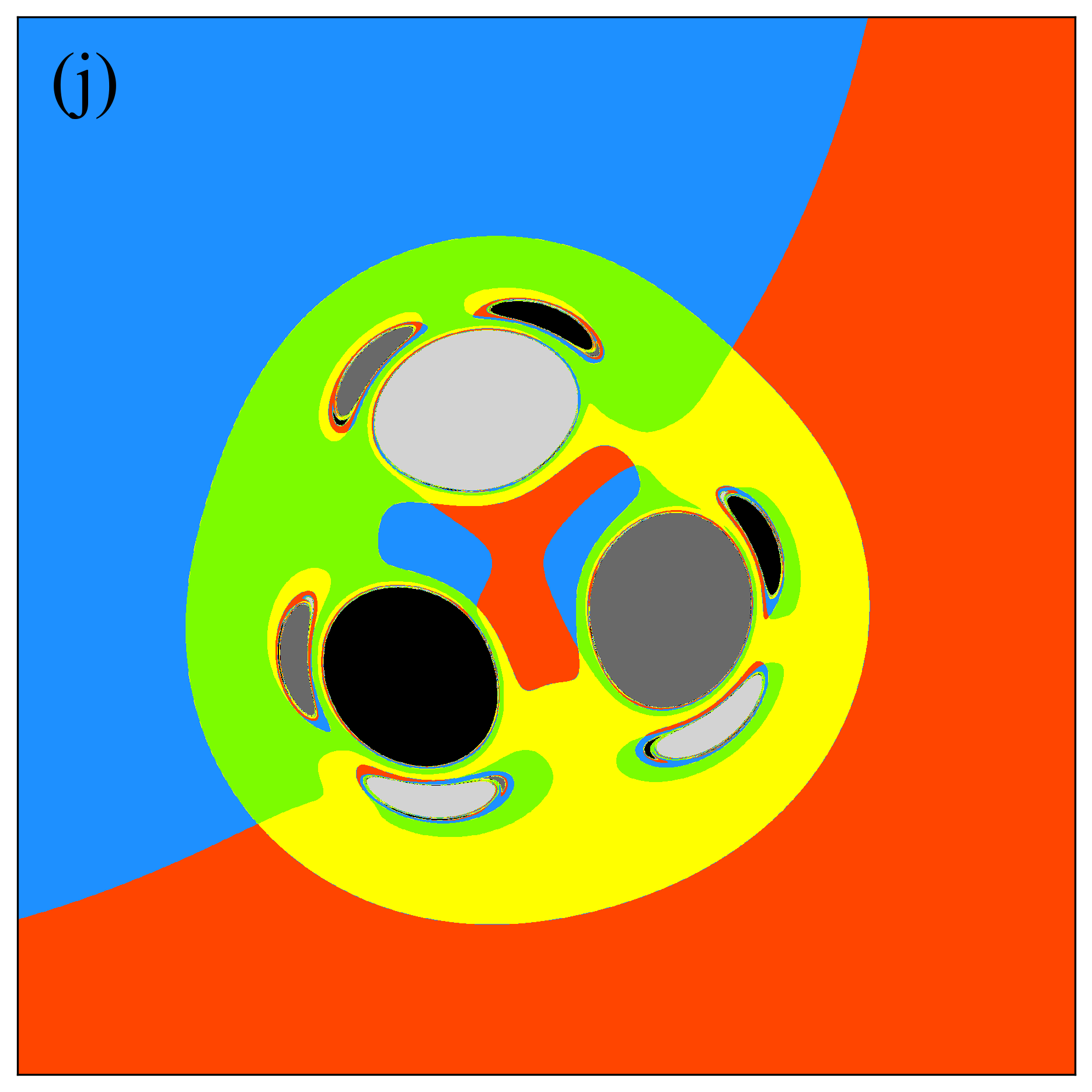}
\includegraphics[width=4.5cm]{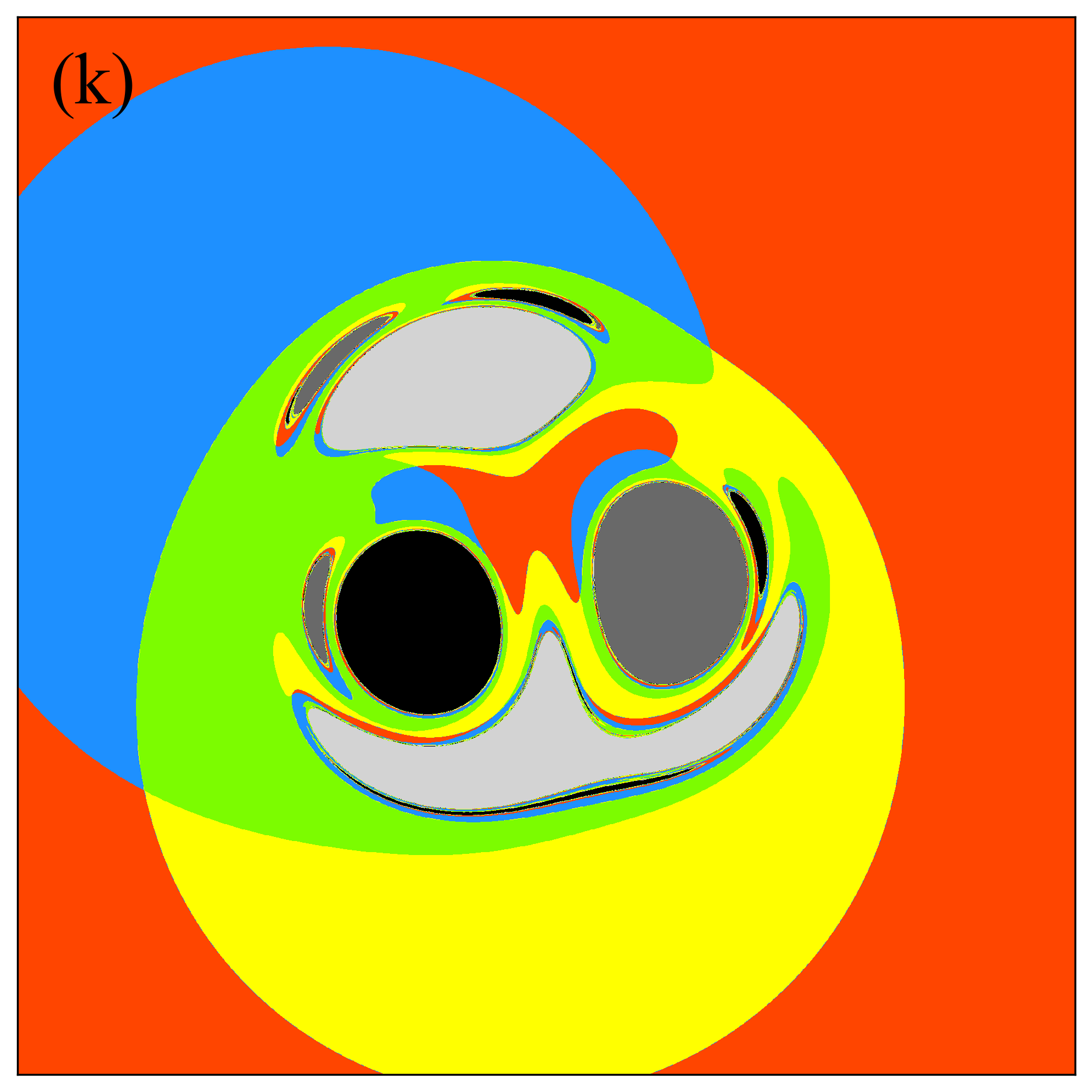}
\includegraphics[width=4.5cm]{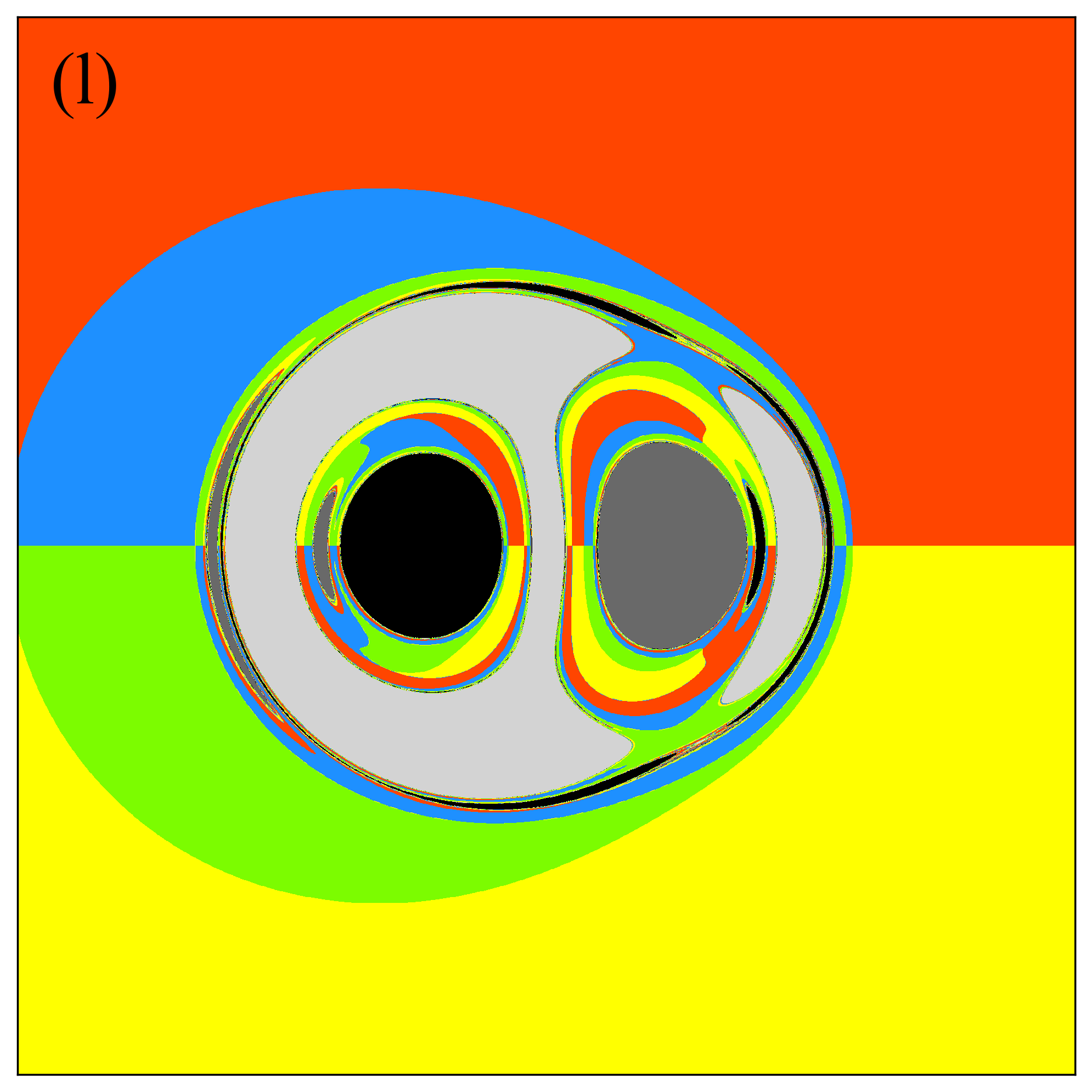}
\includegraphics[width=4.5cm]{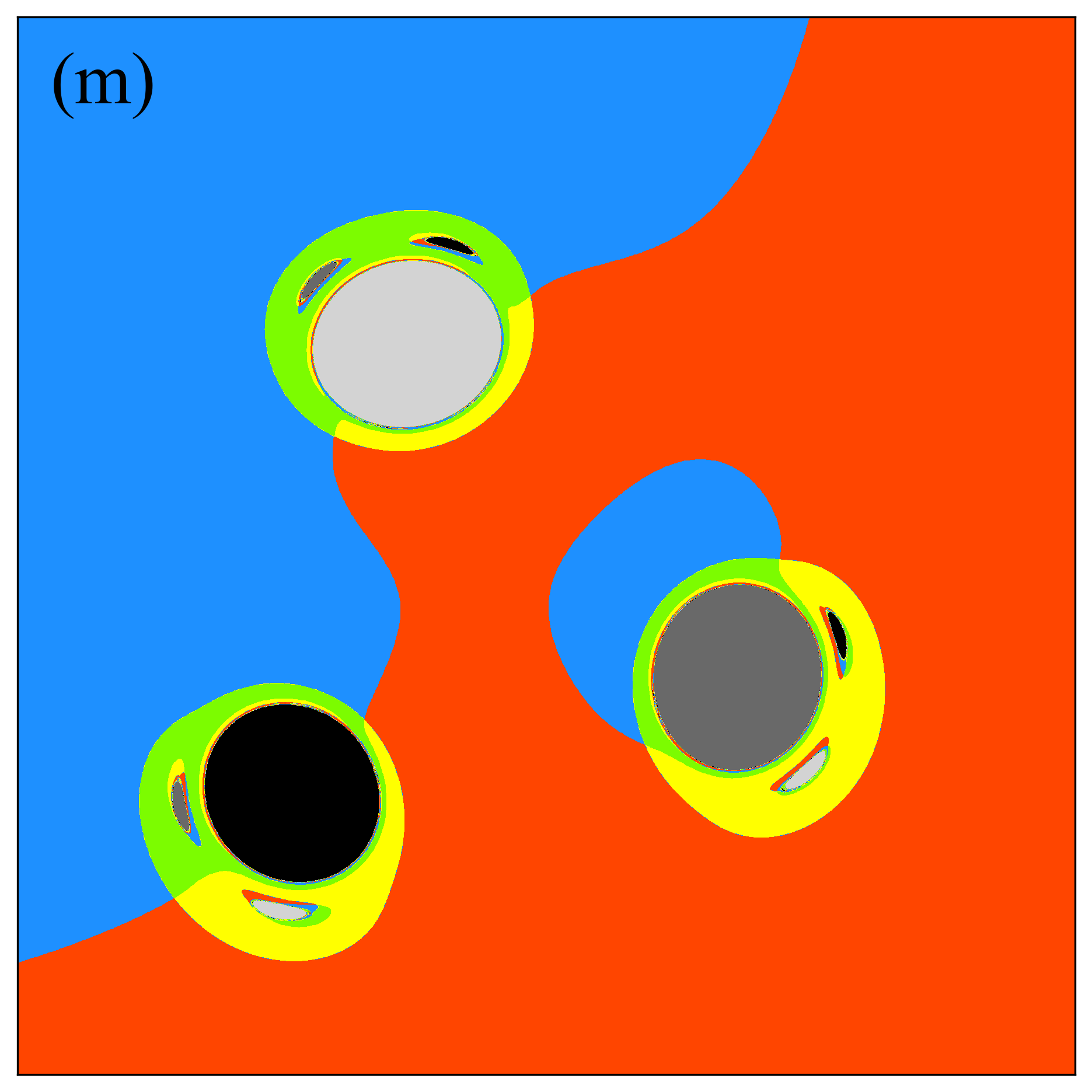}
\includegraphics[width=4.5cm]{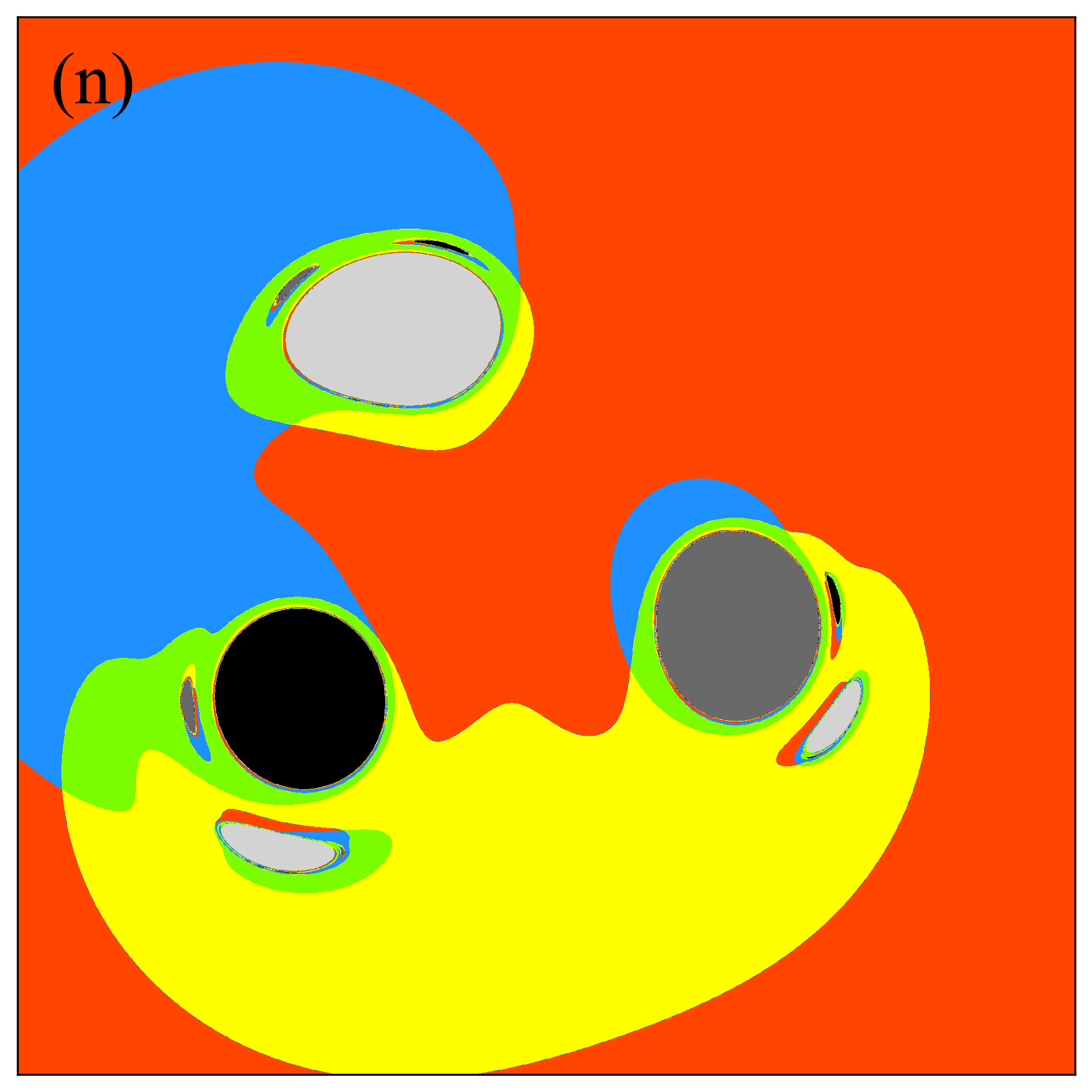}
\includegraphics[width=4.5cm]{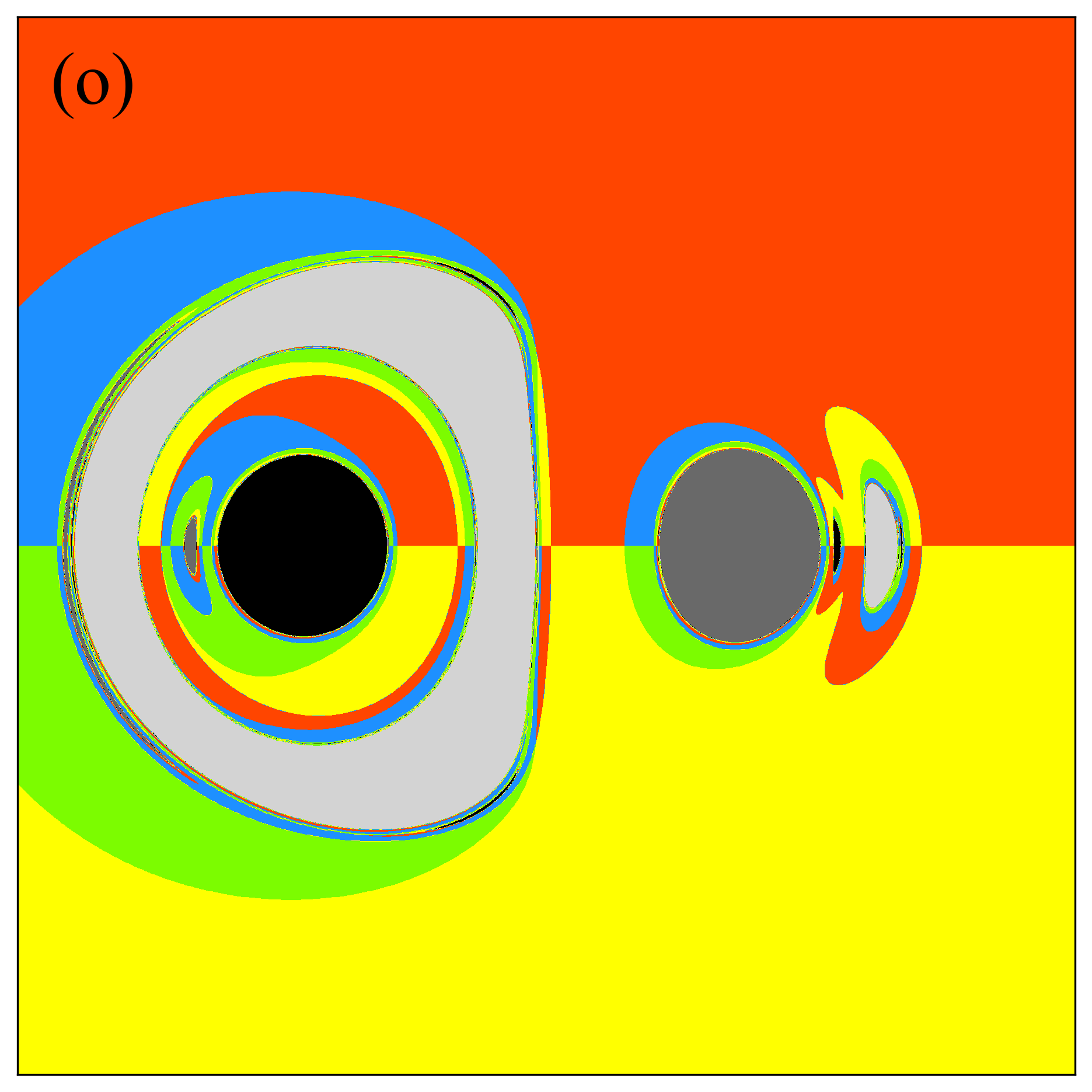}
\caption{Similar to Fig. 8, but for an azimuthal angle of $\Phi = 45^{\circ}$.}}\label{fig9}
\end{figure*}
\section{Conclusions and Discussions}
By utilizing the Majumdar-Papapetrou solution, we numerically simulated the shadows of three black holes in static equilibrium. We found that the contours of the three black holes' shadow are influenced by the black hole separation, spatial arrangement, and observational parameters. When the three black holes are collinear, each black hole exhibits not only a large, nearly circular primary shadow but also at least two secondary shadows with eyebrow-like features. It is noteworthy that eyebrow-like secondary shadows are not unique to triple black hole systems. Such features have also been identified in simulations of binary black holes \cite{Nitta et al. (2011),Yumoto et al. (2012),Bohn et al. (2015),Cunha et al. (2018),Shipley and Dolan (2016),Moreira et al. (2025)}, scalar hairy black holes \cite{Cunha et al. (2015),Cunha et al. (2016),Gyulchev et al. (2024)}, and distorted single black holes \cite{Abdolrahimi et al. (2015),Grover et al. (2018)}. However, the number of eyebrow-like secondary shadows observed in these models is fewer than those reported in our current work. As the black hole separation decreases, both the size and shape of the primary shadows undergo substantial alterations, accompanied by elongation of the secondary shadows. This observation suggests that the shadow morphology can be used to probe three black holes and infer their separation. When the line of sight aligns with the three black holes, shadow rings induced by gravitational lensing are observable, with their size being dependent on the black hole separation. In configurations where the black holes form the vertices of an equilateral triangle, the shadow structure can still exhibit collinear-like features, such as eyebrow-like shadows and ring-shaped shadows. However, substantial differences remain between these images and those arising from collinear configurations. Hence, it is feasible to infer the geometric configuration of a triple black hole system from the observed shadows. Furthermore, regardless of the spatial distribution of the three black holes, we observed self-similar fractal structures at the boundaries of both the primary and secondary shadows, which arise from chaotic scattering of the photon due to the non-integrability of spacetime.

Additionally, we found that when the black hole separation becomes sufficiently small, the shadows of three black holes merge into a standard disk, which can be cast by a single massive, spherically symmetric black hole. This suggests a degeneracy in the light-capturing ability between three sufficiently close extremal RN black holes and a single black hole. This raises the intriguing possibility that an astrophysical black hole observed in the universe might, in fact, be the collective manifestation of multiple closely packed extremal RN black holes.

This study focused primarily on the influence of black hole separation, configuration, and observational parameters on the shadow structure, which not only reveals the connection between parameters and shadows, but also reflects the potential shadow evolution of the three black holes during the process of merging and inspiral. In addition, the Majumdar-Papapetrou solution also incorporates mass parameters, which warrants further exploration of the shadow characteristics of three black holes with varying masses. This will be addressed in future work.
\section*{Acknowledgments}
The authors are very grateful to the referee for insightful comments and valuable suggestions. This research has been supported by the National Natural Science Foundation of China [Grant No. 12403081].

\end{document}